\documentclass[aps,pre,twocolumn,amsmath,amssymb,floatfix,eqsecnum]{revtex4-1}
\pdfoutput=1
\usepackage{bm}
\usepackage{color}
\usepackage{graphicx,epsfig}

\newtheorem{theorem}{Theorem}[section]
\newtheorem{corollary}{Corollary}[section]

\newcommand{\sgn}{\text{sgn}}
\newcommand{\tb}{\mathring{\tau}}
\newcommand{\tl}{\tb}
\newcommand{\tlc}{\tb_{\mathrm{c}}}
\newcommand{\gl}{g}

\newcommand{\gammaE}{\gamma_{\text{E}}}

\newcommand{\rmd}{\mathrm{d}}
\newcommand{\rme}{\mathrm{e}}
\newcommand{\rmi}{\mathrm{i}}

\newcommand{\Tr}{\mathop{\mathrm{tr}}}
\newcommand{\fex}{f^{\mathrm{ex}}}
\newcommand{\fres}{f^{\mathrm{res}}}
\newcommand{\fb}{f_{\mathrm{b}}}
\newcommand{\fs}{f_{\mathrm{s}}}
\newcommand{\Vb}{V_{\mathrm{b}}}
\newcommand{\tc}{T_{\mathrm{c}}}

\newcommand{\catalan}{G}
\newcommand{\Htheta}{\theta}
\newcommand{\spek}{\mathrm{spec}}
\newcommand{\HT}{\mathrm{HT}}
\renewcommand{\Re}{\text{Re}} \renewcommand{\Im}{\text{Im}}
\newcommand{\oneigfct}{\mathfrak{f}}
\newcommand{\noneigfct}{\mathfrak{\varphi}}
\newcommand{\regsol}{\varphi} 

\newcommand{\zm}{\mathsf{z}} \newcommand{\zL}{\mathfrak{z}}
\newcommand{\sk}{\mathsf{k}} \newcommand{\sko}{\mathring{\sk}}
\newcommand{\kL}{\mathfrak{k}}
\newcommand{\normreg}{\varkappa}

\definecolor{dblue}{rgb}{0.1,0.1,0.44}
\definecolor{dgreen}{rgb}{0.2 ,0.54, 0.2}

\newcommand{\mat}[1]{\mathbf{#1}}
\newcommand{\D}[1]{\Delta_{\mathrm{C}#1}}

\newcommand{\C}[1]{{}}

\newcommand{\aseq}{\simeq}
\newcommand{\asprop}{\sim}
\newcommand{\etwa}{\approx}

\newcommand{\ellipK}{K}

\begin{document}

\title{Inverse scattering-theory approach to the exact large-$n$ solutions of  $O(n)$ $\phi^4$ models on films and semi-infinite systems bounded by free surfaces }

\author{Sergei B. Rutkevich and H.~W. Diehl}
\affiliation{Fakult\"at f\"ur Physik, Universit\"at Duisburg-Essen, D-47058 Duisburg, Germany}

\date{\today}

\begin{abstract}
The $O(n)$ $\phi^4$ model on a strip bounded by a pair of planar free surfaces at  separation $L$ can be solved exactly in the large-$n$ limit in terms of the eigenvalues and eigenfunctions of a  self-consistent one-dimensional Schr\"odinger equation. The scaling limit of a continuum version of this model is considered. It is shown that the self-consistent potential can be eliminated in favor of scattering data by means of appropriately extended methods of inverse scattering theory. The scattering data (Jost function) associated with the self-consistent potential are determined for the ${L=\infty}$ semi-infinite case in the scaling regime for all values of the temperature scaling field $t=(T-\tc)/\tc$ above and below the bulk critical temperature $\tc$. These results are used in conjunction with semiclassical and boundary-operator expansions and a trace formula to derive exact analytical results for a number of quantities such as two-point functions, universal amplitudes of two excess surface quantities, the universal amplitude difference associated with the thermal singularity of the surface free energy, and potential coefficients. The asymptotic behaviors of the scaled eigenenergies and eigenfunctions of the self-consistent Schr\"odinger equation as function of $x= t(L/\xi_+)^{1/\nu}$  are determined for $x\to-\infty$.  In addition, the asymptotic ${x\to -\infty}$ forms of the  universal finite-size scaling functions $\Theta(x)$ and $\vartheta(x)$ of the residual free energy and the Casimir force are computed exactly to order $1/x$, including their $x^{-1}\ln|x|$ anomalies. 
\end{abstract}

\maketitle

\section{Introduction}

In the vicinity of critical points of systems undergoing  continuous phase transitions fluctuations occur on length scales ranging from microscopic separations up to the correlation length $\xi$. If the dimensionality of the system is below the upper critical dimension $d^*$ above which the Ginzburg-Levanyuk criterion \cite{Lev59,Gin60}  holds arbitrarily close to the critical point, the fluctuations on all such scales  affect the long-distance behavior in a nontrivial fashion \footnote{For background on critical behavior and the renormalization group, see e.g.\ \cite{Fis83,WK74,Fis74,DG76,Fis98}.}. \nocite{Fis83,WK74,Fis74,DG76,Fis98} The renormalization group (RG) has provided an appropriate conceptual framework for dealing with such problems involving many length scales and led to the development of powerful calculational tools for their quantitative investigation \cite{Fis83,WK74,Fis74,DG76,Fis98,PV02,MZ03}.

When such fluctuations in a near-critical medium are confined by external boundaries (walls) or macroscopic bodies immersed into it, effective forces are induced between the walls and these objects. The theory of such fluctuation-induced critical forces (critical ``Casimir forces'' \footnote{For background on critical Casimir forces and lists of references, see e.g.\ \cite{Kre94,KG99,BDT00,Gam09}.}) \nocite{Kre94,KG99,BDT00,Gam09} is substantially harder than the theory of bulk critical behavior because it involves a number of additional challenges. First and foremost, beyond bulk critical behavior, boundary and finite-size critical behaviors must be treated in an adequate manner. Second, difficult dimensional crossovers are typically encountered, which perturbative RG approaches do not normally capture \cite{KD91,KD92a,DGS06,MGD07,ZSRKC07,GD08,SD08,DG09,DS11,Doh14}. A particularly demanding case is that of three-dimensional systems whose large-scale physics can be represented by a $\phi^4$ model with an $O(n)$ symmetric Hamiltonian in a geometry that involves a crossover to two-dimensional behavior, such as a slab of size $\infty^2\times L$ whose width $L$ is finite. Provided the boundary conditions along the finite direction do not explicitly break the $O(n)$ symmetry, we know from the Mermin-Wagner theorem \cite{MW66,JF71,MW94} that for finite $L$ the system cannot exhibit long-range order at temperatures $T>0$. Thus the low-temperature behavior strongly interferes with the dimensional crossover. 

This combination of hard problems has hampered the design and application of satisfactory analytical theories. Exact solutions of appropriate models can --- and have --- provided helpful guidance and benchmarks for approximations. An example is Danchev's exact solution of the $O(n)$ $\phi^4$ model on a cylinder of circumference $L$, i.e., a slab of thickness $L$ subject to periodic boundary conditions 
\cite{Dan96,Dan98}. Its incompatibility with the $n$-dependence obtained for the critical Casimir force by naive extrapolation of the $\epsilon$-expansion results of \cite{KD91,KD92a} to $d=3$ dimensions \cite{BDT00,GD08} strongly hinted at a breakdown of the $\epsilon$ expansion \cite{DGS06,DS11}.

Unfortunately, real experimental systems of finite size usually involve free rather than periodic boundary conditions (pbc). In the present paper we shall be concerned with the exact $n\to\infty$ solution of the $O(n)$ $\phi^4$ theory on a slab of size $\infty^2\times L$ subject to free boundary conditions along the finite direction (called $z$-direction henceforth).  Unlike the case of pbc, where the $n\to\infty$ limit leads to a translation-invariant constraint Gaussian model equivalent to the spherical model \cite{Sta68}, the breakdown of translation invariance along the $z$-direction due to the free boundary conditions (fbc) entails that the exact $n\to\infty$ solution involves a self-consistent Schr\"odinger equation with a $z$-dependent potential $V(z)$. Upon solving this self-consistency problem by numerical means, very precise results for the temperature-dependent scaling functions of the excess surface free energy  and the Casimir force could be obtained in \cite{DGHHRS12,DGHHRS14}.

The aim of this paper is to explore the potential of inverse scattering-theory methods \cite{FS63,CS89} for obtaining exact analytical information about the self-consistent potential $V(z)$ and the above-mentioned scaling functions for temperatures $T\ne \tc$ both for the semi-infinite case $L=\infty$ and for finite $L$. 

To put things in perspective, recall that  the available  exact analytical knowledge about ${n\to\infty}$~solutions for fbc is rather scarce. Bray and Moore \cite{BM77a,BM77c} managed to  determine the self-consistent potential $V(z)$ for the semi-infinite case precisely at the critical point $T=\tc$. Building on this work, we computed in \cite{DR14} the correction to $V(z;{L=\infty},{t=0})$ linear in $t\propto (T-\tc )/\tc$ and combined it with results deduced via boundary-operator  and operator-product expansions to work out a few other exact analytical properties. These were accurately confirmed by the numerical results of \cite{DGHHRS12,DGHHRS14}, along with the exact analytical information about the low-temperature behavior of the Casimir force derived in the latter one of these two papers.

The basic idea of our subsequently developed approach based on inverse scattering-theory methods is to eliminate the potential $V(z)$ in favor of scattering data. Since the self-consistency equation for $V(z)$ corresponds to a stationarity condition for the free-energy density, one can exploit the latter to determine the scattering data from the corresponding variational equations. Proceeding in this way in the semi-infinite case enables us to get the scattering data for all temperatures $t\gtreqqless 0$. In this manner, the determination of $V(z)$ can be by-passed, although $V(z;{L=\infty},t)$ can in principle be reconstructed as the solution to an integral equation analogous to those of Gelfand, Levitan, and Marchenko \cite{FS63,CS89}. From Bray and Moore's solution \cite{BM77a,BM77c} it follows that the potential becomes singular at the boundary planes \cite{DR14}. Although singular potentials were considered in some inverse scattering-theory investigations \cite{CS89,FY05}, the particular kind of boundary singularities that the self-consistent potential $v(z)$ exhibits at $d=3$ has not yet been investigated and requires appropriate modifications of the established inverse scatterin theory. The necessary extensions of the latter are described in a separate, accompanying paper \nocite{RD14b} \cite{*[{Accompanying  paper: }] [{, referred to as II.}] RD14b} (referred to as II henceforth). Here the results required from II will simply be stated and applied.

The remainder of this paper is organized as follows. Although our ultimate interest is in the $n\to\infty$ solution of the continuum $\phi^4$ model on a strip of size $\infty^2\times L$, we start in the next section with a discretized version of it, namely, the lattice $\phi^4$  model called ``model B'' in  the numerical analyses of \cite{DGHHRS12,DGHHRS14}. The chosen discretization serves to avoid any ultraviolet (UV) (bulk and surface) singularities. We then recall the exact ${n\to\infty}$ solution of the model in terms of the eigenenergies and eigenstates of a self-consistent one-dimensional Schr\"odinger equation,  discuss the simplifications  that can be achieved by taking the limit  $g\to\infty$ of the $\phi^4$ coupling constant and the addition of appropriate irrelevant interactions, and reformulate the self-consistency equation via Green's functions. In Sec.~\ref{sec:scall} we turn to the analysis of the continuum limit of the model and its self-consistency equation, recall known properties of the self-consistent scaling solutions for the potential and required background on the thermal singularities of the surface free energy and excess energy, their logarithmic anomalies, and an associated universal amplitude difference. In Sec.~\ref{sec:ist}, we first provide the necessary background on inverse scattering theory, next  use it to reformulate the self-consistency equations in terms of scattering data rather than the potential, and then determine the scattering data (scattering amplitudes and phase shifts) for temperatures $t>0$, $t=0$, and $t<0$.

In Sec.~\ref{sec:istres2pt} these data are exploited to obtain exact results for the asymptotic large-distance scaling forms of the two-point order-parameter correlation function at, above, and below $\tc$. In Sec.~\ref{eq:potrelq} the exact scattering data are used in conjunction with a trace formula, semiclassical expansions, and perturbative asymptotic solutions of the Schr\"odinger equation to determine the limiting value $v_0$ of the regular part of the self-consistent potential at the boundaries, universal amplitudes associated with the thermal singularities of the surface excess energy, and the surface excess squared order parameter for $t<0$.

Sec.~\ref{sec:aslowt} deals with the asymptotic behavior of the universal scaling functions $\Theta(x)$ and $\vartheta(x)$ of the $L$-dependent part of the surface free energy per boundary area and the critical Casimir force in the low-temperature scaling limit $x=t(L/\xi_+)^{1/\nu}\to-\infty$, where $\nu=1$ is the critical exponent of the bulk correlation length $\xi$  and $\xi_+$  its nonuniversal amplitude for $t\to 0+$. This requires the determination of the ${x\to-\infty}$ behavior of the eigenenergies and the eigenfunctions of the self-consistent Schr\"odinger equation. Subsequently, the universal amplitudes of  the logarithmic anomalies $x^{-1}\ln|x|$ and the $x^{-1}$ terms are computed. The first agree with the results previously derived from a nonlinear sigma model in \cite{DGHHRS14}; the latter is new. Section~\ref{sec:SumConcl} contains a brief summary of our main results and concluding remarks. Finally, there are 7 appendixes describing technical details of some of the required calculations.

\section{The lattice $\phi^4$ model and its $n\to\infty$ limit }\label{sec:latmod}
\subsection{Definition of the lattice model}

We consider a lattice $\phi^4$ model on a three-dimensional slab  with the $O(n)$-symmetric Hamiltonian
\begin{equation} \label{eq:Hl}
\mathcal{H}_{\mathrm{l}}=\sum_{\bm{j}}\bigg[
\frac{1}{2} \sum_{i=1}^3  (\bm{\phi}_{\bm{j}+\bm{e}_i}- \bm{\phi}_{{\bm{j}}})^2
+\frac{\tl }{2} \bm{\phi}_{{\bm{j}}}^2
+\frac{\gl}{4! n}
 |\bm{\phi}_{{\bm{j}}}|^4\bigg],
\end{equation}
where ${\bm{j}}=(j_1,j_2,j_3)\in\mathbb{Z}^3$ with $1\le j_i\le N_i$, $i=1,2,3$, labels the sites of a finite simple cubic (sc) lattice whose lattice constant we denote as $a$ \footnote{Note that the interaction constants $\tb$ and $g$ of this lattice model are dimensionless, just as the lattice field $\Phi_{\bm{j}}$ is. These interaction constants must not be confused with their dimensionful analogs of the continuum model considered in \cite{DGHHRS14}.}. For the sake of brevity, we will write $\bm{j}_\parallel =(j_1,j_2)$,  $j_3=j$, and $N_3=N$ henceforth.   Each $\bm{\phi}_{{\bm{j}}}$ is an $n$-vector spin $\in\mathbb{R}^n$ of unconstrained length, and the $\bm{e}_i$ represent orthonormal unit vectors pointing along the principal directions $i$ of the lattice.

Along the first two directions we choose pbc: 
\begin{subequations}\label{eq:lbc}
\begin{equation}\label{eq:lpbc12}
\bm{\phi}_{{\bm{j}}+N_i\bm{e}_i}=\bm{\phi}_{{\bm{j}}}\quad\text{for } i=1,2.
\end{equation}
Along the third one, we impose Dirichlet boundary conditions, requiring
\begin{equation} \label{eq:lDbc}
\bm{\phi}_{{\bm{j}_\parallel},j}=\bm{0} \quad\text{ for } j=0,N+1.
\end{equation}
\end{subequations}

In conjunction with the chosen interactions in $\mathcal{H}_{\mathrm{l}}$, the prb~\eqref{eq:lpbc12}  imply  that the model has translation invariance with respect to lattice translations along the directions $i=1,2$. By contrast, the Dirichlet boundary conditions~\eqref{eq:lDbc} break the corresponding discrete translation invariance along  the  direction ($i=3$) normal to the boundary planes $j=0,N$. 

The latter breakdown of translation invariance generically occurs for models with free surfaces perpendicular to the $3$-direction. As a simple and natural generalization one might want to consider analogs of our model~\eqref{eq:Hl} where the strength of the coupling between $\bm{\phi}_{\bm{j}}$ and $\bm{\phi}_{\bm{j}'}$ on nearest-neighbor (NN) sites has been changed from the uniform value $K=1$ to different ones $K_1$ and $K_2$ in the layers $j=1$ and $j=N$, respectively (cf.\ the standard semi-infinite lattice models reviewed in \cite{Bin83,Die86a}). However, this is unnecessary. Since a phase with long-range surface order is ruled out  in the bulk limit $N_i\to\infty$, $i=1,2,3$, for all temperatures $T>0$ and arbitrary finite values of $K_1/K$ and $K_2/K$  by extensions of the Mermin-Wagner theorem \cite{MW66,JF71,MW94}, only the ordinary surface transition \cite{Bin83,Die86a,Die97} remains for this generalized three-dimensional model in the semi-infinite case. It is well established \cite{DD80,DD81a,DD81b,DD83a,Die86a,Die97,DDE83,DS94,DS98} that the  universal surface critical behavior at this transition is described by a continuum field theory satisfying large-scale Dirichlet boundary conditions. Deviations from this boundary condition induced by modified surface interactions (which entail Robin boundary conditions for the continuum theory) are irrelevant in the RG sense. As is expounded in \cite{DGHHRS14}, this irrelevance can  be checked explicitly in the large-$n$ theory for a three-dimensional film of finite thickness $L=Na$. For the sake of simplicity, we will therefore restrict ourselves to the above-defined model with $K=K_1=K_2$ and the boundary conditions~\eqref{eq:lbc}.

For later use, we introduce the partition function of the model
\begin{equation}\label{eq:Z}
\mathcal{Z}=\int\mathcal{D}[\bm{\phi}]\,\rme^{-\mathcal{H}[\bm{\phi}]}
\end{equation}
and define the $n\to\infty$ limit of the free energy per number of boundary sites and number of components by
\begin{equation} \label{eq:fNdef}
f_{N}=-\lim_{n\to\infty}\,\lim_{N_1,N_2\to\infty}\frac{\ln\mathcal{Z}}{n\, N_1\,N_2}\;.
\end{equation}

We also introduce the associated bulk, excess, surface,  and residual free energy densities by
\begin{eqnarray}\label{eq:fdefs}
\fb(T)&\equiv &\lim_{N\to\infty}f_N(T)/N,\nonumber\\
\fex_N(T)&\equiv& f_N-N\fb(T),\nonumber\\
\fs(T)&\equiv&\fex_\infty(T)/2,\nonumber\\
\fres_N(T)&\equiv&\fex_N(T)-2f_{\mathrm{s}}(T),
\end{eqnarray}
respectively, and the Casimir force 
\begin{equation}\label{eq:betaFCdef}
\beta \mathcal{F}_{\mathrm{C}}(T,N)=-\frac{\partial}{\partial N}\fres_N(T)=-\frac{\partial}{\partial N}\fex_N(T).
\end{equation}

Near the bulk critical temperature $\tc$, the asymptotic behaviors of the singular parts of the above $N$-dependent functions on long length-scales are described by familiar scaling forms. Let us introduce a dimensionless temperature variable $t=\text{const}\,(T/\tc-1)$ and fix the proportionality constant by absorbing in it the nonuniversal amplitude $\xi_+$ of the bulk correlation length \footnote{Following the conventions of \cite{DGHHRS12} and \cite{DGHHRS14}, we use the symbols $\aseq$ and $\asprop$ to denote asymptotic equality and asymptotic proportionality, respectively.} 
\begin{equation}
\xi(t)\aseq \xi_+(T/\tc-1)^{-\nu},\quad T\ge \tc,
\end{equation}
in the disordered phase, defining
\begin{equation}
t=\mathrm{sgn}(t)[\xi(|t|)/\xi_+]^{-1/\nu}
\end{equation}
and the scaling variable
\begin{equation}
x=tN^{1/\nu}.
\end{equation}

Following the conventions of \cite{KD91,DGHHRS12,DGHHRS14}, we write the scaling forms of the residual free energy and the Casimir force as
\begin{equation}\label{eq:fresscf}
\fres_N(T)\aseq N^{-(d-1)}\,\Theta(x)\,.
\end{equation}
and
\begin{equation}\label{eq:FCscalf}
\beta \mathcal{F}_\mathrm{C}(T,L) \aseq N^{-d}\,\vartheta(x).
\end{equation}
The scaling function $\vartheta(x)$ is related to $\Theta(x)$ via (see, e.g., \cite{KD91,SD08,DS11,DGHHRS12,DGHHRS14})
\begin{equation}\label{eq:relvarthetaTheta}
\vartheta(x)=(d-1)\,\Theta(x)-\frac{x}{\nu}\,\Theta'(x).
\end{equation}
Furthermore, the value of the  function $\Theta(x)$ at $x=0$, which measures the strength of the critical Casimir force, is the so-called Casimir amplitude
\begin{equation} \label{eq:DeltaCdef}
\D{} = \Theta(0).
\end{equation}
Since we have fixed the scale of $x$ and no other (e.g., surface-related) macroscopic lengths are present, the functions $\Theta$ and $\vartheta$ are universal \footnote{Deviations from Dirichlet boundary conditions give rise to corresponding surface scaling fields for either surface plane. In the $d=3$~dimensional case with $n=\infty$, such deviations are irrelevant since no special transition occurs \cite{DGHHRS12,DGHHRS14}.}.

\subsection{Large-$n$ self-consistency equations}

The techniques for deriving the equations that govern the $n\to\infty$ limit of models such as ours are well established (see, e.g., \cite{MZ03}, \cite{Eme75}, \cite{Ami89}, Appendix B of \cite{BDS10}, and \cite{DGHHRS12}). For the case of the lattice model defined above, they have been explicitly given in \cite{DGHHRS14}. Hence we can be brief and list them in the slightly different notation preferred here. 

Let us disregard for the moment the possibility that the symmetry is spontaneously broken in the bulk limit $N_i\to\infty$, $i=1,2,3$, focusing on the disordered phase. Then, in the limit $n\to\infty$, the lattice model~\eqref{eq:Hl} is equivalent to $n$ copies of a constrained Gaussian model for a one-component field $\Phi_{\bm{j}}$ with the Hamiltonian
\begin{equation} \label{eq:HG}
\mathcal{H}_{\text{G}}=\frac{1}{2}\sum_{\bm{j}}\bigg[
\sum_{i=1}^3  ({\Phi}_{{\bm{j}}+\bm{e}_i}- {\Phi}_{{\bm{j}}})^2
+V_j\Phi_{{\bm{j}}}^2-\frac{3}{ \gl}\,(V_{{j}}-\tl  )^2
\bigg].
\end{equation}
Here ${\Phi}_{\bm{j}}\equiv{\phi}_{{\bm{j}},1}$   and $V_{j}$ is a self-consistent potential satisfying the constraint
\begin{equation} \label{eq:Vs}
V_j=\tl  +\frac{\gl}{6 } \langle \Phi_{\bm{j}_\|,j}^2\rangle=\tl  +\frac{\gl}{6} \langle 
\Phi_{\bm{0},j}^2\rangle.
\end{equation}

Upon taking the limits $N_1,N_2\to\infty$, the self-consistency equation implied by Eqs.~\eqref{eq:HG} and \eqref{eq:Vs} becomes
\begin{equation} \label{eq:Vsce}
\tl  -V_{j}=-\frac{\gl}{6 }\, \sum_{\nu=1}^N W_2(\varepsilon_\nu)\,|\oneigfct_\nu(j)|^2=-\frac{g}{6}\,\langle  j|W_2(\mat{H})|j\rangle,
\end{equation} 
where $W_2(\lambda)$  is a particular one of the Watson integrals  \cite{JZ01}
\begin{equation}\label{eq:Wddef}
W_d(\lambda)\equiv\int_0^\pi\frac{\rmd{q}_1}{\pi}\dotsm
\int_0^\pi\frac{\rmd{q}_d}{\pi}\frac{1}{\lambda+4\sum_{i=1}^d\sin^2(q_i/2)}.
\end{equation}
The index $\nu=1,\dotsc,N$ in Eq.~\eqref{eq:Vsce} labels the eigenvalues $\varepsilon_\nu$ and orthonormalized eigenstates $\oneigfct_\nu(j)=\langle j|\nu\rangle$ of the discrete Sturm-Liouville problem 
\begin{widetext}
\begin{equation} \label{eq:StL}
\sum_{j'=1}^NH_{j,j'}\oneigfct_\nu(j')\equiv V_{j}\oneigfct_\nu(j)+\left[2\oneigfct_\nu(j)-\oneigfct_\nu(j-1)-\oneigfct_\nu(j+1)\right]=
\varepsilon_\nu \oneigfct_\nu(j), \quad
j=1,\ldots,N.
\end{equation}
\end{widetext}
The latter must satisfy the Dirichlet boundary conditions
\begin{equation}\label{eq:DbcH}
\oneigfct_\nu(0)= \oneigfct_\nu(N+1)=0. 
\end{equation}
We choose them real-valued, so that their orthonormality relations become 
\begin{equation}\label{eq:orthonorm}
\langle\nu|\nu'\rangle=\sum_{j=1}^{N}\oneigfct_\nu(j)\,\oneigfct_{\nu'}(j)=\delta_{\nu,\nu'}.
\end{equation}
The coefficients $H_{j,j'}$ in Eq.~\eqref{eq:StL} correspond to the elements of an $N\times N$ tridiagonal matrix ``Hamiltonian'', namely 
\begin{equation}\label{eq:Hmat}
\mat{H}=\begin{pmatrix}
2+V_1&-1&&\\
-1&\ddots&\ddots&\\
&\ddots&\ddots&-1\\
&&-1&2+V_{N}
\end{pmatrix}.
\end{equation}
In order that the free energy $f_N$ be well defined, $\mat{H}$ must be positive definite, i.e., we must have $\varepsilon_\nu>0$ for all $\nu=1,\dotsc,N$.

Note that the trace of $\mat{H}$ is simply related to the trace of the diagonal potential matrix $\mat{V}=(V_j\,\delta_{j,j'})$. One has
\begin{equation}\label{eq:TrHV}
\Tr\mat{H}=2N+\Tr\mat{V}.
\end{equation}
Further, by symmetry, the self-consistent potential $V_{j}$ must be even under reflection about the midplane, i.e., 
\begin{equation}\label{mirror}
V_{j}=V_{N+1-j}.
\end{equation}

Straightforward evaluation of the free energy~\eqref{eq:fNdef} gives
\begin{equation} \label{eq:fN1}
f_{N}=\frac{1}{2}\sum_{\nu=1}^{N}U_2(\varepsilon_\nu)-
\frac{3}{2 \gl}\sum_{j=1}^{N}(V_{j}-\tl  )^2+f_{N}^{(0)}, 
\end{equation}
where $f_{N}^{(0)}$ is a trivial background term independent of $V_{j}$, which can be eliminated by a shift and is henceforth dropped. The function $U_d(\lambda)$ means the antiderivative
\begin{equation}\label{eq:U2def}
U_d(\lambda)=\int_0^\pi\frac{\rmd{q}_1}{\pi}\dotsm
\int_0^\pi\frac{\rmd{q}_d}{\pi}\ln\left\{\lambda+4\sum_{i=1}^d\sin^2(q_i/2)\right\}
\end{equation}
of the Watson integral~\eqref{eq:Wddef}. The explicit form of its derivative for $d=2$, $W_2(\lambda)$, in terms of a complete elliptic integral is given in Eq.~\eqref{eq:W2lambda} of Appendix~\ref{app:W2U2}. Integrating this result yields the explicit form of  $U_2(\lambda)$ given in Eq.~\eqref{eq:U2lambda}  (cf.\ Eq.~(48) of \cite{HGS11}, and \cite{Gut10}). We shall not work with the explicit forms of these functions, but will make use of some of their properties  below. We postpone a discussion of these properties for the time being.

As has been shown in \cite{DGHHRS14}, the free energy $f_N$ (with $f_{N}^{(0)}$ omitted) is given by the global  maximum of the functional $f_N[\mat{V}]$ of the potential $\mat{V}=(V_j)$ defined by the first two terms on the right-hand side of Eq.~\eqref{eq:fN1}. That is
\begin{equation}\label{eq:fNmax}
f_N=\max_{\mat{V}}f_N[\mat{V}]
\end{equation}
with
\begin{equation}\label{eq:fNV}
f_N(\mat{V})\equiv \frac{1}{2}\Tr\{U_2[\mat{H}(\mat{V})]\}-\frac{3}{2g}\Tr [(\mat{V}-\tb)^2],
\end{equation}
where we have explicitly indicated the dependence of the matrix Hamiltonian~\eqref{eq:Hmat} on $\mat{V}$. That the self-consistent potential corresponds to the  global maximum of $f_N(\mat{V})$ is a consequence of two facts: (i) $f_N(\mat{V})$ is concave in $\mat{V}$ because it is a difference of a concave and a convex function of $\mat{V}$ \footnote{Recall that a strictly concave  function $g(t)$ satisfies $g[\alpha t_1+(1-\alpha)t_2]>\alpha\, g(t_1)+(1-\alpha)\,g(t_2)$ for $0<\alpha<1$. The statements about the concavity of $\Tr U_2(\mat{H}[\mat{V}])$ and convexity of $\Tr (\mat{V}-\tl)^2$ hold in the domain of $\mat{V}$ where $\mat{V}$ and $\mat{H}[\mat{V}]$ are positive definite.}; (ii) the self-consistency Eq.~\eqref{eq:Vsce} is equivalent to the extremum condition
\begin{equation}\label{eq:fNxtrcond}
\frac{\partial f_{N}(\mat{V})}{\partial V_{j}}=0.
\end{equation}

Taking the bulk limit $N\to\infty$ in Eq.~\eqref{eq:Vsce} shows that the bulk critical value of $\tl$ is given by
\begin{equation}\label{tbulk}
\tlc=-\frac{\gl}{6}\,W_3(0).
\end{equation}

Following \cite{DGHHRS12,DGHHRS14}, we define a temperature variable $t$ into which the amplitude $\xi_+=\gl/(24\pi)$ of the bulk correlation length $\xi\aseq\xi_+/(\tl-\tlc)$ for deviations $ \tl-\tlc\to 0+$ is absorbed by
\begin{equation}\label{eq:tdef}
t=\frac{24\pi}{\gl}(\tl-\tlc)=24\pi\left[\frac{\tl}{\gl}+\frac{1}{6}\,W_3(0)\right].
\end{equation}
Owing to this definition, the bulk critical point is located at $t=0$.

\subsection{Simplifications of the equations}
\subsubsection{due to the limit $g\to\infty$}

Further simplifications can be achieved by taking the limit $\gl\to\infty$. In this limit, the model~\eqref{eq:Hl} goes over into a layered spherical model in which $\sum_{\bm{j}_\|}\langle\Phi_{\bm{j}_\|,j}^2\rangle/(N_1N_2)=\langle\Phi_{\bm{0},j}^2\rangle$ fulfills a separate constraint for each layer $j=1,\dotsc,N$. As we know from \cite{DGHHRS12,DGHHRS14}, taking the limit $\gl\to\infty$ causes a suppression of corrections to scaling, making it easier to extract the universal large-scale behavior.
To take this limit, we add a contribution regular in $t$ and define
\begin{eqnarray} \label{eq:finfNV}
f^{(\infty)}_{N}(\mat{V}, t)&\equiv&\lim_{\gl\to\infty}\bigg[f_N(\mat{V})+N\frac{\gl}{4!}\bigg(\frac{t}{4\pi}-W_3(0)\bigg)^2\bigg]\nonumber\\ 
&=&\frac{1}{2}\Tr U_2[\mat{H}(\mat{V})]+\frac{1}{2}\Big[\frac{t}{4\pi}-W_3(0)\Big]\Tr\mat{V},\nonumber\\
\end{eqnarray}
where we have explicitly  indicated the $t$-dependence of the limiting function, and the superscript $^{(\infty)}$ reminds us that $\gl$ has been set to $\infty$.

The ${\gl=\infty}$~analog of the self-consistency condition \eqref{eq:Vsce} follows from $\partial f^{(\infty)}_N(\mat{V},t)/\partial V_j=0$; it reads
\begin{equation} \label{eq:Scc1}
\frac{t}{4\pi}-W_3 (0)=-\langle j|W_2(\mat{H})|j\rangle.
\end{equation} 

Substitution of the solution $\mat{V}\equiv\mat{V}(t,N)$ of Eq.~\eqref{eq:Scc1} for given $t$ and $N$ that maximizes the functional~\eqref{eq:finfNV} gives us the ${\gl=\infty}$~analog $f^{(\infty)}_N(t)=f_N^{(\infty)}[\mat{V}(t,N),t]$ of the free energy $f_N$ in Eq.~\eqref{eq:fNmax}. Taking its temperature derivative yields the useful equation
\begin{equation}\label{eq:dfNdt}
\frac{\rmd f^{(\infty)}_{N}( t)}{\rmd  {t}}=\frac{1}{8\pi}
\Tr\mat{V}(t,N).
\end{equation}

Let us also list some known results \cite{DGHHRS14} for bulk quantities that will be needed below. The bulk value of the self-consistent potential, $\Vb(t)=\lim_{N\to\infty}V_N(t,N)$, maximizes the bulk analog
\begin{equation}\label{eq:fbVbt}
f_{\mathrm{b}}^{(\infty)}(\Vb,t)=\frac{1}{2}\,U_3[\Vb]+\Vb\left[\frac{t}{8\pi}-\frac{W_3 (0)}{2}\right]
\end{equation}
of the function $f^{(\infty)}[\mat{V},t]$, where we temporarily restrict ourselves to the disordered bulk phase $t\ge 0$.     When $t<0$ and $N=\infty$, one must allow for spontaneous symmetry breaking. Since the corresponding bulk results can be found in textbooks such as \cite{Ami89} and elsewhere (see, e.g., \cite{MZ03,DG76,DDG06}), there is no need to rederive known bulk results for $t<0$ here. Instead, we shall incorporate them into the $t\gtreqless 0$ results for $\Vb$ and $\fb^{(\infty)}$ given in Eqs.~\eqref{eq:Vbas} and \eqref{eq:fbsing} below.

The necessary condition $\partial \fb^{(\infty)}[\Vb,t]/\partial \Vb=0$ yields
\begin{equation}\label{eq:beqst}
W_3(\Vb)-W_3(0)=-\frac{t}{4\pi},\quad t\ge 0.
\end{equation}
The function $W_3(\lambda)$ is known to behave as \cite{Gut10}
\begin{equation}\label{eq:W3smalllambda}
W_3(\lambda)=W_3(0)-\frac{1}{4\pi}\sqrt{\lambda} +O(\lambda)
\end{equation}
for small $\lambda>0$. Using this together with the fact that 
$\Vb$ for $t<0$ corresponds to the inverse transverse bulk susceptibility (which vanishes on the coexistence line), one concludes that
\begin{equation} \label{eq:Vbas}
\Vb(t)= t^2\, \Htheta(t) +O(t^3),
\end{equation}
where $\Htheta(t)$ is the Heaviside step function. The familiar result \cite{MZ03,DGHHRS12,DGHHRS14,DR14} 
\begin{equation}\label{eq:fbsing}
\fb^{\text{sing}}(t)\aseq\frac{1}{24\pi}\,t^3\,\Htheta(t)
\end{equation}
for the leading thermal singularity  of $\fb$ can be recovered by integrating Eq.~\eqref{eq:W3smalllambda} to obtain $U_3(\lambda)$ for small $\lambda$ and substituting Eq.~\eqref{eq:Vbas} into Eq.~\eqref{eq:fbVbt}.

\subsubsection{due to an appropriate choice of $U_2$}

The above Eq.~\eqref{eq:finfNV} for the free-energy function and the self-consistency condition~\eqref{eq:Scc1} can be simplified further. This is because the functions $W_2$ and $U_2$ are expected to contain contributions that are irrelevant in the sense that their omission does not change the asymptotic large-scale behavior such as the leading thermal singularities of the bulk and surface free energies, and the scaling functions $\Theta(x)$ and $\vartheta(x)$ of the residual free energy and the Casimir force. Hence we should be able to replace $W_2$ and $U_2$ by appropriate simplified functions $\tilde{W}_2$ and $\tilde{U}_2$ that differ from $W_2$ and $U_2$ by such irrelevant contributions. To justify the choice of  $\tilde{W}_2$ and $\tilde{U}_2$  we are going to make below, we need some properties of their exact counterparts $W_2$ and $U_2$, which are established in Appendix~\ref{app:W2U2} and will now be discussed.

From Eq.~\eqref{eq:Wddef} or the closed-form expression~\eqref{eq:W2lambda}  one sees that $W_2(\lambda)$ is analytic in the complex $\lambda$-plane except for the branch cut $[-8,0]$. Arguments given in Appendix~\ref{app:W2U2} show that it can be written for small $|\lambda|$ as
\begin{equation}\label{eq:W2form}
W_2(\lambda)=-w_2(\lambda)\ln\lambda+R_2(\lambda),
\end{equation}
where both $w_2(\lambda)$ and $R_2(\lambda)$ are analytical functions at small enough $|\lambda|$. 
The former one is given by the spectral function
\begin{eqnarray}\label{eq:w2def}
w_2(\lambda)&\equiv&\frac{1}{2\pi\rmi}[W_2(\lambda-\rmi 0)-W_2(\lambda+\rmi 0)]\nonumber\\&=&-\frac{1}{\pi}\Im W_2(\lambda+\rmi 0),
\end{eqnarray}
which for real $\lambda$ characterizes the singularity across the branch cut. Integration of Eq.~\eqref{eq:W2form} shows that $U_2$ can be written as
\begin{equation}\label{eq:U2form}
U_2(\lambda)=\frac{\lambda(1-\ln\lambda)}{4\pi}[1+\lambda\,A(\lambda)]+B(\lambda)
\end{equation}
with regular functions $A(\lambda)$ and $B(\lambda)$. 

Let us introduce the analogs  
\begin{equation}
\tilde{W}_2(\lambda)=1-\frac{\ln\lambda}{4\pi},
\end{equation}
\begin{equation}
\tilde{w}_2(\lambda)=\frac{1}{4\pi},
\end{equation}
and
\begin{equation}
\tilde{U}_2(\lambda)=\lambda+\frac{1}{4\pi}\,\lambda(1-\ln\lambda)-2
\end{equation}
of the functions  $W_2$, $w_2(\lambda)$, and $U_2$ corresponding to the substitutions
\begin{equation}
\left.\begin{array}{l}A(\lambda)\\B(\lambda)\end{array}\right\}\to\left\{\begin{array}{l} \tilde{A}(\lambda)\equiv 0,\\ \tilde{B}(\lambda)\equiv \lambda-2,\end{array}\right.
\end{equation}
in Eqs.~\eqref{eq:W2form}--\eqref{eq:U2form}. Then the free energy function~\eqref{eq:finfNV} can be decomposed as
\begin{equation}\label{eq:finftydec}
f^{(\infty)}_N(t,\mat{V})=\tilde{f}_N(t,\mat{V})+\Tr R(\mat{H})
\end{equation}
into the contribution
\begin{equation}\label{eq:fNtilde}
\tilde{f}_N(t,\mat{V})=\frac{1}{8\pi}\left\{\Tr[\mat{H}(1-\ln\mat{H})]+t\Tr\mat{V}\right\}
\end{equation}
associated with $\tilde{U}_2$ and the remainder
\begin{equation}
R(\lambda)=\frac{1}{8\pi}A(\lambda)\lambda^2(1-\ln\lambda)+\frac{B(\lambda)}{2}+\frac{W_3(0)}{2}\,(2-\lambda).
\end{equation}
In the derivation of $\tilde{f}_N(t,\mat{V})$ we used Eq.~\eqref{eq:TrHV} and   anticipated that $\tilde{W}_3(\lambda)$, the analog of the function $W_3(\lambda)$, has the property
\begin{equation}\label{eq:W3tilde1}
\tilde{W}_3(0)=1.
\end{equation}
The latter follows from the explicit expression given below in Eq.~\eqref{eq:W3tilderes}.

The power series expansions of $A$ and $B$ can be determined in a straightforward fashion. Explicit results to low powers of $\lambda$ can be found in Eqs.~\eqref{eq:Aexp} and \eqref{eq:Bexp} of Appendix~\ref{app:W2U2}.

The functions  $\tilde{U}_3(\lambda)$ and $\tilde{W}_3(\lambda)$ associated with $\tilde{U}_2(\lambda)$ and $\tilde{W}_2$ can be obtained via the analog of the relation
\begin{eqnarray}\label{eq:U32}
U_3(\lambda)\equiv\int_{-\pi}^{\pi}\,\frac{\rmd{p}}{2\pi} \, U_2[\lambda+4\sin^2 (p/2)].
\end{eqnarray}
This gives
\begin{eqnarray}
\tilde{U}_3(\lambda)&=&\lambda+\frac{\sqrt{\lambda(4+\lambda)}}{4\pi}\nonumber\\&&\strut +\frac{2+\lambda}{4\pi}\ln\frac{2}{2+\lambda+\sqrt{\lambda(4+\lambda)}}
   \end{eqnarray}
and
\begin{equation}\label{eq:W3tilderes}
\tilde{W}_3(\lambda)=\tilde{U}'_3(\lambda)=1+\frac{1}{4\pi}\ln\frac{2}{2+\lambda+\sqrt{\lambda(4+\lambda)}}.
\end{equation}

Furthermore, the stationarity condition for $\mat{V}$ corresponding to the omission of $R(\mat{H})$ in Eq.~\eqref{eq:finftydec}, namely $\partial \tilde{f}_N(t,\mat{V})/\partial V_j=0$, takes the simple form
\begin{equation}\label{eq:simpsceV}
\langle j|\ln\mat{H}|j\rangle=t
\end{equation}
known from the analysis of model A in \cite{DGHHRS12,DGHHRS14}.

\subsection{Green's function reformulation of the self-consistency equation}

Being interested in the universal large-length-scale properties of the solutions to the above equations, we consider their behavior in an appropriate continuum (scaling) limit $a\to 0$. To this end,  their reformulation in terms of Green's functions turns out to be helpful. 

\begin{figure}[htbp]
\begin{center}
\includegraphics[width=\columnwidth]{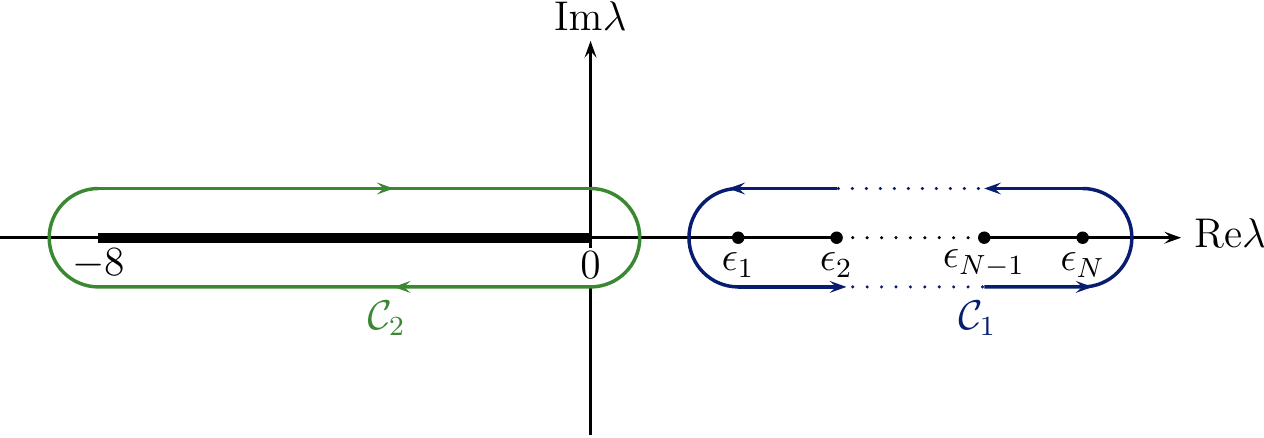}
\caption{\label{paths} Integration paths $\mathcal{C}_1$ and $\mathcal{C}_2$ in the complex $\lambda$-plane.}
\label{fig:paths}
\end{center}
\end{figure}

Let us introduce the resolvent
\begin{equation} \label{eq:resolv}
\mat{G}(\lambda)=(\lambda\mat{1}-{\mat H})^{-1}=\sum_{j,j'=1}^NG_{j,j'}(\lambda)|j\rangle\langle j'|
\end{equation}
for $\lambda\in\mathbb{C}\setminus\spek(\mat{H})$, where $\spek(\mat{H})=\{\varepsilon_\nu,\forall \nu\}$ denotes the spectrum of $\mat{H}$ and a self-explanatory Dirac notation is used on the right-hand side.  
Equation~\eqref{eq:Scc1}  can be rewritten as
\begin{equation} \label{eq:VC1}
\frac{t}{4\pi}-W_3(0)=-\oint_{\mathcal{C}_1} \frac{\rmd\lambda}{2\pi i} \,W_2(\lambda) G_{j,j}(\lambda),
\end{equation}
where $\mathcal{C}_1$ encircles the spectrum in a counterclockwise fashion
as depicted in Fig.~\ref{fig:paths}. 

Since the function $W_2(\lambda)$ is analytic in the complex $\lambda$-plane except for the branch cut 
$[-8,0]$, we can deform the contour $\mathcal{C}_1$ into $\mathcal{C}_2$ 
(see Fig.~\ref{fig:paths}) and recast Eq.~\eqref{eq:VC1} as
\begin{equation} \label{eq:VC2}
\frac{t}{4\pi}-W_3(0)=\int_{-8}^0 \rmd\lambda\, w_2(\lambda)\, G_{j,j}(\lambda),
\end{equation}
where $w_2$ is the spectral function defined in Eq.~\eqref{eq:w2def}. 

It is convenient to rewrite the term $W_3(0)$ on the left-hand side also as an integral involving $w_2$. Using the analog of Eq.~\eqref{eq:U32} for $W_3$ gives \footnote{The value of $W_3(0)$ is explicitly known from \cite{JZ01}. One has $
W_3(0)=(\sqrt{3}-1)\pi^{-3}[
\Gamma(1/24)\,
  \Gamma(11/24)]^2/192$.}
\begin{eqnarray}\label{eq:W30int}
W_3(0) &=&\int_{-\pi}^{\pi}\frac{\rmd p}{2\pi}W_2(2-2\cos p)=
\int_0^4\frac{\rmd \lambda}{\pi}
\frac{W_2(\lambda)}{\sqrt{\lambda(4-\lambda)}}\nonumber\\ &=&\int_{-8}^0d \lambda\,
\frac{w_2(\lambda)}{\sqrt{\lambda(\lambda-4)}}.
\end{eqnarray}
This can now be subtracted from Eq.~\eqref{eq:VC2} to recast Eq.~\eqref{eq:VC2} as
\begin{equation} \label{eq:VscG}
\frac{t}{4\pi}=\int_{-8}^0 \rmd\lambda\,\, w_2(\lambda) 
\left[G_{j,j}(\lambda)+\frac{1}{\sqrt{\lambda(\lambda-4)}}\right].
\end{equation}

The second term inside the square brackets of Eq.~\eqref{eq:VscG}  corresponds to minus the bulk critical analog of the first one. Thus it must be given by the diagonal element $G^{\mathrm{b,c}}_{j,j}(\lambda)$ of the critical bulk Green's function
\begin{equation}
\mat{G}^{\mathrm{b,c}}(\lambda)\equiv \mat{G}(\lambda)|_{\mat{V}=\mat{0},N=\infty}.
\end{equation}
To verify this, note that $\mat{G}$ for given $\mat{V}$ and finite $N$ is the solution to $(\lambda\mat{1}-\mat{H})\cdot\mat{G(\lambda)}=\mat{1}$ subject to the Dirichlet boundary conditions 
\begin{equation}\label{eq:DbcG}
G_{0,j'}(\lambda)=G_{N+1,j'}(\lambda)=0
\end{equation}
implied by Eq.~\eqref{eq:DbcH}. Hence $\mat{G}^{\mathrm{b,c}}(\lambda)$ is the solution to  the difference equation
\begin{align} \label{eq:Gbc}
(\lambda-2) \,G^{\mathrm{b,c}}_{j,j'}(\lambda)&+G^{\mathrm{b,c}}_{j+1,j'}(\lambda)\nonumber\\ &+
G^{\mathrm{b,c}}_{j-1,j'}(\lambda)=\delta_{j,j'},\quad j,j'\in\mathbb{Z},
\end{align}
that vanishes as $|j-j'|\to\infty$, namely
\begin{equation} \label{eq:Gbcres}
G^{\mathrm{b,c}}_{j,j'}(\lambda)=	-\frac{1}{\sqrt{\lambda(\lambda-4)}}\exp\left[-q(\lambda)|j-j'|\right]
\end{equation}
with 
\begin{equation} \label{eq:qlambda}
q (\lambda)=2\, {\rm{arcsinh}}\,  \sqrt{-\lambda/4}\,.
\end{equation}
The result proves the above statement made below Eq.~\eqref{eq:VscG} about its diagonal element . 

Note that all equations given in this subsection remain valid upon replacement of the functions $W_2$ and $W_3$ by their analogs with a tilde, except that the lower integration limit must be changed from $-8$ to $-\infty$ in Eqs.~\eqref{eq:VC2}--\eqref{eq:VscG}.

\section{Scaling limit}\label{sec:scall}
\subsection{Scaling limit of the self-consistency equation}\label{sec:sclsce}

We now turn to the study of the appropriate continuum scaling limit $a\to 0$ of the ${n=\infty}$ self-consistency problem. Equation~\eqref{eq:VscG} is a convenient starting point since the subtracted bulk term eliminates the UV singularities. To this end, we introduce the dimensionful quantities
\begin{align}\label{eq:scaledqies}
y_1&=j_1a &y_2&=j_2 a, &z&= j a, \nonumber\\
L&=  N a, &E&=\lambda a^{-2}, &m &=t a^{-1},
\end{align}
and
\begin{align}
v_a(z)&= V_{j}\,a^{-2},
&\mathcal{G}_a(z,z';E)&= G_{j,j'}(\lambda)\,a.  
\end{align}
Note that the variable $m$, which reduces for $t\ge 0$ to  the inverse of the bulk correlation length $\xi$, becomes negative for $t<0$.

Equation~\eqref{eq:VscG} changes into
\begin{equation} \label{eq:mJa}
m=J_a(z),
\end{equation}
with
\begin{eqnarray}\label{eq:Jadef}
J_a(z)
&\equiv&
4\pi \int_{-8/a^2}^0 \rmd{E}\,w_2(E a^2) 
\bigg[\mathcal{G}_a(z,z;E)\nonumber\\ &&\strut \phantom{\bigg[\mathcal{G}_a(z,z;E)}+\frac{1}{\sqrt{E(E a^2-4)}}\bigg]
\end{eqnarray}
when expressed in these variables. We are interested in the limit $a\to 0$ at fixed $L$ and $m$. Since the widths of the nonuniversal boundary regions $|z|\lesssim a$ and $|L-z|\lesssim a$ shrink to zero in this limit, we must be prepared to encounter singular behavior at the boundary planes and UV singular contributions with support on the  boundaries. 
This means that $J_a(z)$ should be regarded as a distribution and the ${a\to0}$ limit of Eq.~\eqref{eq:mJa} interpreted accordingly.

Before we turn to this issue, let us first discuss the continuum limits of $v_a(z)$ and the Green's function $\mathcal{G}_a$. Dimensional considerations imply that $v_a(z)$ can involve, on the one hand, $a$-independent terms that diverge as $\sim z^{-2}$ and $\sim(L-z)^{-2}$ as $z\to 0+$ and $z\to L-$, respectively, and boundary singularities of the form $a^{-1}\delta'(z)$ and $a^{-1}\delta'(z-L)$, on the other hand. Let us disregard possible  UV boundary singularities of the latter kind $\propto \delta'$ for the moment by restricting ourselves to the open interval $(0,L)$ when studying the ${a\to 0}$ limit of $v_a(z)$ and assuming that the limit
\begin{equation}\label{eq:vas}
\lim_{a\to0}{v}_a(z)=v(z),\quad z\in(0,L),
\end{equation}
exists. That is to say, we consider $v(z)$ as a function rather than a distribution. For  the Green's function $\mathcal{G}_a$ such caution is not necessary. Owing to the  Dirichlet boundary condition~\eqref{eq:DbcG} and the fact that its engineering dimension is $a$, it should have a smooth ${a\to0}$ limit for $z$ and $z'$ in the entire closed interval $[0,L]$, so that the limit
\begin{equation}\label{eq:Gas}
\lim_{a\to0}\mathcal{G}_a(z,z';E)=\mathcal{G}(z,z';E),\quad z,z'\in[0,L],
\end{equation}
should exist. 

The  differential equation that the limiting function $\mathcal{G}$ satisfies follows in a straightforward fashion. Noting  that the matrix
Hamiltonian~\eqref{eq:Hmat}, scaled as $\mat{H}/a^2$, approaches  the operator 
\begin{equation}\label{eq:Hv}
\mathcal{H}_v=-\partial_z^2+v(z)
\end{equation}
 as $a\to 0$, we arrive at
 \begin{equation}\label{eq:cGeq}
 [E+\partial_z^2-v(z)]\mathcal{G}(z,z';E)=\delta(z-z').
 \end{equation}
 
 We shall also need the continuum analogs of the orthonormality relation~\eqref{eq:orthonorm} and the  spectral representation~\eqref{eq:VC1} of the Green's function. They read
\begin{equation}\label{eq:onspsi}
 \langle \oneigfct_\nu|\oneigfct_{\nu'}\rangle=\int_0^L dz \,\oneigfct^*_\nu(z)\,\oneigfct_{\nu'}(z)=\delta_{\nu,\nu'}
\end{equation}
and
\begin{equation} \label{eq:Gd}
\mathcal{G}(z,z';E)=\sum_{\nu=1}^{\infty}\,
\frac{\oneigfct_\nu(z)\,\oneigfct^*_\nu(z')}{E-E_\nu},
\end{equation}
where the asterisk on $\oneigfct_{\nu}^*$ may be dropped since we will  again choose real-valued eigenfunctions.
 
 If $v(z)$ were not singular at the boundaries, we would find that both Eq.~\eqref{eq:cGeq}  and the corresponding Schr\"odinger equation 
\begin{equation} \label{eq:StLc}
\mathcal{H}_v\psi(z)=E\,\psi(z)
\end{equation}
on the interval $[0,L]$ must be subjected to standard Dirichlet boundary conditions. The latter boundary conditions would then ensure the self-adjointness of the Hamiltonian $\mathcal{H}_v$. However, it can easily be seen that the self-consistent potential $v(z)\equiv v(z;L,m)$ must be singular at the boundary planes  $z=0$ and $z=L$. For $d=3$, it must vary asymptotically as
\begin{equation}\label{eq:vsurfsing}
v(z;L,m)\aseq \frac{-1}{4}\begin{cases}z^{-2},& z\to 0+,\\(L-z)^{-2},&z\to L-.
\end{cases}
\end{equation}

An easy way of seeing this is to recall Bray and Moore's exact solution \cite{BM77a,BM77c} 
\begin{equation}\label{eq:vord}
v_\infty(z;0)=\frac{(d-3)^2-1}{4z^2},\quad 2<d<4,
\end{equation}
for the semi-infinite critical $d$-dimensional case $L=\infty$ and $m=0$. Here we have introduced the notation
\begin{equation}\label{eq:vinftydef}
v_\infty(z;m)\equiv v(z;\infty,m).
\end{equation}

The solution~\eqref{eq:vord} refers to  the so-called ordinary surface transition \cite{Bin83,Die86a,Die97}.
When ${3<d<4}$, a second self-consistent solution with the same $z$-dependence $\propto z^{-2}$ but a different amplitude exists \cite{BM77a,BM77c}, which pertains to the so-called special surface transition. Since we will exclusively be concerned with the $({d=3})$-dimensional case, we shall not consider the latter transition and restrict ourselves to the ordinary one.

Clearly, for general ${m\ne 0}$ and ${L<\infty}$,  the behavior of this potential $v(z;L,m)$  at distances from the boundary planes $z=0$ and $z=L$ much smaller than  $1/|m|$ and $L$  must comply with the  exact solution~\eqref{eq:vord}. This dictates the singular short-distance behavior~\eqref{eq:vsurfsing}. 

The result can also be understood within the framework of the boundary operator expansion (BOE)  \cite{DD83a,Die86a,Die97}.  Since the application of the BOE to $v(z)$ has been discussed in some detail in a recent paper \cite{DR14},  we can be brief.  Its central idea is that scaling operators $\mathcal{O}(\bm{y},z)$ with scaling dimensions $\Delta[\mathcal{O}]$ can be expanded for small distances $z$ from the boundary plane $z=0$ as 
\begin{equation}\label{eq:BOE}
\mathcal{O}(\bm{y},z)\mathop{=}_{z\to 0}\sum_j z^{\Delta^{(\rm{s})}_j-\Delta[\mathcal{O}]}\,
C_{\mathcal{O}\/,j}(m z,z/L)\,\hat{\mathcal{O}}_j(\bm{y}),
\end{equation}
where $\hat{\mathcal{O}}_j$ are surface operators with scaling dimensions $\Delta^{(\rm{s})}_j$. The potential $v(z)$ corresponds to the energy density operator $\Phi^2$  whose scaling dimension is $({1-\alpha)/\nu} = d-1/\nu$, which is $2$ for $2<d\le 4$ when  $n=\infty$. The asymptotic behavior~\eqref{eq:vsurfsing} results from the contribution of the unity operator $\hat{\openone}$.

A similar reasoning can be used to clarify which boundary conditions must be  imposed on the eigenfunctions $\oneigfct_\nu(z)$. The leading contribution to the BOE of the order parameter $\Phi(\bm{y},z)\sim\hat{\Phi}(\bm{y})$ originates from the  boundary operator $\hat{\Phi}$. Since the associated scaling dimensions of these two operators are $(d-2+\eta)/2$ and $(d-2+\eta_\|)/2$, respectively, one has  $\Phi(\bm{y},z)\sim z^{(\eta_\|-\eta)/2}\hat{\Phi}(\bm{y})$ for $z\to 0$, where $\eta=0$ and $\eta_\|=1$ when $n=\infty$ and $2<d<4$. Consequently the boundary conditions subject to which the Schr\"odinger Eq.~\eqref{eq:StLc} must be solved are
\begin{equation}\label{eq:BCcpsi}
\oneigfct_\nu(z)\propto
\begin{cases}
\sqrt{z},& \text {for }  z\to 0+,\\
\sqrt{L-z}& \text {for }  z\to L-.
\end{cases}
\end{equation}

Again, this conclusion is in complete accord with the findings of Bray and Moore \cite{BM77a,BM77c} for the semi-infinite critical case. When $L\to\infty$, the spectrum $\spek(\mathcal{H}_v)$ becomes dense and continuous. At $m=0$, it is given by the interval $[0,\infty)$. The corresponding (improper) eigenvalues $E_k$ and eigenfunctions $\oneigfct_k(z)$  can be labeled by a nonnegative variable $k$, rather than by a discrete index $\nu$. They read \cite{BM77a,BM77c}
\begin{equation}\label{eq:eigvf}
\oneigfct_k(z)=\sqrt{z} \,J_0(kz), \quad E_k=k^2,\;\;\;(L=\infty,m=0).
\end{equation}
The corresponding  Green's function $\mathcal{G}_{\infty,\text{c}}(z,z';E)\equiv \mathcal{G}(z,z';E;L=\infty,m=0)$ for $E<0$ may be read off from Bray and Moore's result for the pair correlation function. One has
\begin{equation}\label{eq:Gr0}
\mathcal{G}_{\infty,\text{c}}(z,z';E)=
-\sqrt{z z'}\,I_0\big(\sqrt{-E}z_<\big)K_0\big(\sqrt{-E}z_>\big),
\end{equation}
where $z_<$ and $z_>$ denote the smaller and larger values of $z$ and $z'$, respectively.

Note that for a given potential $v(z)$ which exhibits the surface singular behavior specified in Eq.~\eqref{eq:vsurfsing}, the matrix elements
\begin{equation}
\langle f|\mathcal{H}_v g\rangle \equiv\int_0^Lf^*(z) \mathcal{H}_v\,g(z)\rmd{z}
\end{equation}
of the Hamiltonian~\eqref{eq:Hv} are well-defined for all complex-valued functions $f$ and $g$ belonging to the subspace of the Hilbert space $L^2([0,L])$ of square integrable functions satisfying the boundary condition~\eqref{eq:BCcpsi}. Furthermore, we have $\langle f|\mathcal{H}_v g\rangle=\langle\mathcal{H}_v   f|g\rangle$, so that $\mathcal{H}_v$ is a symmetric operator on this subspace.

We are now ready to return to the continuum limit of the self-consistency Eq.~\eqref{eq:VscG}, \eqref{eq:mJa}, \eqref{eq:Jadef}. We  claim that $J_a(z)$ with $a\to 0$ is a representation of the distribution
\begin{equation}\label{eq:Jdistr}
J(z)=J_{\text{sm}}(z)-\frac{1}{2}[\delta(z)+\delta(z-L)],
\end{equation}
whose ``smooth part'' $J_{\mathrm{sm}}(z)$ is given by
\begin{equation}\label{eq:Jsm}
J_{\mathrm{sm}}(z)=\int_{-\infty}^0 \rmd{E}\,\left[\mathcal{G}(z,z;E)+\frac{1}{2\sqrt{-E}}\right],\quad 0<z<L.
\end{equation}
As has been shown in \cite{DR14}, Eq.~\eqref{eq:Jdistr} simplifies to $J(z)=-\delta(z)/2$ in the critical semi-infinite case (see  Eq.~(3.8) of \cite{DR14}). That $J_{\mathrm{sm}}(z>0;{L=\infty,m=0})=0$ is known from \cite{BM77a,BM77c}. The consistency with Eq.~\eqref{eq:Jdistr} can be verified by inserting Eq.~\eqref{eq:Gr0} into it and computing the integral. The result for the semi-infinite critical case just mentioned implies that the second boundary plane contributes a term $-\delta(z-L)/2$. Power counting rules out  other contributions localized on the boundary planes. Hence, to complete the proof of Eq.~\eqref{eq:Jdistr}, it remains to show that $J_a(z)$ converges  to $J_{\mathrm{sm}}(z)$ for all $z\in(0,L)$. The leading contribution  to  the integral on the right-hand side of Eq.~\eqref{eq:Jadef}  for small $a$ comes from the region of small values of $Ea^2$. Therefore, we can insert the small-$\lambda$ expansion of 
\begin{equation} \label{eq:was}
w_2(\lambda)=\frac{1}{4\pi}\left[1-\frac{\lambda}{8}+\frac{5 \lambda^2}{256}\right]+
O(\lambda^3)
\end{equation}
of the spectral function which one finds  from its explicit form given in Eq.~\eqref{eq:w2} to conclude that the integral converges to $J_{\mathrm{sm}}$ as $a\to 0$ provided $z\in(0,L)$.

The upshot is that the self-consistency equation becomes
\begin{align}\label{eq:vsc}
m&+\tfrac{1}{2}[\delta(z)+\delta(L-z)]\nonumber\\ =&\int_{-\infty}^0 \rmd{E}\left[\mathcal{G}(z,z;E)+\frac{1}{2\sqrt{-E}}\right]
\end{align}
in the $a\to 0$ continuum limit considered.

Equations~\eqref{eq:cGeq} and \eqref{eq:StLc}, in conjunction with  the boundary conditions   \eqref{eq:BCcpsi}, define a Sturm-Liouville problem for potentials with the singular behavior~\eqref{eq:vsurfsing}. In order to determine the exact $n\to\infty$ solution of the model in the continuum scaling limit specified at the beginning of this subsection, it must be solved self-consistently with  Eq.~\eqref{eq:vsc}. Finding closed-form analytical solutions to these equations for noncritical temperatures $m\ne 0$ and finite thicknesses $L$ is a major challenge, which may well turn out to be too difficult to master. In \cite{DGHHRS12,DGHHRS14} this problem was bypassed by resorting to numerical solutions of discretized equations. This has yielded precise results for the universal scaling functions $\Theta(x)$ and $\vartheta(x)$. 

Regrettably, not much exact information on the self-consistent potential $v(z;L,m)$ is available beyond the exact solution~\eqref{eq:vord} for the critical semi-infinite case $L=\infty$, $m=0$. To our knowledge, it is limited to what has been provided by us in a recent paper \cite{DR14}. Before we turn to the issue of how the situation can be improved through the use of inverse scattering-theory methods, it will be helpful to recall these exactly known properties. This is done in the next subsection, where we also express quantities such as the excess energy density in terms of $v(z;\infty,m)$. 

\subsection{The self-consistent potential}\label{sec:BOEscpot}

The self-consistent potential $v(z)$ is proportional to the energy density $\langle\Phi^2(\bm{y},z)\rangle$, where $\Phi(\bm{y},z)$ means the continuum analog of the lattice field $\Phi_{\bm{j}_\|,j}$ derived from it via the scaling $\Phi_{\bm{j}_\|,j}=a^{d/2-1}\Phi(\bm{y},z)$. Therefore, $v(z;L,m)$ must exhibit the corresponding scaling behavior 
\begin{eqnarray}
v(z;L,m)&=&z^{-(1-\alpha)/\nu}\,\Upsilon_\pm^{(d)}( |t|^{\nu}z/a,z/L)\nonumber\\
&\mathop{=}\limits_{n=\infty,\, d=3}&z^{-2}\,\Upsilon_\pm^{(3)}(|m| z,z/L)
\end{eqnarray}
on long length scales \cite{DD81c,KED95} where $\pm$ refers as usual to $m=t/a\gtrless 0$.  Although our ultimate interest is in the case $d=3$, we give here  --- and where appropriate below  --- both results for general values of $n$ and $d\in (2,4)$, and  for the case of $n=\infty$ and $d=3$. The latter case is special; it involves degeneracies, which imply logarithmic anomalies in the leading thermal singularities of the surface free energy $\fs$ and related quantities \cite{DR14}. The results for general $d$ we are going to present here serve to illustrate the special features of the ${d=3}$ case. 

As is discussed in \cite{DR14} and elsewhere (see, e.g., \cite{DD83a,Die86a,Die97,Car90b,EKD93,MO95}), useful information about properties of the scaling function $\Upsilon_d$ can be obtained from the BOE. For the case of the ordinary transition we are concerned with here, the leading contributions to the BOE arise from the unity operator $\hat{\openone}$ and the $zz$-component $\hat{T}_{zz}$ of the stress tensor on the boundary. One obtains
\begin{eqnarray}
\Upsilon_\pm^{(d)}(\zm,\zL)&\mathop{=}\limits_{z\to 0}&A_d^{\mathrm{or}}\left[X_\pm^{(d)}(\zm,\zL)+\zm^d\,Y_\pm^{(d)}(\zm,\zL)+\ldots\right]\nonumber\\
&\mathop{=}\limits_{d=3}&\frac{-1}{4}\left[X_\pm^{(3)}(\zm,\zL)+\zm^3\,Y_\pm^{(3)}(\zm,\zL)+\ldots\right],\nonumber\\
\end{eqnarray}
where $\zm$ and $\zL$ denote the dimensionless distances
\begin{equation}\label{eq:zmzLdef}
\zm=|t|^\nu z/a\mathop{=}_{d=3}|m|z,\quad\zL=z/L,
\end{equation}
while 
 $A_d^{\mathrm{or}}$ is a normalization factor ensuring 
\begin{equation}
X_\pm^{(d)}(0,0)=1.
\end{equation}

Since  BOE expansion coefficients such as  $X_\pm^{(d)}$ and $Y_\pm^{(d)}$ are short-distance properties, they are expected to be regular in the temperature variable. This implies 
\begin{eqnarray}
X_\pm^{(d)}(\zm,0 )&\mathop{=}\limits_{\zm\to 0} &1\pm a_1(d)\zm^{1/\nu}+a_2(d)\zm^{2/\nu}+\ldots\nonumber\\ &\mathop{=}\limits_{d=3,\,n=\infty}&1+a_1(3)\,mz+a_2(3)\,m^2z^2+\ldots,\nonumber\\
\end{eqnarray}
where the value
\begin{equation}\label{eq:a13}
a_1(3)=-\frac{16}{\pi^2}
\end{equation}
is known from \cite{DR14}. The coefficients $a_j(d)$, $j=1,2,\ldots$, are independent of the sign of $m$ because
they can be expressed in terms of derivatives of the energy density at $m=0$ and $L=\infty$. The coefficient $a_2(3)$ is determined exactly in Sec.~\ref{sec:v0calc}. We shall prove there that it has the value
\begin{equation}\label{eq:a23res}
a_2(3)=-\frac{224}{\pi^4}\,\zeta(3)
\end{equation}
independent of the sign of $m$, where $\zeta(x)$ is the Riemann zeta function.

Returning to the case of general $d\in(0,4)$ and $n$, consider the analogous expansion of $Y_\pm^{(d)}(\zm,0 )$, 
\begin{eqnarray}\label{eq:Yd}
Y_\pm^{(d)}(\zm,0 )&\mathop{=}\limits_{\zm\to 0} &b_{0,\pm}(d)\pm b_{1,\pm}(d)\,\zm^{1/\nu}+\ldots\nonumber\\
&\mathop{=}\limits_{d=3,\,n=\infty}&b_{0,\pm}(3)+b_{1,\pm}(3)\,mz+\ldots.\;\;\;
\end{eqnarray}
As indicated, the expansion coefficients here must be expected to have different values for $m\gtrless 0$. In fact, the term $\propto b_{0,\pm}$ yields the leading singularity $\sim |m|^{2-\alpha}$ of the surface energy density and the ratio $b_{0,+}(d)/b_{0,-}(d) $ should take the same universal value as its analog for the bulk free energy \cite{BD94,Die94a,Die97}. In results for $d=3$ and $n=\infty$ given in the second line of Eq.~\eqref{eq:Yd} we have utilized the fact that $b_{0,\pm}(d)$ does not have a pole at $d=3$ (see the discussion in  \cite{DR14}).

As has been discussed in \cite{DR14}, the above BOE expansion can also be applied to the $L<\infty$ case at $\tc$ to gain information about the distant-wall correction $\asprop (z/L)^d$ to the potential. The associated amplitude is proportional to the Casimir amplitude $\D{}$. The proportionality factor is known from \cite{DR14} for the $({d=3})$-dimensional case. This led to the prediction
\begin{equation}\label{eq:vsmallzL}
\frac{v(z;L,0)}{v(z;\infty,0)}\mathop{=}\limits_{d=3}1-\frac{1024\,\D{}}{\pi}\,\frac{z^3}{L^3}+o\big[(z/L)^{-3}\big],
\end{equation}
which is in conformity with the numerical solution of the self-consistency equation \cite{DGHHRS14}.

Focusing on the case of $d=3$, we can expand the scaling function $\Upsilon_\pm^{(3)}(\zm,\zL)$ about $\zL =0$ and match with the above equations to conclude that\begin{eqnarray}\label{eq:vinftyplusLmin3}
v(z;L,m)&\mathop{=}\limits_{z/L\to 0}&\frac{1}{z^2}\left[A_\pm(|m|z)+\frac{z^3}{L^3}\,B_\pm(|m|z)+\ldots\right]\quad\nonumber\\
&=&v(z;\infty,m)+\frac{z}{L^3}\,B_\pm(|m|z)+\ldots,\quad
\end{eqnarray}
where the functions $A_\pm(\zm)$ and $B_\pm(\zm)$ must vary for small $\zm$ as
\begin{equation}\label{eq:Apmsmallzm}
A_\pm(\zm)\mathop{=}\limits_{\zm\to 0}\frac{-1}{4}\pm\frac{4\zm}{\pi^2}+\frac{56\zeta(3)}{\pi^4}\,\zm^2+o(\zm^2)
\end{equation}
and
\begin{equation}\label{eq:Bpmsmallzm}
B_\pm(\zm)\mathop{=}\limits_{\zm\to 0}\frac{256\,\Delta_{\text{C}}}{\pi}+O(\zm),
\end{equation}
respectively. In the large-$\zm$ limit, $A_\pm(\zm)$ and $B_\pm(\zm)$ must decay on the scale of $1/|m|$. Depending on whether $m>0$ or $m<0$, the latter length corresponds to the correlation length or  Josephson length and the decay of these functions is exponential or algebraic, where the algebraic decay is due to the presence of Goldstone modes in the ordered bulk phase $(m<0,L=\infty)$. In fact, results derived in Sec.~\ref{sec:asspecprop} yield the asymptotic behaviors
\begin{equation}\label{eq:Aminlargezm}
A_-(\zm)\mathop{=}\limits_{\zm\to\infty}\frac{-1}{2\zm}-\frac{\alpha_2}{\zm^2}+o(1/\zm^2)
\end{equation}
and
\begin{equation}\label{eq:Bminlargezm}
B_-(\zm)\mathop{=}\limits_{\zm\to\infty}\frac{-\zeta(3)}{\zm}-\frac{\beta_2}{\zm^2}+o(1/\zm^2)
\end{equation}
with
\begin{equation}\label{eq:beta2res}
\beta_2=\zeta(3)/2
\end{equation}
and the unknown coefficient $\alpha_2$  whose value we have not determined and shall not need below.

For use below, let us also briefly mention the  behavior of $v(z;\infty,m)$ for large $z$. In this limit, $v(z;\infty,m)$ approaches the bulk value $v_{\mathrm{b}}(m)$,
\begin{equation}
 v(z;\infty,m)\mathop{=}\limits_{z\to\infty}v_{\mathrm{b}}(m)=m^2\,\Htheta(m),
\end{equation}
given by the square of the inverse correlation length when $m>0$, and zero in the ordered bulk phase on the coexistence curve. The approach to the limiting value $v_{\mathrm{b}}(m)$ is exponential $\sim \rme^{-2mz}$ or algebraic  $\sim (|m|z)^{-3}$, depending on whether  $m>0$ or $m<0$. Evidently, $v_{\mathrm{b}}(m)$
is  the continuum limit of the scaled bulk lattice potential $a^{-2}V_{\mathrm{b}}({t=ma})$, where 
 \begin{equation}
V_{\mathrm{b}}(t)\equiv\lim_{j\to\infty}V_j({N=\infty},t)= \lim_{N\to\infty}\frac{1}{N}\Tr\mat{V}(N,t)
\end{equation}
is the bulk potential given in Eq.~\eqref{eq:Vbas}.

\subsection{Excess surface energy and its leading thermal singularity}\label{sec:surfexcessen}

In Section~\ref{sec:sclsce} we have seen that the continuum limit of the self-consistent equation, Eq.~\eqref{eq:vsc}, is well-defined and free of UV singularities. However, quantities such as the excess energy density, and bulk and free energies still involve UV singularities, which must be properly subtracted to gain the desired information about universal quantities such as universal amplitude combinations and scaling functions. We first consider this problem for the excess surface energy density. To this end, we introduce the $N\times N$ matrix
\begin{equation}\label{eq:Vtilde}
\tilde{\mat{V}}(N,t)\equiv \mat{V}(N,t)-V_{\mathrm{b}}(t)\,\mat{1},
\end{equation}
in terms of which the excess energy density  
\begin{equation}\label{eq:excE}
\mathcal{E}_{\mathrm{s}}(t)\equiv \frac{\rmd}{\rmd{t}}\fs(t)
\end{equation}
 of our lattice model can be expressed with the aid of Eq.~\eqref{eq:dfNdt} as
\begin{equation}\label{eq:EsVtilde}
\mathcal{E}_{\mathrm{s}}(t)=\frac{1}{8\pi}\lim_{N\to\infty}\Tr\tilde{\mat{V}}(N,t).
\end{equation}

For general dimension $d\in(2,4)$ with $d\ne 3$, the leading thermal singularity of the surface free energy $\fs$ for $t\to\pm 0$ is of the form 
\begin{equation}
\fs-\fs^{(\mathrm{reg})}=\fs^{(\mathrm{sing})}\aseq A^{(\mathrm{s})}_\pm(d)\,|t|^{2-\alpha-\nu}=A^{(\mathrm{s})}_\pm(d)\,|t|^{\frac{d-1}{d-2}},
\end{equation}
where $\fs^{(\mathrm{reg})}=\sum_{k=0}f_k^{(\mathrm{s})}(d)t^k$ is a regular background term. As has been explained in \cite{DR14}, the degeneracy of the singular and regular terms $\propto t^2$ at $d=3$ entail by a standard mechanism \cite{CK86} possible logarithmic temperature anomalies. The amplitudes  $A^{(\mathrm{s})}_\pm(d)$ and $f_2^{(\mathrm{s})}(d)$ generically are expected to have Laurent expansions of the forms
\begin{equation}
A^{(\mathrm{s})}_\pm(d)=-\frac{f^{(\mathrm{s})}_{2,-1}}{d-3}+A^{(\mathrm{s})}_{0,\pm}-f^{(\mathrm{s})}_{2,0}+O(d-3)
\end{equation}
and
\begin{equation}
f_2^{(\mathrm{s})}(d)=\frac{f^{(\mathrm{s})}_{2,-1}}{d-3}+f^{(\mathrm{s})}_{2,0}+O(d-3).
\end{equation}
These yield the limit
\begin{equation}\label{eq:fssing}
\lim_{d\to 3}\left[A^{(\mathrm{s})}_\pm(d)|t|^{\frac{d-1}{d-2}}+f^{(\mathrm{s})}_2(d)\,t^2\right]=t^2\Big[A_{0,\pm}^{(\mathrm{s})}-f^{(\mathrm{s})}_{2,-1}\ln |t|\Big],
\end{equation}
which in turn implies the following small-$t$  expansion  for the surface free energy 
  $f_s(t)$ in the three-dimensional system $d=3$
  near the bulk critical point,
  \begin{eqnarray}\label{eq:fs3}
  f_s(t)\big|_{d=3}&=&\left[f_0^{(s)}(3)+f_1^{(s)}(3)\, t+ f^{(s)}_{2,0}\, t^2 +O(t^3)\right]\nonumber \\
  &&\strut +\left[
  A_{0,\pm}^{(\mathrm{s})}-f^{(\mathrm{s})}_{2,-1}\ln |t|
  \right] t^2+O(t^3 \ln|t|).\qquad\quad
  \end{eqnarray}
  Accordingly, one obtains for the singular part of the excess energy density \eqref{eq:excE} 
\begin{equation}\label{eq:EssingDelA0}
\mathcal{E}^{\mathrm{sing}}_{\mathrm{s}}(t)+\mathcal{E}^{\mathrm{sing}}_{\mathrm{s}}(-t)\mathop{\aseq}_{t\to 0+ }2\,\Delta A_0^{(\mathrm{s})}\,|t|,
\end{equation}
where 
\begin{equation}
\Delta A_0^{(\mathrm{s})}\equiv A_{0,+}^{(\mathrm{s})}-A_{0,-}^{(\mathrm{s})}
\end{equation}
is a universal amplitude difference  \footnote{The universality of $\Delta A_0^{(\mathrm{s})}$ follows also from the fact that both  $f^{(\mathrm{s})}_{2,-1}$ and the ratio $A^{(\mathrm{s})}_+/A^{(\mathrm{s})}_-=\Delta A^{(\mathrm{s})}_0/f^{(\mathrm{s})}_{2,-1}+O(d-3)$ are universal \cite{DR14}.}. The residue $f^{(\mathrm{s})}_{2,-1}$ is also universal and according to  \cite{DR14} given by\begin{equation}
f^{(\mathrm{s})}_{2,-1}=\frac{1}{4\pi^3}.
\end{equation}

Upon studying the continuum scaling limit of $\mathcal{E}_{\mathrm{s}}$, we will now show that this amplitude difference can be expressed as
\begin{equation}\label{eq:DelA0pmres}
\Delta A_0^{(\mathrm{s})}=\frac{\alpha_++\alpha_-}{16\pi}
\end{equation}
in terms of the following finite integrals involving the self-consistent potential $v_\infty(z;m)$ of the continuum theory:
\begin{eqnarray}\label{eq:alphapmdef}
\alpha_\pm&=&\int_0^\infty\rmd{\zm}\bigg[v_\infty(\zm;\pm1)-v_{\mathrm{b}}(\pm1)+\frac{1}{4\zm^2}\nonumber\\&&\strut \mp\frac{4}{\pi^2\zm}\,\Htheta(1-\zm)
\bigg].
\end{eqnarray}

These integrals are finite because the two last terms in the square brackets subtract the boundary singularities near $\zm=0$, while the subtraction produced by the second term along with the restriction of the last term to the interval $(0,1)$ implied by the theta function ensures the integrability at  the upper integration limit $\zm=\infty$. Note that the coefficient of the subtracted $\zm^{-1}$ term has been chosen in accordance with the result for $a_1(3)$ given in Eq.~\eqref{eq:a13}.

To prove the asserted representation of $\Delta A_0^{(\mathrm{s})}$ in terms of $v_\infty$, we start from Eq.~\eqref{eq:EsVtilde}, express $\tilde{V}_j(\infty,t)$ in terms of its continuum analog $\tilde{v}_\infty(z;m)=m^2\,\tilde{v}_\infty(|m|z ;\pm1)$, and use the Euler-McLaurin formula 
\begin{equation}\label{eq:EulerMcL}
a\sum_{j=1}^\infty f(ja)=\int_a^\infty\rmd{z}\,f(z)+\frac{a}{2}\,f(a)+O(a^2)
\end{equation}
to find
\begin{equation}
8\pi \, \mathcal{E}_s(t)=a|m|\int_{a|m|}^\infty\tilde{v}_\infty(\zm;\pm)\,\rmd{\zm}+\frac{1}{2}\,\tilde{V}_1(\infty,t)+\ldots
\end{equation}
The second term is regular in $t$ and does not contribute to  $\mathcal{E}^{\mathrm{sing}}_{\mathrm{s}}$.  From the integral we split off the contribution
\begin{equation}
\int_{a|m|}^\infty\rmd{\zm}\left[\frac{1}{4\zm^2}\mp\frac{4}{\pi^2\zm}\,\Htheta(1-\zm)\right]=\frac{1}{4a|m|}\pm\frac{4}{\pi^2}\ln(a |m|).
\end{equation}
that diverges as $a\to 0$, obtaining
\begin{eqnarray}
8\pi\,\mathcal{E}_{\mathrm{s}}&=&a|m|\int_{a|m|}^\infty\rmd{\zm}\left[\tilde{v}_\infty(\zm|\pm)+\frac{1}{4\zm^2}\mp\frac{4}{\pi^2\zm}\,\Htheta(1-\zm)\right]\nonumber\\&&\strut 
\mp\frac{4a|m|}{\pi^2}\ln(a |m|)+\text{reg},
\end{eqnarray}
where ``reg'' represents contributions regular in $t$. Hence the leading thermal singularity of $\mathcal{E}_{\mathrm{s}}$ is given by
\begin{equation}\label{eq:Essing}
\mathcal{E}^{\mathrm{sing}}_{\mathrm{s}}(t)\mathop{\aseq}_{t=a m\to 0+} \frac{a|m|}{8\pi}\left[\alpha_\pm\mp\frac{4}{\pi^2}\ln(a|m|)\right].
\end{equation}
Comparing with Eqs.~\eqref{eq:fssing} and \eqref{eq:EssingDelA0} then yields the result for $\Delta A_{0}^{(\mathrm{s})}$ given in Eq.~\eqref{eq:DelA0pmres}. 

To determine the universal number $\Delta A_{0}^{(\mathrm{s})}$ in an analytic manner from this equation, one would have to know the self-consistent potential $v_\infty(\zm ;\pm 1)$ for all values of $\zm\in (0,\infty)$. Unfortunately, the latter is known only in numerical, but not in closed analytical, form. The inverse scattering-theory techniques developed in Sec.~\ref{sec:ist} in conjunction with the semiclassical expansions described in Appendix~\ref{app:semiclass}  will enable us to obtain an exact analytic result for $\alpha_-$  and to express the quantities $\alpha_+$ and $\Delta A_{0}^{(\mathrm{s})}$ in terms of a single, numerically computable integral. The results are given in Eqs.~\eqref{eq:alphaplusres}--\eqref{eq:DelA0anexp} below [see Sec.~\ref{sec:alphapm}].

\section{Inverse scattering theory}\label{sec:ist}

\subsection{Preliminaries}\label{sec:istintro}

The aim of inverse scattering theory \cite{FS63,CS89,Fad95} is to reconstruct the potential of the stationary Schr\"odinger equation $v(r)$  from scattering data. Usually, either Schr\"odinger equations for three-dimensional systems with radially symmetric potentials or one-dimensional Schr\"odinger problems are considered. In the first case, one must deal with a radial Schr\"odinger equation involving an effective potential that differs from $v(r)$ by a centrifugal term $l(l+1)r^{-2}$, where $l=0,1,2,\ldots$ are orbital angular momentum quantum numbers. Provided the effective potential satisfies certain conditions, such as absolute integrability of $V(r)$ over intervals $(b,\infty)$ with $b>0$ and  integrability of $r|V(r)|$ over integrals $(0,a)$ with $a>0$, various classes of inverse scattering problems are known to have a unique solutions so that the potential can be determined --- at least, in principle --- from scattering data as solutions to certain integral equations such as the Gel'fand-Levitan or Marchenko  integral equations \cite{CS89}.

As we have seen above, the Schr\"odinger equation we are concerned with here involves a self-consistent potential $v(z;L,m)$ that becomes singular at the boundary planes and has a leading near-boundary singularity of the form specified in Eq.~\eqref{eq:vsurfsing}. At $L=\infty$ and $m=0$, it corresponds formally to a radial Schr\"odinger equation with angular momentum quantum number  $l=-1/2$.  Although there exists some recent work on inverse scattering problems involving  singular potentials  $\sim \kappa /r^{2}$ with $\kappa>-1/4$ \cite{FY05}, the case of $\kappa=-1/4$, corresponding to the marginal value below which the particle is supposed to fall into the center according to  Landau and Lifshitz \cite{LL58},  requires appropriate modifications of the theory which will become clear as we describe our procedure.

Let us begin by noting that we are not faced with a usual inverse scattering problem here where scattering data are given from which the potential is to be reconstructed. Rather, we are dealing with a self-consistent Schr\"odinger problem. An obvious way to attack the problem is to exploit the relation of the potential with scattering data to reformulate the self-consistency equation in terms of scattering data and then determine the latter from it. Upon expressing quantities of interest through the scattering data, one can bypass the determination of the self consistent potential. This is the strategy we will pursue, focusing on the semi-infinite case $L=\infty$ and considering both the disordered phase $m\ge0$ as well as the ordered one $m<0$.

Scale invariance enables us to scale the temperature variable $m$ to $\pm 1$. Introducing
\begin{equation}
v_\pm(\zm)\equiv  v_\infty(\zm;\pm1)
\end{equation}
and noting that 
\begin{equation}\label{eq:vinf}
v_\pm(\infty)=\delta_{\pm1,1},
\end{equation}
we see that  we must study  the Schr\"odinger problem
\begin{equation}\label{eq:SEhalfline}
[-\partial_\zm^2+v_\pm(\zm)-\delta_{\pm 1,1}]\psi(\zm,\sk)=\sk^2\,\psi(\zm,\sk)
\end{equation}
with
\begin{equation}\label{eq:Eksquared}
\sk^2=E-\delta_{\pm 1,1}
\end{equation}
on the half-line $[0,\infty)$ for potentials that vary as
\begin{equation}\label{eq:vnearboundary}
v_\pm(\zm)\mathop{=}_{\zm\to 0}-\frac{1}{4\zm^2}+v_{-1}^\pm\frac{1}{\zm} +v^\pm_0+O(\zm)
\end{equation}
and approach the limiting values $\delta_{\pm1,1}$   sufficiently fast as $\zm\to\infty$ (exponentially or as a power, depending on the sign $\pm$) \footnote{We shall show in Sec.~\ref{sec:asspecprop} that $v_-(\zm\to\infty)\aseq -\zm^{-3}/2$; cf.\ Eq.~\eqref{eq:vmininfas}.}.

Since $v(z;L,m)$ corresponds to the energy density whose leading thermal singularity in the near-boundary region originates from the contribution $\asprop m^3$, the terms linear and quadratic in  $m$  must both be regular in $m$. Therefore  the coefficients  $\pm v^\pm_{-1}$ and $v^\pm_0$ must be independent of the sign of $m$. According to Eqs.~\eqref{eq:a13} and \eqref{eq:vinftyplusLmin3}-- \eqref{eq:Bpmsmallzm}, we have indeed
\begin{equation}\label{eq:vm1}
v_{-1}^\pm=\mp\frac{1}{4} a_{-1}(3)=\pm\frac{4}{\pi^2}
\end{equation}
and
\begin{equation}\label{eq:v0res}
v^\pm_0\equiv v_0=-\frac{a_2(3)}{4}=\frac{56}{\pi^4}\,\zeta(3).
\end{equation}

The spectrum  of the associated Sturm-Liouville operator $\mathcal{H}_{v_\pm}$ on $[0,\infty)$ should be continuous and equal to $\spek(\mathcal{H}_{v_\pm})=[\delta_{\pm1,1},\infty)$. 
Following a standard approach, we introduce two Jost solutions of Eq.~\eqref{eq:SEhalfline} satisfying the boundary conditions
\begin{equation}\label{eq:flargez}
f(\zm,\pm \sk)\mathop{\aseq}_{\zm\to\infty}\rme^{\pm\rmi \sk \zm}
\end{equation}
and normalize the so-called ``regular solution'' of Eq.~\eqref{eq:SEhalfline} such that
\begin{equation}\label{eq:varphinormal}
\regsol(\zm,\sk)=\sqrt{\zm}\,[1+O(\zm)].
\end{equation}
The latter behaves asymptotically as
\begin{equation}\label{eq:varphiassigma}
\regsol(\zm,\sk)\mathop{\aseq}_{\zm\to\infty}\frac{A(\sk)}{\sk}\sin[\sk\zm+\eta(\sk)],\quad A(\sk)=\rme^{\sigma(\sk)},
\end{equation}
which defines the scattering amplitude $A(\sk)$ and  phase shift $\eta(\sk)$. 
The regular solution can be expressed in terms of the Jost solutions as
\begin{equation}\label{eq:regsolJost}
\regsol(\zm,\sk)=\frac{1}{2\rmi \sk}\,[F(-\sk)\,f(\zm,\sk)-F(\sk)\,f(\zm,-\sk)],\quad \sk>0,
\end{equation}
where $F(\sk)$ is a complex-valued function, the Jost function. For real  $\sk$ it can be written as 
\begin{equation}\label{eq:FJostk}
F(\sk)=\begin{cases}
\rme^{\sigma(\sk)-\rmi\eta(\sk)},&\sk>0,\\
\rme^{\sigma(-\sk)+\rmi\eta(-\sk)},&\sk<0.
\end{cases}
\end{equation}
For later use  let us also note that $F(\sk)$ can written as a  Wronskian of two functions $f_1$ and $f_2$,
\begin{equation}\label{eq:Wronskiandef}
W[f_1(\zm),f_2(\zm)]=f_1(\zm)\frac{\partial f_2(\zm)}{\partial\zm}-\frac{\partial\ f_1(\zm)}{\partial\zm}f_2(\zm).
\end{equation}
One has
\begin{equation}\label{eq:FWronskian}
F(k)= W[f(\zm,\sk),\regsol(\zm,\sk)].
\end{equation}
This known result can be verified in a straightforward fashion by substituting \eqref{eq:regsolJost} for $\regsol(\zm,\sk)$ in the Wronksian and evaluating the latter for $\zm\to\infty$, making use of its  independence  of $\zm$.

For the sake of notational conciseness, we have refrained here from adding subscripts $\pm$ to the above quantities $\regsol$, $F$, \ldots, $\eta$. But it should be remembered that they all differ for the cases $\pm$ of the self-consistent potentials $v_\pm$.

In the ordered phase $m<0$, $L=\infty$, the divergence of the susceptibility on the coexistence curve implies that the Schr\"odinger Eq.~\eqref{eq:SEhalfline} must have an $E=0$ state $\regsol_0(\zm)$ that approaches a nonzero value at $\zm=\infty$. It satisfies
\begin{equation}\label{eq:zerophi}
\regsol_0''(\zm)=v_-(\zm)\,\regsol_0(\zm).
\end{equation}

 This is precisely the equation for the order-parameter profile $\langle\Phi(\bm{y},z)\rangle_{L=\infty,m=-1}$ in the presence of an infinitesimal magnetic field oriented along a fixed direction. The square of the  spontaneous bulk magnetization is known to be given by $|m|/4\pi$ \cite{MZ03}. Hence, if the eigenfunction $\regsol_0(\zm)$ is normalized such that
\begin{equation}\label{eq:varphiinf}
\lim_{\zm\to\infty}\regsol_0(\zm)=1,
\end{equation}
then the spontaneous order-parameter profile at $L=\infty$ is given by
\begin{equation}\label{eq:G1squscal}
\langle\Phi(\bm{y},z)\rangle_{L=\infty,m<0}^2=\frac{|m|}{4\pi}\,\regsol_0^2(|m|z).
\end{equation}
The small-$\zm$ behavior $\regsol_0({\zm\to 0}) \aseq\text{const}\, \sqrt{\zm}$ of $\regsol_0$ is similar to Eq.~\eqref{eq:varphinormal}. At this stage we do not yet  know that  the proportionality constant ``const'' is exactly $1$ and that $\regsol_0$ agrees with the $\sk\to 0$ limit of $\regsol_0(\zm,\sk)$, i.e.,
\begin{equation}\label{eq:varphi0klim}
\lim_{\sk\to 0}\regsol(\zm,\sk)=\regsol_0(\zm),
\end{equation}
so that  $\regsol_0$ indeed satisfies the boundary condition~\eqref{eq:varphinormal}. However, our inverse scattering analysis in Sec.~\ref{sec:hssdmm} will confirm the validity of both statements, i.e., that Eqs.~ \eqref{eq:varphiinf} and \eqref{eq:varphi0klim} hold and $\regsol_0$ fulfills the boundary condition~\eqref{eq:varphinormal}.

Owing to the limiting behavior~\eqref{eq:varphiinf}, $\regsol_0$  is not square-integrable. Such a zero-momentum state, which is finite at $z=\infty$ but does not decay sufficiently fast to be square-integrable is called ``half-bound state'' \cite{Ma06}. It occurs in the semi-infinite case we consider here because the lowest eigenvalue $E_1(L,-1)$ of the strip tends exponentially to zero as its thickness $L\to\infty$; one has \cite{DGHHRS12,DGHHRS14}
\begin{equation}\label{eq:E1lim}
\lim_{L\to\infty}L^{-1}\ln E_1(L,m)=m.
\end{equation}
The half-bound state arises for $m<0$ from the eigenfunction $\oneigfct_1(z;L,m)$ in the limit $L\to\infty$. Since the latter is orthonormalized and $\oneigfct_1\sqrt{L}$ is dimensionless, the normalization~\eqref{eq:varphiinf} of $\regsol_0$ implies that
 \begin{equation}\label{eq:varphipsi1}
 \lim_{L\to\infty}\oneigfct_1(z;L,m)\,\sqrt{L}=\regsol_0(|m|z).
 \end{equation}

The presence of an order-parameter profile for $m<0$ and $L=\infty$ entails an additional contribution to the $L=\infty$ analog of the self-consistency Eq.~\eqref{eq:vsc}. The latter reads
\begin{subequations}\label{eq:Gscepmges}
\begin{eqnarray}\label{eq:Gscepm}
\lefteqn{\int_{-\infty}^0\rmd{E}\bigg[\mathcal{G}_\pm(\zm,\zm;E)+\frac{1}{2\sqrt{-E}}\bigg]} &&\nonumber\\ &=&\frac{1}{2}\,\delta(\zm)\pm 1+\delta_{\pm1,-1}\,\regsol_0^2(\zm)
\end{eqnarray}
with
\begin{equation}\label{eq:Gpm}
\mathcal{G}_\pm(\zm,\zm;E)\equiv \mathcal{G}(\zm,\zm;E;\infty,\pm 1).
\end{equation}
\end{subequations}

Equation~\eqref{eq:Gscepm} is a well-suited starting point for applying inverse scattering theory. To this end we proceed as follows. Let $v_\pm(\zm)$ be the above specified self-consistent potential for $m=\pm 1$ and $L=\infty$, $\mathcal{G}_\pm$ the associated Green's function \eqref{eq:Gpm}, and $\regsol_0(\zm)\equiv\regsol_0(\zm;-1)$ the solution to Eq.~\eqref{eq:zerophi} for $m=-1$, normalized according to Eq.~\eqref{eq:varphiinf}. Hence Eqs.~\eqref{eq:Gscepmges} hold for these quantities $\mathcal{G}_\pm$ and $\regsol_0(\zm)$. We  now consider variations 
\begin{equation}\label{eq:deltav}
v_\pm(\zm)\to v_\pm(\zm)+\delta v(\zm).
\end{equation}
It will be sufficient and convenient to restrict the variations $\delta v$ to functions with the properties
\begin{align}\label{eq:delvprop}
&\text{(i)}&&\delta v(\zm)= \text{ regular for }\zm>0,\nonumber \\
&\text{(ii)}&&\lim_{\zm\to\infty}\delta v(\zm)=0,\nonumber\\
&\text{(iii)}&&\lim_{\zm\to 0}\delta v(\zm)=\delta v_0,\nonumber\\
&\text{(iv)}&&\int_0^\infty\rmd{\zm}\,\delta v(\zm)=0.
\end{align}

Equation~\eqref{eq:Gscepm} simply corresponds to the continuum limits of the self-consistency conditions~\eqref{eq:Scc1} and \eqref{eq:simpsceV}. Since the latter two are nothing else but stationarity conditions such as  $\partial f_N(t,\mat{V})/\partial V_j=0$, Eq.~\eqref{eq:Gscepm} may be read as the stationarity condition $\delta f[v]/\delta v(\zm)=0$ for a free-energy functional $f[v]$, which is the continuum analog of $f_{N=\infty}(t,\mat{V})$. Hence the linear variation 
\begin{equation}\label{eq:delGpm}
\delta f[v,\delta v]\equiv\int_0^\infty\rmd{\zm}\,\frac{\delta f[v]}{\delta v(\zm)}\,\delta v(\zm)
\end{equation}
of this functional must vanish when evaluated at the self-consistent potential $v_*(\zm)$, where here and below an asterisk is used to mark potentials $v_*$ and Green's functions $\mathcal{G}_*$ that solve the self-consistency equations. To express Eq.~\eqref{eq:delGpm} in terms of the Green's function, we must simply multiply Eq.~\eqref{eq:Gscepm} by $\delta v(\zm)$ and integrate $\zm$ from $0$ to $\infty$. We then proceed by expressing the result  in terms of scattering data. 

To this end it is convenient to introduce the energy-dependent scattering phase $\eta_E$ by
\begin{equation}\label{eq:etaEdef}
\eta_E=\eta(\sk)|_{\sk=+\sqrt{E-\delta_{\pm1,1}}},\quad E\in(\delta_{\pm1,1},\infty),
\end{equation}
for the cases $m=\pm 1$. This is because both the  Green's function and $\delta v(\zm)$ can be eliminated in favor of  $\eta_E$ and its variation  $\delta \eta_E$ induced by $\delta v(\zm)$ by exploiting the relations 
\begin{equation}\label{eq:etaE}
\frac{\rmd\eta_E}{\rmd E}=-\int_0^\infty \rm{d}\zm\, 
\left[
\Im\, \mathcal{G}(\zm,\zm;E+\rmi 0)+\frac{1}{2\sqrt{E}}
\right]
\end{equation}
and
\begin{equation}\label{eq:deletav}
\delta\eta_E=\int_0^\infty\rmd{\zm}\,\delta v(\zm)\,\Im\,\mathcal{G}_{\pm,*}(\zm,\zm;E+\rmi 0).
\end{equation}

The first relation, Eq.~\eqref{eq:etaE}, may be viewed as a special case of the Lifshitz-Krein trace formula   \cite{Lif52,Kre53}
\begin{equation}\label{eq:Kr}
{\text {Tr}}\,[h(\mathcal{H})-h(\mathcal{H}^{(0)})]=-\frac{1}{\pi}\int_{-\infty}^\infty dE\, \eta_E\, h'(E)
\end{equation}
for the trace of the difference due to a change of Hamiltonians  $\mathcal{H}^{(0)}\to\mathcal{H}$. In the original Lifshitz-Krein formula~\eqref{eq:Kr} $h(E)$ stands for an arbitrary function $h(E)$. Remembering that $-\Im \Tr[E+\rmi 0-\mathcal{H}]^{-1}/\pi$ gives the density of states of $\mathcal{H}$ one sees that Eq.~\eqref{eq:etaE} follows formally from Eq.~\eqref{eq:Kr} if we choose $h(E)=\delta(E-E')$. In Appendix~\ref{app:scatphase} we give a direct proof of Eq.~\eqref{eq:etaE}. The second relation, Eq.~\eqref{eq:deletav}, is a direct consequence of Eq.~\eqref{eq:etaE} and also derived in this appendix.

Use of the above relations will enable us to derive from Eq.~\eqref{eq:delGpm} an equation involving $\eta_E$, $\delta \eta_E$, and $\delta v_0$. The first two quantities can be expressed in a straightforward manner in terms of the phase shift $\eta(\sk)$ and its variation $\delta\eta(\sk)$. Relating $\delta v_0$ to scattering data is more subtle but easily achieved with the aid of the following corollary of a trace formula proved in II \footnote{This  corollary follows directly from the trace formula  for the ${L=\infty}$ case of a semi-infinite system formulated as Theorem~\ref{thm:traceform} below.}.
\begin{corollary}
Let $v(\zm)$ and $\tilde{v}(\zm)$ be two continuous potentials on the half-line $(0,\infty)$ with the  following properties: 
\begin{enumerate} \item[(i)] They vanish sufficiently fast for $\zm\to\infty$ so that the integral
\begin{equation}
\int_{\zm_0}^\infty\rmd{z}\,|v(\zm)|<\infty 
\end{equation}
and its analog for $\tilde{v}(\zm)$ exist for all $\zm_0>0$.
\item[(ii)] They have the same singular behavior at $\zm=0$ specified in Eq.~\eqref{eq:vnearboundary}, with identical coefficients $v_{-1}$ and $\tilde{v}_{-1}$ though possibly different limiting values $v_0$ and $\tilde{v}_0$ of their regular parts.
\item[(iii)] The Schr\"odinger equation $\mathcal{H}_v\psi(\zm)=\varepsilon\,\psi(\zm)$ subject to the boundary condition $\psi(\zm\to 0)=O(\sqrt{\zm})$ and its analog with $v\to\tilde{v}$ have no bound-state solutions.
\end{enumerate} 
Then the following relation holds between the difference of the latter coefficients and  $\sigma(\sk)$, the logarithm of the scattering amplitude introduced in Eq.~\eqref{eq:varphiassigma}:
\begin{equation}\label{eq:vvtildesigma}
v_0-\tilde{v}_0=\frac{4}{\pi}\int_0^\infty\rmd{\sk}\,\sk^2\left[\rme^{-2\tilde{\sigma}(\sk)}-\rme^{-2\sigma(\sk)}\right].
\end{equation}
\end{corollary}

We set $\tilde{v}=v+\delta v$ and $\tilde{\sigma}=\sigma+\delta \sigma$ in Eq.~\eqref{eq:vvtildesigma} and linearize in $\delta \sigma$. Taking into account that $\sigma(\sk)$ is an even function of $\sk$, we thus arrive at the relation
\begin{equation}\label{eq:delv0sigma}
\delta v_0=\frac{4}{\pi}\int_{-\infty}^\infty\rmd{\sk}\,\sk^2\rme^{-2\sigma(\sk)}\,\delta\sigma(\sk).
\end{equation}

The implementation of the above strategy requires separate considerations in the cases $m=\pm 1$. We begin with the $m=+1$ case.

\subsection{Scattering data for the half-space problem with $m=+1$}\label{sec:hssdmp}
\begin{figure}[t]
\includegraphics[width=0.9\columnwidth]{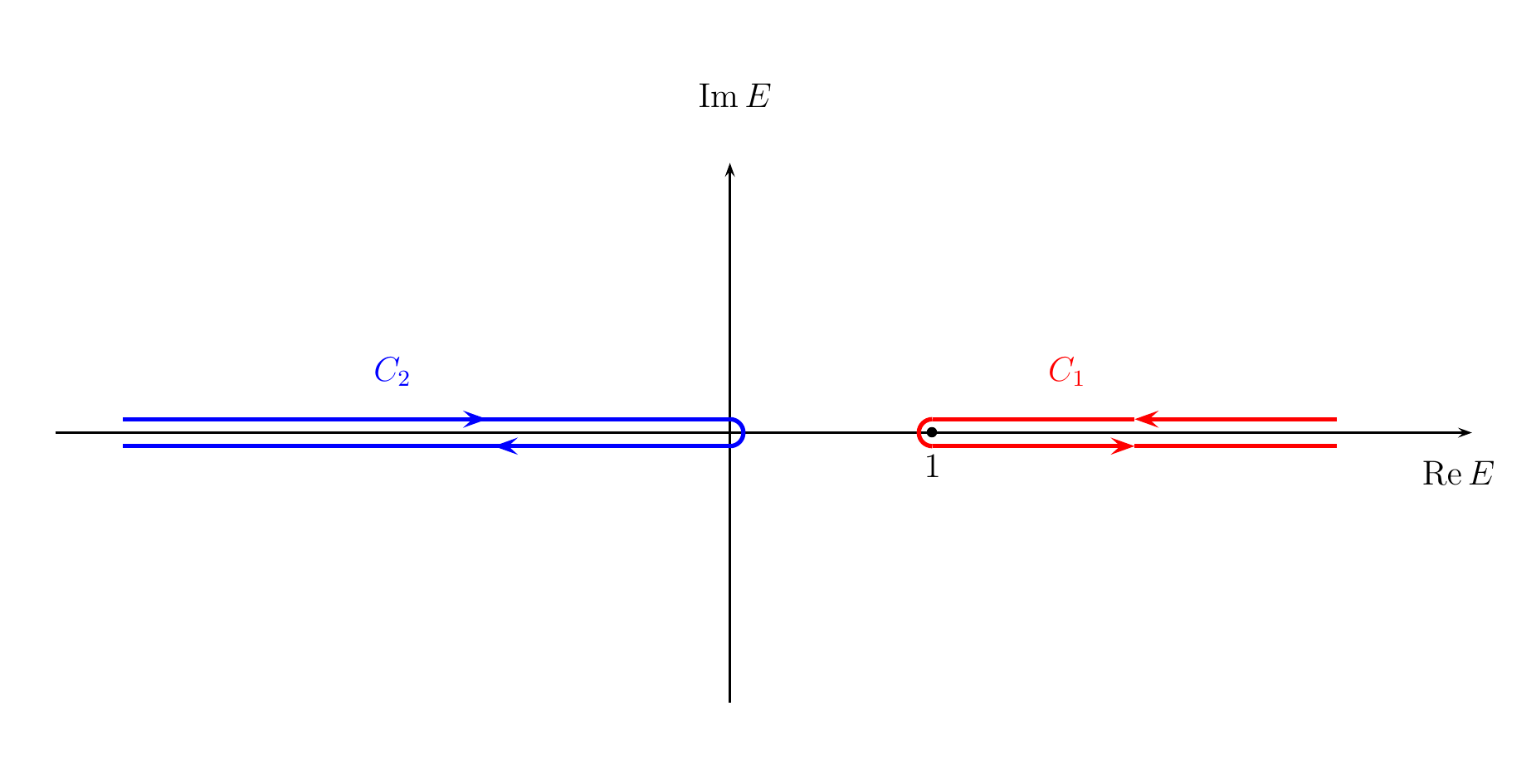}
\caption{Integration paths in Eq.~\eqref{eq:SCE2}.}
\label{fig:Int7}
\end{figure}

We choose the plus sign in Eq.~\eqref{eq:Gscepm} and rewrite the integral as a contour integral along the path $\mathcal{C}_2$  shown in Fig.~\ref{fig:Int7}, obtaining
\begin{eqnarray}\label{eq:SCE2}
\frac{\delta(\zm)}{2}&=&\int_{-\infty}^0 \rmd{E}\left[
\mathcal{G}_{+,*}(\zm,\zm,E)+\frac{1}{2 \sqrt{1-E}}\right]\nonumber\\
&=&\int_{\mathcal{C}_2} \frac{\rmd E}{2\pi \rmi}\ln E \left[
\mathcal{G}_{+,*}(\zm,\zm,E)+\frac{1}{2 \sqrt{1-E}}\right].\qquad
\end{eqnarray}
The contour $\mathcal{C}_2$ can be deformed into $\mathcal{C}_1$, which gives
\begin{equation}
\int_{1}^\infty \frac{\rmd E}{2\pi }\ln E\left[
\Im\,\mathcal{G}_{+,*}(\zm,\zm,E+\rmi 0)+\frac{1}{2 \sqrt{E-1}}\right]=-\frac{\delta(\zm)}{4}.
\end{equation}
We then multiply by $\delta v(\zm)$, integrate over $\zm$, use the properties (iii) and (iv) specified in Eq.~\eqref{eq:delvprop}, change the order of integrations in the double integral over $\zm$ and $E$, and  make use of Eq.~\eqref{eq:deletav}
to perform the integration over $\zm$.  Making the change of integration variable $E=\sk^2+1$, we then rewrite the resulting integral $\int_0^\infty\rmd{\sk^2}$ as an integral $\int_{-\infty}^\infty\rmd{\sk}$, express $\eta_E$ and $\delta\eta_E$ in terms of the phase shifts $\eta(\sk)$ and $\delta\eta(\sk)$, and take into account that the latter two quantities are odd in $\sk$. One thus arrives at
\begin{equation}\label{eq:deletaeq1}
\int_{-\infty}^\infty\frac{\rmd{\sk}}{2\pi}\,\sk\ln(\sk^2+1)\,\delta\eta(\sk)+\frac{1}{4}\,\delta v_0=0.
\end{equation}

For $\delta v_0$ we can substitute Eq.~\eqref{eq:delv0sigma}. However, the variations $\delta\eta(\sk)$ and $\delta\sigma(\sk)$ are not independent. To see this, recall from Eq.~\eqref{eq:FJostk} that $\sigma(\sk)$ and $-\eta(\sk)$ are the real and imaginary parts, respectively,  of the logarithm of the Jost function, $\ln F(\sk)$. Since the Hamiltonian $\mathcal{H}_{v_+}$ has no discrete spectrum, $F(\sk)$ has no zeros in the upper half-plane $\Im\, \sk>0$. Therefore, $\ln F(\sk)$ and $\delta\ln F(\sk)$ are analytic there. Moreover, $\delta\ln F(\sk)$ can be shown to vanish faster than $|\sk|^{-1}$ as $\sk\to\infty$ in the upper half-plane: we show in Appendix~\ref{app:semiclass} that \footnote{Here property (iv) enters because it is required for $\alpha$ not to change.}
\begin{equation}\label{eq:dellnFas}
\delta\ln F(\sk)\mathop{=}_{\substack{\sk\to\infty\\[0.1em] \Im\, \sk>0}}O\big(|\sk|^{-3}\big);
\end{equation}
see Eq.~\eqref{eq:Fplargek}. It follows that $\delta\sigma(\sk)$ and $\delta\eta(\sk)$ are related via the Kramers-Kronig relations
\begin{subequations}\label{eq:KKsigmaeta}
\begin{equation}
\delta\eta(\sk)=\HT[\delta\sigma](\sk)\equiv\frac{1}{\pi}\mathcal{P}\int_{-\infty}^\infty\rmd{\sk'}\,\frac{\delta\sigma(\sk')}{\sk'-\sk}\label{eq:KKsigmaetaa}
\end{equation}
and
\begin{equation}
\delta\sigma(\sk)=\HT^{-1}[\delta\eta](\sk)=-\frac{1}{\pi}\mathcal{P}\int_{-\infty}^\infty\rmd{\sk'}\,\frac{\delta\eta(\sk')}{\sk'-\sk}\,,\label{eq:KKsigmaetab}
\end{equation}
\end{subequations}
where $\mathcal{P}$ denotes the Cauchy principal value and $\HT[f]$ is the Hilbert transform of $f$ (see, e.g.,  \S 4.2 of \cite{MF53}).

Another consequence of Eq.~\eqref{eq:dellnFas} is the constraint
\begin{equation}\label{eq:deletakconstr}
\int_{-\infty}^\infty\rmd{\sk}\,\sk\,\delta\eta(\sk)=0.
\end{equation}
This follows from the fact that the integral on the left-hand side can be written as  $-\Im\int_{-\infty}^\infty\rmd{\sk}\,\sk\,\delta\ln F(\sk)$, which vanishes because the integration contour along the real axis can be completed to a closed loop in the upper complex half-plane by a half-circle of infinite radius.

We now insert Eq.~\eqref{eq:KKsigmaetaa} into Eq.~\eqref{eq:delv0sigma}. Upon exploiting the antisymmetry of $\delta\eta(\sk)$ together with the constraint~\eqref{eq:deletakconstr}, one sees that $\delta v_0$ can be written as
\begin{equation}\label{eq:delv0sigmares}
\delta v_0=\frac{4}{\pi^2}\int_{-\infty}^\infty\rmd{\sk}\,\sk^2\,\delta\eta(\sk)\,\mathcal{P}\int_{-\infty}^\infty\rmd{\sk'}\,\frac{\rme^{-2\sigma(\sk')}}{\sk'-\sk}.
\end{equation}
The result can be inserted into Eq.~\eqref{eq:deletaeq1}, but we must take into account the constraint~\eqref{eq:deletakconstr}, which we do by means of a Lagrange multiplier $\Lambda/2\pi$. Equating the coefficient of $\delta\eta(\sk)$ to zero then gives us an integral equation for $\sigma_+(\sk)$, namely
\begin{equation}\label{eq:inteqsigmap}
 \frac{\ln(\sk^2+1)}{\sk}+\frac{2}{\pi}\,\mathcal{P}\int_{-\infty}^\infty\rmd{\sk'}\,\frac{\rme^{-2\sigma_+(\sk')}}{\sk'-\sk}+\frac{\Lambda}{ \sk}=0,
\end{equation}
where we have reintroduced the subscript $+$ to remind us that $\sigma_+(\sk)$ refers to the $m=+1$ case.

We claim that the solution to this equation is
\begin{equation}\label{eq:sigmasol}
\sigma_+(\sk)=-\frac{1}{2}\ln\big(\sk^{-1}\arctan \sk\big),
\end{equation}
which implies the remarkably simple result 
\begin{equation}\label{eq:Asol}
A_+(\sk)=\sqrt{\sk/\arctan \sk}
\end{equation}
for the scattering amplitude. 

To show this, let us define the functions
\begin{equation}
g_\gtrless(\sk)\equiv \frac{1}{2\pi\rmi}\int_{-\infty}^\infty\rmd{\sk'}\,\frac{\rme^{-2\sigma_+(\sk')}}{\sk'-\sk}\;\; \text{ for }\Im\, \sk\gtrless 0.
\end{equation}
The functions $g_>(\sk)$ and $g_<(\sk)$ are analytical for $\Im\, \sk>0$ and $\Im\, \sk<0$, respectively, and vanish as $\sk\to\infty$ in the respective upper or lower complex half-plane. For real $\sk$ one has
\begin{equation}\label{eq:gpmgm}
g_>(\sk)-g_<(\sk)=\rme^{-2\sigma_+(\sk)}
\end{equation}
and
\begin{equation}
2\rmi [g_>(\sk)+g_<(\sk)]=\frac{2}{\pi}\mathcal{P}\int_{-\infty}^\infty\rmd{\sk}'\,\frac{\rme^{-2\sigma_+(\sk')}}{\sk'-\sk}.
\end{equation}
Using the latter equation to eliminate the principal value integral in Eq.~\eqref{eq:inteqsigmap}, we are led to
\begin{equation}\label{eq:raentiref}
\frac{\ln(1-\rmi \sk)}{\sk}+2\rmi\,g_>(\sk)+\frac{\Lambda}{\sk}=-\frac{\ln(1+\rmi \sk)}{\sk}-2\rmi\,g_<(\sk).
\end{equation}

The left-hand and right-hand sides of this equation are analytical in the upper and lower complex half-plane, respectively, and vanish there as $\sk\to\infty$. Hence both sides define a bounded entire function which must be identical to zero, just as $\Lambda=0$.  Hence
\begin{equation}
g_\gtrless(\sk)=-\frac{\ln(1\mp\rmi \sk)}{2\rmi \sk},
\end{equation}
from which the claimed solution~\eqref{eq:sigmasol} follows at once via Eq.~\eqref{eq:gpmgm}. As a consistency check, one can insert Eq.~\eqref{eq:gpmgm} for $\sigma_+(\sk)$, set $\Lambda=0$ in Eq.~\eqref{eq:inteqsigmap}, and compute the principal value integral to confirm that this equation is fulfilled.

Since the Kramers-Kronig relations~\eqref{eq:KKsigmaeta} carry over to  $\eta_+'(\sk)$ and $\sigma_+'(\sk)$, we can compute the derivative of the phase shift $\eta_+(\sk)$ from the result~\eqref{eq:sigmasol}. One finds
\begin{equation}\label{eq:etapp}
\eta_+'(\sk)=\frac{1}{2\pi}\mathcal{P}\int_{-\infty}^\infty\rmd{\sk'}\,\frac{1}{\sk'-\sk}\bigg[\frac{1}{\sk^\prime}-\frac{1}{(\sk^{\prime 2}+1)\arctan \sk'}\bigg].
\end{equation}

The integral $\mathcal{P}\int_{-\infty}^{\infty}\ldots\rmd{\sk'}$ equals the real part of  $\int_{-\infty+\rmi 0}^{\infty+\rmi 0}\ldots\rmd{\sk'}$.  We now transform to the variable $\beta=\arctan \sk'$. Then the latter integral transforms into an integral in the complex $\beta$-plane along a path infinitesimally above the real axis from $-\pi/2+\rmi 0$ to $-\pi/2+\rmi 0$. This path can be deformed such that it becomes the sum of two  integrals parallel to the imaginary axis, one  from $(-\pi/2+\rmi 0)$ to $(-\pi/2+\rmi \infty)$ and another one from $(-\pi/2+\rmi 0)$ to $(-\pi/2+\rmi \infty)$, and a vanishing contribution at infinity. It follows that
\begin{equation}\label{eq:etaprimempir}
\eta_+'(\sk)=\frac{1}{2}\int_0^\infty\frac{\coth u}{\sk^2+\coth^2 u}\,\frac{\rmd{u}}{u^2+\pi^2/4},
\end{equation}
where $u=\Im\, \beta$. 

Below we shall need the limiting behavior of $\eta'(\sk)$ for $\sk\to\infty$. In order to determine this, we split off the analytically computable term
\begin{equation}
I_1(\sk)=\frac{1}{2}\int_0^\infty\frac{1/u}{\sk^2+1/u^2}\,\frac{\rmd{u}}{u^2+\pi^2/4}=\frac{2\ln(\sk\pi/2)}{\sk^2\pi^2-4}
\end{equation}
from the integral on the right-hand side of Eq.~\eqref{eq:etaprimempir}. In the remaining integral, one can expand the integrand in powers of $\sk^{-2}$.  One thus obtains 
\begin{equation}\label{eq:etaprimempas}
\eta_+'(\sk\to\infty)=\frac{2}{\sk^2\pi^2}\ln\frac{\sk\pi}{2} +\frac{J_+}{2\sk^2} +O(\sk^{-4}\ln \sk)
\end{equation}
with
\begin{equation}\label{eq:Jdef}
J_+=\int_0^\infty\rmd{u}\,\frac{\coth(u)-1/u}{u^2+\pi^2/4}.
\end{equation}
The integral $J_+$ can be shown to have the series representation \footnote{To prove the series expansion~\eqref{eq:Jseries} we replace $\cosh(u)-1/u$ in the integrand by the series $\sum_{j=1}^\infty \frac{2u}{u^2+\pi^2j^2}$  and integrate termwise. The integration is straightforward and yields the result.}
\begin{equation}\label{eq:Jseries}
J_+=\frac{8}{\pi^2}\sum_{\sk=1}^\infty\frac{\ln(2\sk)}{4\sk^2-1}.
\end{equation}
Its value 
\begin{equation}\label{eq:Jplusnum}
J_+=0.474572605201526198964\ldots
\end{equation}
can be determined in a straightforward fashion  by numerical computation of either the series or the integral using {\sc Mathematica} \cite{Mathematica10}.

In order to determine $\eta_+(\sk)$, we integrate its derivative given in Eq.~\eqref{eq:etaprimempir}, using $\eta_+(0)=0$. Upon changing the order of integrations and performing the integration over $\sk$, we are  led to
\begin{equation}\label{eq:etapfres}
\eta_+(\sk)=\int_0^\infty\rmd{u}\,\frac{2\arctan(\sk\tanh u)}{4u^2+\pi^2}.
\end{equation}

The remaining integral can be evaluated numerically. The result is plotted below  in Fig.~\ref{fig:eta}  together with the corresponding phase shift $\eta_-(\sk)$ for $m=-1$, whose calculation we present in the next subsection. 

From the results for $\sigma_+(\sk)$ and $\eta_+(\sk)$,  one can also derive a helpful representation of the logarithm of the Jost function for $\Im\, \sk>0$, namely
\begin{equation}\label{eq:logFplus}
\ln F_+(\sk)=\int_0^\infty\frac{\rmd{\sk'}}{\rmi\pi}\bigg[\frac{1}{\sk'}-\frac{1}{(\sk^{\prime 2}+1)\arctan \sk'}\bigg]\text{arctanh}\frac{\sk}{\sk'}.
\end{equation}
We have benefited from it in our analysis of the asymptotic $x\to+\infty$ behavior of the free-energy scaling function $Y(x)$, which is introduced and discussed in Sec.~\ref{sec:freeenergyfunc}.

To obtain this representation, we computed $\sigma'_+(\sk)$ from Eq.~\eqref{eq:sigmasol}, used the result for $\eta'_+(\sk)$ given in Eq.~\eqref{eq:etapp}, inserted these expressions into  $\rmd\ln F_+(\sk)/\rmd\sk=\sigma_+'(\sk)-\rmi\eta_+'(\sk)$, and combined both contributions into an integral of the form $\int_0^\infty\frac{\rmd{\sk'}}{2\pi\rmi}\,[\ldots]/(\sk'-\sk-\rmi 0)$. If one then chooses instead of the wavevector $\mathsf{K}=\sk+\rmi 0$ the purely imaginary one $\mathsf{K}=\rmi q$, $(\ln F_+(k))'|_{\sk=\rmi q}$ becomes real-valued and can be integrated to obtain $\ln F_+(\mathsf{K}=\rmi q)$. Equation~\eqref{eq:logFplus} finally follows by analytic continuation of the result to complex values of $\mathsf{K}$.

\subsection{Scattering data for the half-space problem at $m=-1$}\label{sec:hssdmm}

In the case of $m=-1$, we start again from Eq.~\eqref{eq:Gscepm}. Now the spectrum of the Hamiltonian becomes gapless, $\spek(\mathcal{H}_{v_-})=[0,\infty)$,  and the contribution $\propto \regsol_0^2(z)$ must be taken into account. The presence of the half-bound state $\regsol_0(z)$ implies that the Jost function $F(\sk)$ now has a zero at the origin, $F(0)=0$. To proceed, we rewrite the integral in Eq.~\eqref{eq:Gscepm} as $\lim_{\epsilon\to 0+}\int_{-\infty}^{-\epsilon}\rmd{z}$,  transform it into
\begin{equation*}
-\lim_{\epsilon\to 0+}\int_\epsilon^\infty\frac{\rmd{E}}{\pi}\ln E\left[\Im\, \mathcal{G}_-(z,z;E+\rmi 0)+E^{-1/2}/2\right],
\end{equation*}
and take the limit $\epsilon\to 0+$. We then multiply by $\delta v(z)$ and integrate over $z$. We thus arrive at
\begin{align}\label{eq:delvintmm}
&\int_0^\infty\rmd{z}\,\delta v(z)\int_0^\infty\frac{\rmd{E}}{2\pi}\ln E\left[\Im\,\mathcal{G}_-(z,z;E+\rmi 0)+\frac{1}{2\sqrt{E}}\right]\nonumber\\ &=\frac{1}{2}\int_0^\infty\rmd{z}\,[1-\regsol_0^2(z)]\delta v(z)-\frac{1}{4}\,\delta v_0.
\end{align}

The function $\delta v(z)$ is  required to have the properties (i)--(iv) known from the ${m=+1}$~case and specified in  Eq.~\eqref{eq:delvprop}. However, we also require that $\delta v$ does not shift the eigenenergy $E_1=0$ associated with the half-bound state $\regsol_0(z)$ to linear order in $\delta v$. This means that the expectation value $\langle\regsol_{0} |\delta v|\regsol_{0}\rangle$ must vanish, i.e., the fifth property $\delta v(z)$ must satisfy is
\begin{align}\label{eq:delvprop5}
&\text{(v)}&\int_0^\infty\rmd{z}\,\delta v(z)\,\regsol_0^2(z)=0.
\end{align}
Thus the integral on the right-hand side and the contribution from $1/(2\sqrt{E})$ on the left-hand side of Eq.~\eqref{eq:delvintmm}  both vanish. Setting $E=\sk^2$ and using Eqs.~\eqref{eq:deletav} and \eqref{eq:delv0sigma} along with the Kramers-Kronig relation~\eqref{eq:KKsigmaetaa}, we obtain
\begin{align}\label{eq:eqdeletamm}
&\int_{-\infty}^\infty\frac{\rmd{\sk}}{2\pi}\,\delta\eta_-(\sk)\,\sk \ln \sk^2=\frac{-1}{4}\,\delta v_0\nonumber\\ &=\frac{1}{\pi^2}\int_{-\infty}^\infty\rmd{\sk'}\,\sk^{\prime 2}\,\rme^{-2\sigma_-(\sk')}\mathcal{P}\int_{-\infty}^\infty\rmd{\sk}\,\frac{\delta\eta_-(\sk)}{\sk-\sk'}. 
\end{align}
Just as in the $m=+1$ case, we could  replace the  factor $\sk^{\prime 2}$ by a factor $\sk^2$ in the principal value integral. However, the subsequent reasoning that either side of Eq.~\eqref{eq:raentiref} equals an entire function is not applicable here because the integrand involves the function $\ln \sk^2$ rather than $\ln(\sk^2+1)$, which is not analytic at $\sk=0$. 

We assert that Eq.~\eqref{eq:eqdeletamm} has solutions of the simple form
\begin{equation}\label{eq:sigmsol}
\rme^{-2\sigma_-(\sk)}=\frac{\Lambda_-}{\sk^2}+\frac{\pi}{2|\sk|},
\end{equation}
where $\Lambda_-\in \mathbb{R}$ is a parameter that needs to be determined. In order to show this, we substitute this equation into Eq.~\eqref{eq:eqdeletamm} and replace $\int_{-\infty}^\infty\rmd{\sk'}$ by $\lim_{K\to\infty}\int_{-K}^K\rmd{\sk'}$. Since $\delta\eta_-(\sk)=O(\sk^{-3})$ as $\sk\to\infty$ by Eq.~\eqref{eq:dellnFas}, its Hilbert transform should exist and behave such that the remaining $\sk'$ integral is convergent. Our rewriting of the $\sk'$-integral as a principal value was necessary to enable us to interchange the order of the two integrations. This leads us to
\begin{equation} \label{eq:delv0rewritten}
\frac{-\delta v_0}{4}=\lim_{K\to\infty}\int_{-\infty}^\infty\frac{\rmd{\sk}}{\pi^2}\,\delta\eta_-(\sk)\,\mathcal{P}\int_{-K}^K\rmd{\sk'}\,\frac{\Lambda_-+\pi|\sk'|/2}{\sk-\sk'}.
\end{equation}
The second integral gives
\begin{eqnarray}
\lefteqn{\mathcal{P}\int_{-K}^K\rmd{\sk'}\,\frac{\Lambda_-+\pi|\sk'|/2}{\sk-\sk'}}&&\nonumber\\ &=&2\Lambda_-\,\mathrm{artanh}\frac{\sk}{K}+\frac{\sk\pi}{2}\ln\frac{\sk^2}{K^2-\sk^2}\nonumber\\
&=&-\pi \sk\ln K +\frac{\pi \sk}{2}\ln \sk^2+O(K^{-1}).
\end{eqnarray}
Owing to the constraint~\eqref{eq:deletakconstr}, the $\sk \ln K$ term does not contribute to $\int_{-\infty}^\infty\rmd{\sk}\,\delta\eta_-(\sk)\ldots$. Hence the limit $K\to\infty$ of Eq.~\eqref{eq:delv0rewritten} indeed yields the left-hand side of Eq.~\eqref{eq:eqdeletamm}.

It remains to determine the constant $\Lambda_-$. To do this, we compare the solution~\eqref{eq:sigmsol} with the asymptotic form 
\begin{equation}\label{eq:emin2sigminlargek}
\rme^{-2\sigma_-(\sk)}\mathop{=}_{\sk\to \infty} \frac{\pi}{2\sk}-\frac{\pi^2}{4\sk^2}\,v^{-}_{-1}+O(1/\sk^4).
\end{equation}
for large $\sk$ derived in Appendix~\ref{app:semiclass}. Insertion of the value of $v_{-1}^-$ given in Eq.~\eqref{eq:vm1} then yields
\begin{equation}\label{eq:Lambdam}
\Lambda_-=1.
\end{equation}
Hence our results for $\sigma_-(\sk)$  become
\begin{eqnarray}\label{eq:sigmsolf}
\rme^{-2\sigma_-(\sk)}&=&\frac{1}{\sk^2}+\frac{\pi}{2|\sk|}\,,\nonumber\\
\sigma_-(\sk)&=&\frac{1}{2}\ln\frac{\sk^2}{1+\pi|\sk|/2}=\ln A_-(\sk)\,,\nonumber\\
\sigma'_-(\sk)&=&\frac{1}{\sk}-\frac{\pi\,\text{sgn}(\sk)}{2\pi |\sk|+4}\,.
\end{eqnarray}

To compute $\eta_-(\sk)$ we first determine $\eta_-'(\sk)$ using the Kramers-Kronig relation~\eqref{eq:KKsigmaetaa}. The Hilbert transform of the contribution $1/\sk$ to $\sigma_-'(\sk)$, considered as a distribution, gives $\HT[1/\sk](\sk)=\pi\,\delta(\sk)$. For $\sk\ne 0$, it has no support and can be dropped. The Hilbert transform of the remaining term, $\sigma_-'(\sk)-1/\sk$, can be computed in a straightforward fashion. One gets
\begin{equation}\label{eq:etamprimefres}
\eta_-'(\sk)=\frac{\ln(\pi^2\sk^2/4)}{4-\pi^2\sk^2}.
\end{equation}

This can be integrated in a straightforward fashion. To fix the integration constant, we use the fact that a variant of Levinson's theorem exists for the Sturm-Liouville problem we are concerned with which predicts that $\eta_-(0\pm)=\pm\pi/2$ in the presence of the half-bound state $\regsol_0(z)$ (see Eq.~(4.43) of \cite{Ma06}). It follows that for real $\sk$
\begin{align}\label{eq:etamfres} 
 \eta_-(\sk)=\sgn(\sk)\Bigg\{&\frac{\pi}{2}+\frac{1}{2\pi}\bigg[\mathrm{Li}_2(-\pi |\sk|/2)-\mathrm{Li}_2(\pi |\sk|/2)\nonumber\\ 
\strut &-\ln\bigg(\frac{\pi |\sk|}{2}\bigg)\ln\frac{2-\pi |\sk|}{2+\pi |\sk|}\bigg]\Bigg\},
\end{align}
where $\mathrm{Li}_2(\sk)$ is the dilogarithm (polylogarithm $\mathrm{Li}_s(\sk)$ for $s=2$) \cite{AS72,NIST:DLMF,Olver:2010:NHMF}.

\subsection{The phase shifts $\eta_\pm(\sk)$ for $m=\pm1$ and the scattering data for $m=0$}

\begin{figure}[htbp]
\includegraphics[width=0.9\columnwidth]{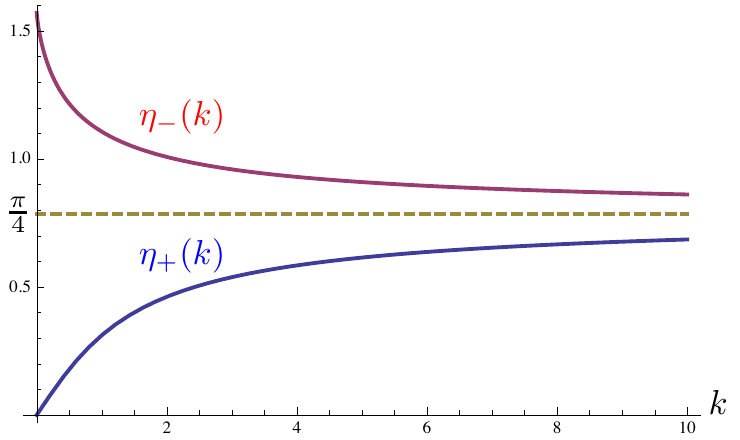}
\caption{The phase shifts $\eta_-(k)$ for $m=-1$ (red)  and $\eta_+(k)$ for 
$m=1$ (blue). The dashed  horizontal line shows the momentum-independent phase shift $\eta_0(k)=\pi/4$ for $m=0$.}
\label{fig:eta}
\end{figure}

Figure~\ref{fig:eta} shows our results for $\eta_\pm(\sk)$ given in Eqs.~\eqref{eq:etapfres}  and \eqref{eq:etamfres}.
To plot $\eta_+(\sk)$, we evaluated the integral on the right-hand side of Eq.~\eqref{eq:etapfres} by numerical means using {\sc Mathematica} \cite{Mathematica10}. We included the $k$-independent result $\eta_0(k)=\pi/4$ for the critical case $m=0$. This follows from the fact that the exactly known eigenfunctions $\regsol(z,k;{L=\infty},{m=0})$  of the critical potential \eqref{eq:vord} for $d=3$ \cite{BM77c} behave as 
\begin{equation}\label{eq:varphicrit}
\regsol(z,k;\infty,0)=\sqrt{z}J_0(kz)\mathop{\aseq}_{z\to\infty}\sqrt{\frac{2}{\pi k}}\sin(kz+\pi/4).
\end{equation}

Comparing with Eq.~\eqref{eq:varphiassigma}, we see that the  associated
scattering amplitude and phase shift are given by
\begin{equation}\label{eq:scatdatm0}
A_0(k)=\rme^{\sigma_0(k)}=\sqrt{\frac{2|k|}{\pi}},\quad\eta_0(\sk)=\frac{\pi}{4}\,\sgn(\sk).
\end{equation}

 \subsection{The half-bound state}
 Using the above results, we  can now also verify that the half-bound state $\regsol_0$ agrees with the ${\sk\to 0+}$~limit of the regular solution $\regsol(\zm,\sk)$, as stated  in Eq.~\eqref{eq:varphi0klim}.
Using the analog of Eq.~\eqref{eq:varphicrit} for the regular solution with $m<0$ along with our results for the scattering data  given in Eq.~\eqref{eq:sigmsolf}, we arrive at
 \begin{eqnarray}
 \regsol(\zm,\sk)&\mathop{\aseq}\limits_{\zm\to\infty}&\frac{\rme^{\sigma_-(\sk)}}{\sk}\,\sin[\sk\zm+\eta_-(\sk)]\nonumber\\ &=&(1+\sk\,\pi/2)^{-1/2}\sin[\sk\,\zm+\eta_-(\sk)].
 \end{eqnarray}
 Since $\eta_-(\sk)=\pi/2+O(\sk\ln\sk)$, the sine function becomes $\cos[\sk\,\zm+O(\sk\ln\sk)]$ and approaches $1$ in the limit $\sk\to 0+$. It follows that $\lim_{\sk\to 0}\regsol(\zm,\sk)$ takes the value $1$ at $\zm=\infty$, satisfies the normalization condition~\eqref{eq:varphiinf} of $\regsol_0$, and hence must be identical to it. 
 
 \subsection{Distinct behaviors of the Jost functions $F_\pm(\sk)$ in the complex plane}
 
\begin{figure}[htbp]
\begin{center}
\includegraphics[width=0.6\columnwidth]{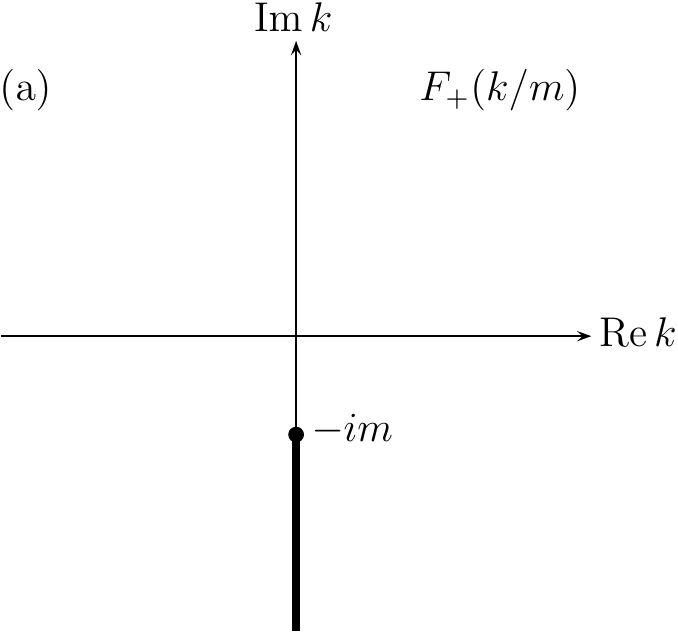}\\[2em]
\includegraphics[width=0.6\columnwidth]{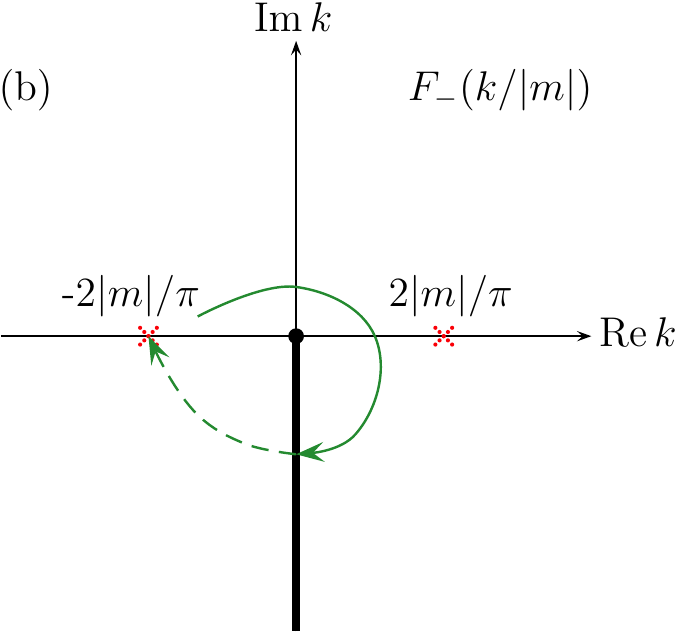}
\caption{Branch cuts of the Jost functions $F_\pm(k/|m|)$ for $m>0$ (a) and $m<0$ (b). The red crosses in (b) indicate the poles and zeros that occur successively on the  higher Riemann sheets.}
\label{fig:Fplusmin}
\end{center}
\end{figure}

 When we determined above the scattering data for the cases $m>0$ and $m<0$, we saw that the qualitatively distinct spectra of the respective Hamiltonians $\mathcal{H}_{v_\pm}$ entail qualitatively different behaviors of the Jost functions $F_\pm(\sk)$. While $F_+({\sk=k/|m|})$ has no zeros in the  upper half-plane $\Im\,k>0$, $F_-(k/|m|)$ has a zero at the origin, reflecting the presence of the half-bound state discussed in the previous subsection. The difference of the spectra for $m>0$ and $m<0$ manifests itself also in the distinct analytical properties of the functions $F_\pm(k/|m|)$. The gap of $\spek(\mathcal{H}_{v_+})$ implies that $F_+(k/m)$ is analytical in the complex $k$-plane except for a branch cut from $-\rmi m$ to $-\rmi\infty$ along the negative $\Im\,k$-axis  (see Fig.~\ref{fig:Fplusmin}a). The distance $m$ of the end point of this branch cut from the origin corresponds to the inverse of the bulk correlation length $1/m$. When $m<0$, the branch cut runs from $k=0$ to $-\rmi\infty$ along the negative $\Im\,k$ axis. From Eqs.~\eqref{eq:sigmsolf} and \eqref{eq:etamfres} one can see that the Jost function $F_-(k/|m|)$ is analytical on the first (physical) Riemann sheet but has successively poles and zeros  at $\pm2|m|/\pi$ on higher Riemann sheets, depending on the number of $2\pi$ turns (see Fig~\ref{fig:Fplusmin}b). 
The distance of the pole at $\Re\, k>0$ from the origin sets the scale for the algebraic decay of the bulk correlation function and corresponds to the Josephson coherence length $\propto 1/|m|$, as  our results for the two-point functions to be presented in the next section explicitly show.
 
 \section{Results for two-point correlation functions from scattering data}\label{sec:istres2pt}
 \subsection{Relation of two-point functions to scattering data}
 We will now use the scattering data determined in the previous section to calculate the scaling function of the $L=\infty$ two-point function for $m\gtreqless 0$. To this end we must express the scaling function in terms of scattering data. It will be helpful to briefly return to the case of finite thickness $L$.
 
 We define the two-point correlation function and cumulant by
 \begin{eqnarray}
 G^{(2)}(\bm{y},z,z';L,m)&=&\lim_{n\to\infty}\langle\bm{\phi}(\bm{y},z)\cdot\bm{\phi}(\bm{0},z')\rangle/n\nonumber\\ &=&\langle\Phi(\bm{y},z)\,\Phi(\bm{0},z')\rangle,
 \end{eqnarray}
 and 
 \begin{eqnarray}
  \lefteqn{G^{(2)}_{\text{cum}}(\bm{y},z,z';L,m)}&&\nonumber\\ &=& G^{(2)}(\bm{y},z,z';L,m)-\langle\Phi(\bm{y},z)\rangle\langle\Phi(\bm{0},z')\rangle,
 \end{eqnarray}
where $\bm{\phi}(\bm{y},z)$ and $\Phi(\bm{y},z)$ are the continuum analogs of $\bm{\phi}_{\bm{j}}$ and $\Phi_{\bm{j}}$, respectively, and we have taken into account the translation invariance parallel to the boundary planes. Since the $O(n)$ symmetry cannot be spontaneously broken at temperatures $T>0$ when $L<\infty$, the two-point function approaches zero as the separation of the two points tends to infinity; i.e., 
\begin{equation}
\lim_{|\bm{y}|\to\infty}G^{(2)}(\bm{y},z,z';{L<\infty},m)=0.
\end{equation}
On the other hand, if we first take the limit $L\to\infty$ and subsequently the limit $|\bm{y}|\to\infty$, then the correlation function factorizes into a product of nonvanishing expectation values if $m<0$,
\begin{equation}
\lim_{|\bm{y}|\to\infty}G^{(2)}(\bm{y},z,z';\infty,m)=\langle\Phi(\bm{0},z)\rangle\langle\Phi(\bm{0},z')\rangle,
\end{equation}
so that $G^{(2)}_{\text{cum}}(\bm{y},z,z';\infty,m)$ vanishes in this limit.

Let us indicate Fourier transforms with respect to $\bm{y}$ by a tilde, writing
 \begin{equation}
 f(\bm{y})=\int\frac{\rmd^2p}{(2\pi)^2}\,\tilde{f}(\bm{p})\,\rme^{\rmi\bm{p}\cdot\bm{y}}.
 \end{equation}
 Dimensional analysis implies the scaling behavior
 \begin{align}\label{eq:G2scal}
&\tilde{G}^{(2)}_{\text{cum}}(\bm{p},z,z';\infty,m)\nonumber \\
&=\frac{1}{|m|}\tilde{G}^{(2)}_{\text{cum}}(\bm{p}/|m|,|m|z,|m|z';\infty,\pm 1)
\end{align}
for ${m\ne 0}$. The analog of this equation for $\langle\Phi(\bm{0},z)\rangle^2_{{L=\infty},m}$ and $m<0$, given in Eq.~\eqref{eq:G1squscal}, asserts that the half-bound state $\regsol_0$ describes the dependence of this quantity on $|m|z$. 

Before we address this issue, let us first consider the ${L\to\infty}$ limit in the  simpler $m> 0$ case.
 To this end we start from the analog of the spectral decomposition~\eqref{eq:Gd} for $\tilde{G}^{(2)}(\bm{p};z,z';L,m)$, expressing it in terms of eigenfunctions that are normalized such that
 \begin{eqnarray}\label{eq:noneigfctnorm}
 \lim_{z\to 0+}z^{-1/2}\,\noneigfct_\nu(z;L,m)&=&1,\\
  \lim_{z\to L-}(L-z)^{-1/2}\,\noneigfct_\nu(z;L,m)&=&(-1)^{\nu-1},
 \end{eqnarray}
and hence related to the orthonormalized ones, $\oneigfct_\mu$,  via
 \begin{equation}
\oneigfct_\nu(z;L,m)=\frac{\noneigfct_\nu(z;L,m)}{\sqrt{\normreg_\nu(L,m)}}
\end{equation}
with
\begin{equation}\label{eq:alphanu}
\normreg_\nu(L,m)\equiv\langle\noneigfct_\nu |\noneigfct_\nu\rangle=\int_0^L[\noneigfct_\nu(z;L,m)]^2\,\rmd{z}.
\end{equation}
This gives
 \begin{equation}\label{eq:G2L}
 \tilde{G}^{(2)}(\bm{p},z,z';L,m)=\sum_{\nu=1}^\infty\frac{\noneigfct_\nu(z;L,m)\,\noneigfct_\nu(z';L,m)}{\normreg_\nu(L,m)\,[p^2+E_\nu(L,m)]}.
 \end{equation}
 
Using Eq.~\eqref{eq:alphanu}, we can now determine the  limit $L\to\infty$ of $\normreg_\nu/L$. We can replace $\noneigfct_\nu(z;L,m)$ by its asymptotic form~\eqref{eq:varphiassigma} in the integral $L^{-1}\int_0^L\rmd{z}\ldots$  for all layers $z$ that are sufficiently far from the boundary planes. Near the boundary planes, this approximation is not justified. However, the integral of the difference between the asymptotic form and $\noneigfct_\nu$ is restricted to two boundary layers of thickness $\asprop 1/|m|$. Therefore, the error this approximation produces for $\normreg_\nu/L$ is of order $1/L$ and we have
 \begin{equation}\label{eq:alphanuas}
 \lim_{L\to\infty}L^{-1}\normreg_\nu(L,m)=\rme^{2\sigma(k)}\,\frac{1}{2k^2}.
 \end{equation}
 
Since  the spectrum is bounded away from zero when $m>0$, i.e., $E_\nu(L,m)\ge m$ for all $\nu$, we can use Eq.~\eqref{eq:alphanuas} along with $L^{-1}\sum_{\nu=1}^\infty \xrightarrow[L\to\infty]{}\pi^{-1}\int_0^\infty\rmd{k}$ and dimensional considerations to conclude that the $L=\infty$ limit of Eq.~\eqref{eq:G2L} is given by
 \begin{align}\label{eq:G2sigma}
&\tilde{G}^{(2)}(\bm{p},z,z';\infty,m\ge 0)\nonumber\\
&=\frac{1}{m}\int_0^\infty\frac{\rmd{\sk}}{\pi}\,\frac{2\sk^2\rme^{-2\sigma_+(\sk)}\,\noneigfct(mz,\sk;\infty,1)\,\noneigfct(mz',\sk;\infty,1)}{p^2/m^2+1+\sk^2},
 \end{align}
 where $\sk=k/m$.
  
When $m<0$, we must be more careful. Since $E_1\to 0$ as $L\to\infty$ according to Eq.~\eqref{eq:E1lim}, the contribution from the lowest-energy mode $\nu=1$ requires  separate considerations, even though the mode sum $\sum_{\nu=2}^\infty$ of the remaining modes can be treated as before. 

Using Eq.~\eqref{eq:varphipsi1}, we see that the contribution of the ${\nu=1}$ mode to $G^{(2)}(\bm{y},z,z';L,m)$ for large $L$ can be written as 
 \begin{align}
&\lim_{L\to\infty} \frac{1}{L}\int\frac{\rmd^2p}{(2\pi)^2}\,\frac{\rme^{\rmi\bm{p}\cdot\bm{y}}\,\regsol_0(|m|z)\,\regsol_0(|m|z')}{p^2+E_1(L,m)}\nonumber\\ &=\lim_{L\to\infty}\frac{\regsol_0(|m|z)\,\regsol_0(|m|z')}{2\pi L}\,K_0\left[|\bm{y}|\sqrt{E_1(L,m)}\right]\nonumber\\ &=\frac{|m|}{4\pi}\,\regsol_0(|m|z)\,\regsol_0(|m|z')\,.
 \end{align}
Here Eq.~\eqref{eq:E1lim} for $E_1$ and the asymptotic behavior $K_0(y\to 0)\aseq -\gammaE-\ln(y/2)$ of the Bessel function $K_0$ were used, where $\gamma_E$ is the Euler-Macheroni constant. 
 The derivation nicely illustrates the emergence of the  half-bound state $\regsol_0$ as $L\to\infty$ and of the square of the spontaneous magnetization in the (noncommuting) limits $\lim_{|\bm{y}|\to\infty}\lim_{L\to\infty}$ of $G^{(2)}$.  The result confirms Eq.~\eqref{eq:G1squscal} for the spontaneous order-parameter profile. 

Combining it with the contribution from the mode sum $\sum_{\nu=2}^\infty$ then yields
 \begin{align}
&\tilde{G}^{(2)}(\bm{p},z,z';\infty,m) - \tilde{G}_{\text{cum}}^{(2)}(\bm{p},z,z';\infty,m)\nonumber\\ &= \frac{|m|}{4\pi}\,\regsol_0(|m|z)\,\regsol_0(|m|z')\,\delta(\bm{p})
 \end{align}
 with
 \begin{align}\label{eq:Gtilde2cummneg}
&\tilde{G}_{\text{cum}}^{(2)}(\bm{p}, z, z';\infty,m< 0)\nonumber\\ &=\frac{1}{|m|} \int_0^\infty\frac{\rmd{\sk}}{\pi}\bigg[\frac{2\sk^2\rme^{-2\sigma_-(\sk)}}{p^2/m^2+\sk^2}\nonumber\\ &
\qquad\times \noneigfct(|m|z,\sk;\infty,-1)\,\noneigfct(|m|z',\sk;\infty,-1)\bigg]. \end{align}
 
 \subsection{Critical two-point cumulant}\label{sec:crit2ptfct}
 
 As a useful first application of Eq.~\eqref{eq:G2sigma}, let us verify that Bray and Moore's exact result \cite{BM77a,BM77c} for the critical two-point cumulant follows from it. Upon insertion of the scattering amplitude $\rme^{\sigma_0}$ given in Eq.~\eqref{eq:scatdatm0} together with the critical eigenfunctions~\eqref{eq:varphicrit}, we arrive at
 \begin{eqnarray}\label{eq:Ghat2crit}
 \tilde{G}^{(2)}(\bm{p},z,z';\infty,0)&=&\sqrt{zz'}\int_0^\infty\rmd{k}\,\frac{k\,J_0(k z)\,J_0(kz')}{p^2+k^2}\nonumber\\ &=& \sqrt{zz'}\,I_0(pz_<)\,K_0(pz_>),
 \end{eqnarray}
in precise agreement with \cite{BM77a,BM77c}. Transformed to position space, the result 
 \begin{subequations} \label{eq:G2crit}
\begin{eqnarray}\label{eq:cfiG2}
 \lefteqn{G^{(2)}(\bm{y},z,z';\infty,0)}&&\nonumber \\ &=&\frac{1}{2\pi}\frac{\sqrt{zz'}}{\sqrt{y^2+(z-z')^2}\,\sqrt{y^2+(z+z')^2}}\nonumber\\[\medskipamount]
 &=&\frac{1}{\sqrt{4zz'}}\,\mathcal{G}^{(2)}\bigg[\frac{y^2+(z-z')^2}{4zz'}\bigg]
 \end{eqnarray}
 takes the form dictated by conformal invariance under conformal transformations that leave the boundary plane $z=0$ invariant \cite{Car84,Car87}, where the scaling function $\mathcal{G}^{(2)}$ is given by
 \begin{equation}
 \mathcal{G}^{(2)}(\rho )=\frac{1}{4\pi\sqrt{ \rho(1+\rho)}}.
 \end{equation}
 \end{subequations}
 
 These results mean that $G^{(2)}$ complies with the familiar asymptotic  large-distance behaviors of the form
 \begin{equation}
 G^{(2)}(\bm{y},z,z')\asprop \begin{cases}{[y^2+(z-z')^2]}^{-\frac{d-2+\eta_\parallel}{2}},\\{[y^2+(z-z')^2]}^{-\frac{d-2+\eta_\perp}{2}},
 \end{cases}
 \end{equation}
as $|\bm{x}|\to \infty$ along a direction parallel to the surface plane or any nonparallel direction, respectively, where $d=3$, $\eta_\perp=1/2$, and $\eta_\|=1$ in our case \footnote{The series expansion of the surface exponents $\eta_\parallel$ and $\eta_\perp=(\eta_\parallel+\eta)/2$ at the ${d=3}$ ordinary transition are known to $O(1/n)$; see \cite{OO83b}}.\nocite{OO83b}

The above findings suggest the introduction of the surface operator
 \begin{equation}
 \Phi_{\mathrm{s}}(\bm{y})=\lim_{z\to0+}z^{-1/2}\,\Phi(\bm{y},z)
 \end{equation}
 and the associated correlation function
 \begin{equation}\label{eq:G11}
 G^{(1,1)}(\bm{y}-\bm{y}',z)\equiv\langle\Phi(\bm{y},z)\,\Phi_{\mathrm{s}}(\bm{y}')\rangle.
 \end{equation}
 Equation~\eqref{eq:G2crit} then yields
 \begin{equation}
 G^{(1,1)}(\bm{y},z)=\frac{\sqrt{z}}{2\pi(y^2+z^2)}
 \end{equation}
 for this function and the related surface two-point function
  \begin{equation}\label{eq:Gsdef}
 G^{(0,2)}(\bm{y})\equiv\lim_{\substack{z\to 0\\ \bm{y}\ne \bm{0}}}z^{-1/2}\,G^{(1,1)}(\bm{y},z)=\frac{1}{2\pi y^2}.
 \end{equation}
 
Here, the condition $\bm{y}\ne\bm{0}$ in the definition~\eqref{eq:Gsdef} is necessary. To see this, note that the Fourier transform of Eq.~\eqref{eq:G11} exists and agrees with the ${z'\to 0}$ limit of $(z')^{-1/2} \tilde{G}^{(2)}(\bm{p},z,z')$ computed from Eq.~\eqref{eq:G2crit}, namely
 \begin{equation}\label{eq:G11hat}
  \tilde{G}^{(1,1)}(\bm{p},z)=\sqrt{z}\,K_0(pz).
  \end{equation}
  
However, $G^{(0,2)}(\bm{y})$, when considered as an ordinary function rather than a generalized one, does not have a Fourier $\bm{p}$-transform. This is because in the derivation of Eq.~\eqref{eq:Gsdef}, we made use of the requirement $\bm{y}\ne\bm{0}$ and therefore ignored potential contributions $\propto \delta(\bm{y})$. To resolve the problem, we should regard the above Fourier $\bm{p}$-transforms as those of generalized functions in $\bm{y}$-space. 

Generalized functions of the form $y^\lambda=(y_1^2+y_2^2)^{\lambda/2}$ with negative integer exponents $\lambda$ are discussed in detail in \cite{GS64}. Let us denote the generalized function associated with $y^{-2}$ as $\mathcal{P}_2y^{-2}$ \footnote{In \cite{GS64}, the notation $y^{-2}$ is used for the latter generalized function.}, where the subscript $2$ reminds us that $\bm{y}\in\mathbb{R}^2$, and introduce an arbitrary momentum scale $\mu$ to make the integration variables dimensionless. Then $\mathcal{P}_2(\mu y)^{-2}$ is defined via its action on test functions $\upsilon(\mu\bm{y})$:
\begin{equation}
\left(\frac{\mathcal{P}_2}{\mu^2 y^2},\upsilon\right)\equiv\int_{|\hat{\bm{y}}|<1}\rmd^2\hat{y}\,\frac{\upsilon(\hat{\bm{y}})-\upsilon(\bm{0})}{\hat{y}^2}+\int_{|\bm{\hat{y}}|>1}\rmd^2\hat{y}\,\frac{\upsilon(\bm{\hat{y}})}{\hat{y}^2},
\end{equation}
where $\hat{\bm{y}}=\mu\bm{y}$. Its Fourier $\bm{p}$-transform is well defined and given by
\begin{equation}\label{eq:ptransformymin2}
\frac{1}{2\pi}\text{FT}\left[\frac{\mathcal{P}_2}{\mu^2y^2},\bm{p}\right]=-\gammaE-\ln\frac{p}{2\mu}.
\end{equation}

To proceed, note that the limit $z\to 0$ of
\begin{equation}
z^{-1/2}\tilde{G}^{(1,1)}(\bm{p},z)=z^{-1}\tilde{G}^{(2)}(\bm{p},z,z)=K_0(pz)
\end{equation}
does not exist because
\begin{equation}\label{eq:limbehK0}
K_0(pz)=-\gammaE-\ln(pz/2)+O(p^2z^2).
\end{equation}
To define a finite two-point surface correlation function, we must first subtract the logarithmic singularity $\asprop\ln z$ before taking the limit $z\to 0$. We choose this subtraction as
\begin{equation}\label{eq:Smuz}
S(\mu z)=-\gammaE-\ln(\mu z/2)
\end{equation}
and define
 \begin{align}\label{eq:Gshat}
 \tilde{G}^{(0,2)}(\bm{p})&\equiv\lim_{z\to 0}\left[z^{-1/2} \tilde{G}^{(1,1)}(\bm{p},z)- S(\mu z)\right]\nonumber \\ &=-\ln(p/\mu).
 \end{align}
From Eqs.~\eqref{eq:ptransformymin2} and \eqref{eq:Gshat} we see that the Fourier $\bm{p}$-transform of this function is
 \begin{equation}\label{eq:G02ycrit}
 G^{(0,2)}(\bm{y})=\frac{\mathcal{P}_2}{2\pi y^2}+(\gammaE-\ln 2)\,\delta(\bm{y}).
 \end{equation}
 
 \subsection{Two-point cumulant in the disordered phase}\label{sec:2ptdisord}
 
 We next consider the case of $m>0$. To determine the scaling function on the right-hand side of Eq.~\eqref{eq:G2scal}, we substitute the result~\eqref{eq:sigmsol} for $\rme^{-2\sigma_+(k)}$ in Eq.~\eqref{eq:G2sigma} and make a change of variable $\sk\to k=\sk\, m$. This gives
 \begin{align}
 &\tilde{G}^{(2)}(\bm{p},z,z';\infty,m)=\frac{1}{m}\int_0^\infty\frac{\rmd k}{\pi}\bigg[\frac{2k\arctan(k/m)}{p^2+k^2+m^2}\nonumber\\ &\times \,\noneigfct(mz,k/m;\infty,1)\noneigfct(mz',k/m;\infty,1)\bigg].
 \end{align}
 
 Note, first, that our results for the critical case are easily recovered  from the right-hand side by performing the limits
 $\lim_{m\to0}\arctan(k/m)=\pi/2$ and $\lim_{m\to0}m^{-1/2}\,\regsol(mz,k/m;\infty,1)=\sqrt{z}\,J_0(kz)$.
Second, we can divide by $\sqrt{z'}$ and take the limit $z'\to 0$ using the normalization~\eqref{eq:varphinormal} of the regular solution to obtain
 \begin{align}
 &\tilde{G}^{(1,1)}(\bm{p},z;\infty,m)\nonumber\\ &=\frac{1}{\sqrt{m}}\int_0^\infty\frac{\rmd k}{\pi}\,\frac{2k\arctan(k/m)\,\noneigfct(mz,k/m;\infty,1)}{p^2+k^2+m^2}\,.
 \end{align}

We next compute the $m>0$ analog of $\tilde{G}^{(0,2)}$ defined as in Eq.~\eqref{eq:Gshat}. To perform the required $z\to 0$ limit, we represent the subtraction term as
 \begin{equation}
 S(\mu z)=\int_0^\infty \rmd{k}\,\frac{k J_0(kz)}{\mu^2+k^2}+O[(\mu z)^2\ln(\mu z)],
 \end{equation}
 substitute $\arctan(k/m)$ by $\pi/2-\arctan(m/k)$, use  $\lim_{z\to0}(zm)^{-1/2}\regsol(mz,k/m;\infty,1)=1$, and exchange the limit $z\to 0$  with the integration. This gives
 \begin{eqnarray}
 \tilde{G}^{(0,2)}(\bm{p};\infty,m\ge 0)&=&\int_0^\infty \!\rmd{k}\bigg[\frac{k}{p^2+k^2+m^2}-\frac{k}{\mu^2+k^2}\bigg]
 \nonumber\\ &&
\strut -\frac{2}{\pi}\int_0^\infty \rmd{k}\,k\,\frac{\arctan(m/k)}{p^2+k^2+m^2}\nonumber\\
&=&\ln\bigg[\frac{\mu}{m+\sqrt{m^2+p^2}}\bigg].
\end{eqnarray}
 
To determine the Fourier back transform, it is convenient to consider the difference 
\begin{align}
&G^{(0,2)}(\bm{y};\infty,m\ge 0)-G^{(0,2)}(\bm{y};\infty,0)\nonumber\\
&=\int\frac{\rmd^2p}{(2\pi)^2}\,\rme^{\rmi\bm{p}\cdot\bm{y}}\ln\left[\frac{p}{m^2+\sqrt{m^2+p^2}}\right]\nonumber\\ &
=\int_0^\infty\frac{\rmd{p}}{2\pi}\,p\,J_0(py)\,\ln\left[\frac{p}{m+\sqrt{m^2+p^2}}\right]\nonumber\\ 
&=\frac{1}{2\pi y^2}\left(\rme^{-my}-1\right),
\end{align}
where the last line follows from the integral listed as 6.775 in \cite{GR80}. Upon combining the result with Eq.~\eqref{eq:G02ycrit}, we obtain
\begin{equation}\label{eq:G02ympos}
G^{(0,2)}(\bm{y};\infty,m\ge 0)=\frac{\mathcal{P}_2}{2\pi y^2}\,\rme^{-my}+(\gammaE-\ln2)\,\delta(\bm{y}).
\end{equation}
If one assumes $\bm{y}\ne\bm{0}$, then $G^{(0,2)}(\bm{y};\infty,m\ge 0)$ can be considered as an ordinary function and both the symbol $\mathcal{P}_2$ as well as the contributions $\propto \delta(\bm{y})$ be dropped. 
 
 \subsection{Two-point cumulant in the ordered phase}\label{sec:2ptord}
 
Upon inserting the scattering amplitude from Eq.~\eqref{eq:sigmsolf} into Eq.~\eqref{eq:Gtilde2cummneg}, we arrive at
 \begin{align}
& \tilde{G}^{(2)}_{\text{cum}}(\bm{p},z,z';\infty,m<0)\nonumber\\ & 
=|m|^{-1}\int_0^\infty\frac{\rmd{k}}{\pi}\bigg[\frac{2k}{p^2+k^2}\,\left(\frac{|m|}{k}+\frac{\pi}{2}\right)\nonumber\\ &\qquad\times \noneigfct(mz,k/m;\infty,-1)\,\noneigfct(mz',k/m;\infty,-1)\bigg].
 \end{align}
 Just as in the $m>0$ case, we can multiply by $(z')^{-1/2}$ and perform the limit $z'\to0$ to obtain
  \begin{align}
& \tilde{G}^{(1,1)}_{\text{cum}}(\bm{p},z;\infty,m<0)\nonumber\\ & 
=\frac{1}{\sqrt{|m|}}\int_0^\infty\frac{\rmd{k}}{\pi}\bigg[\frac{2k\,\noneigfct(mz,k/m;\infty,-1)}{p^2+k^2}\,\left(\frac{|m|}{k}+\frac{\pi}{2}\right)\bigg].
 \end{align}
 The calculation of the two-point surface cumulant, defined by analogy with Eq.~\eqref{eq:Gshat}, is similarly straightforward. It yields
 \begin{equation}\label{eq:G02tildecumord}
 \tilde{G}^{(0,2)}_{\text{cum}}(\bm{p};\infty,m<0)=\frac{|m|}{p}-\ln\frac{p}{\mu}
 \end{equation}
 and hence
  \begin{equation}\label{eq:G02tildeord}
 \tilde{G}^{(0,2)}(\bm{p};\infty,m<0)=\frac{|m|}{p}-\ln\frac{p}{\mu}+\delta(\bm{p})\,\frac{m^2}{4\pi}.
 \end{equation}
 
 For the corresponding Fourier back transforms one finds
 \begin{align}\label{eq:G02mneg}
 &G^{(0,2)}(\bm{y};\infty,m<0)= G^{(0,2)}_{\text{cum}}(\bm{y};\infty,m<0)+\frac{m^2}{4\pi}\nonumber \\ &=\frac{\mathcal{P}_2}{2\pi y^2}+\frac{|m|}{2\pi y}+\frac{m^2}{4\pi} +(\gammaE-\ln 2)\,\delta(\bm{y}).
 \end{align}
 
In Table~\ref{tab:Gcomp} we compare our exact ${n=\infty}$ results for the surface correlation function $G^{(0,2)}(\bm{y};\infty,m)$ with those known for $G^{(2)}(\bm{y},z,z';\infty,m)$ and the bulk correlation function $G^{(2)}_{\mathrm{b}}(\bm{x};m)=\langle\Phi(\bm{x})\,\Phi(\bm{0})\rangle_{\mathrm{b}}$.
 \begin{table}[htdp]
\begin{center}
\begin{tabular}{c|@{\hspace{2ex}}c@{\hspace{2ex}}c@{\hspace{2ex}}c@{\hspace{2ex}}}
&$G^{(2)}(\bm{y},z,z')$&$G^{(0,2)}(\bm{y})$&$G_{\mathrm{b}}^{(2)}(\bm{x})$\\\hline
$m>0$&&$\frac{1}{2\pi y^2}\,\rme^{-my}$&$\frac{1}{4\pi x}\,\rme^{-mx}$\\\\[\medskipamount]
$m=0$&$\frac{\sqrt{zz'}}{2\pi \sqrt{(y^2+z_+^2)(y^2+z_-^2)}}$&$\frac{1}{2\pi y^2}$&$\frac{1}{4\pi x}$\\\\[\medskipamount]
$m<0$&&$\frac{1}{2\pi y^2}+\frac{|m|}{2\pi y}+\frac{m^2}{4\pi}$&$\frac{1}{4\pi x}+\frac{|m|}{4\pi}$
\end{tabular}
\end{center}
 \caption{Comparison of exact two-point correlation functions $G^{(2)}(\bm{y},z,z';\infty,m)$, $G^{(0,2)}(\bm{y};\infty,m)$, and bulk correlation function $G_{\mathrm{b}}^{(2)}(\bm{x};m)$, where $z_\pm\equiv z\pm z'$ and $\bm{x}=(\bm{y},z)$.}\label{tab:Gcomp}
 \end{table}
 
 \section{Results for some potential-related quantities}\label{eq:potrelq}
 
 \subsection{The potential coefficient $v_0$}\label{sec:v0calc}

In order to compute the potential coefficient $v_0^\pm$, we use the following variant of a trace formula derived in the accompanying paper II: 

\begin{theorem}[Trace formula]\label{thm:traceform}
Let $v(\zm)$ and $\tilde{v}(\zm)$  be two potentials on the half-line $(0,\infty)$ that vanish faster than $\zm^{-1}$ as $\zm\to\infty$ and behave as 
\begin{eqnarray}\label{eq:traceform}
v(\zm\to 0)&=&v^{(\mathrm{sg})}(\zm)+v_0+o(1),\nonumber\\
\tilde{v}(\zm\to 0)&=&v^{(\mathrm{sg})}(\zm)+\tilde{v}_0+o(1),
\end{eqnarray}
with the same singular part
\begin{equation}\label{eq:vsg}
v^{(\mathrm{sg})}(\zm)=-\frac{1}{4\zm^2}+\frac{v_{-1}}{\zm}.
\end{equation}
Let $\regsol_\nu(\zm)$, $\nu=1,\dotsc,n_b$, be the regular solutions to the Schr\"odinger equation $\mathcal{H}_{v}\regsol_\nu(\zm)=\varepsilon_\nu\regsol_\nu(\zm)$ subject to the boundary conditions
\begin{eqnarray}\label{eq:regsolinfbc}
\regsol_\nu(\zm)&\mathop{=}\limits_{\zm\to 0}&\sqrt{\zm}[1+O(\zm)]
\end{eqnarray}
that correspond to bound states (where $n_b$ may be zero). Denote by
\begin{equation}\label{eq:normbs}
 \normreg_\nu=\int_0^\infty\regsol^2_\nu(\zm)\,\rmd{\zm}
 \end{equation}
the squares of the $L^2([0,\infty))$ norms of these (real-valued) functions and by $A(\sk)=\rme^{\sigma(\sk)}$ the scattering amplitude for $\sk>0$. Furthermore, let $\tilde{\varphi}_{\tilde{\nu}}$, $\tilde{n}_b$, $\tilde{\normreg}_{\tilde{\nu}}$, and $\tilde{A}(\sk)=\rme^{\tilde{\sigma}(\sk)}$ be the  analogous quantities pertaining to the potential $\tilde{v}(\zm)$. Then the following relation holds 
 \begin{align}\label{eq:tfv0}
 v_0-\tilde{v}_0&=\frac{4}{\pi}\int_0^\infty\rmd{\zm}\,\sk^2\left[\rme^{-2\tilde{\sigma}(\sk)}-\rme^{-2\sigma(\sk)}\right]\nonumber\\ &\strut +\sum_{\substack{\text{bound}\\ \text{ states }\tilde{\nu}}}\frac{2}{\tilde{\normreg}_{\tilde{\nu}}}-\sum_{\substack{\text{bound}\\ \text{ states }\nu}}\frac{2}{\normreg_\nu}.
 \end{align}
 \end{theorem}

We shall use this relation for the $({m=\pm 1,L=\infty})$ choices of  self-consistent potentials $v(\zm)\equiv v_\infty^\pm(\zm)-\delta_{\pm 1,1}$ with the singular parts
\begin{equation}\label{eq:vspmdef}
v^{(\text{sg})}(\zm)\equiv v^{\text{sg},\pm}\equiv -\frac{1}{4\zm^2}\pm\frac{4}{\pi^2\zm}
\end{equation}
and $\tilde{v}(\zm)=v^{\text{sg},\pm}(\zm)$. The potentials  $v^{\text{sg},\pm}(\zm)$ weakly violate the required $\zm\to\infty$ behavior $\tilde{v}(\zm) =o(1/\zm)$,  where the ``little-$o$ symbol''  ${f=o(g)}$ means that $\lim_{x\to-\infty}|f(x)/g(x)|=0$. If fulfilled, the condition would ensure (together with the presumed boundedness and smoothness of $\tilde{v}$ for $\zm>0$) that $\int_{\zm_l>0}^\infty |\tilde{v}(\zm)|\rmd \zm<\infty$. The consequence of the slow $O(1/\zm)$ decay is known from the Coulomb potential: the phase shift $\tilde{\eta}(\sk)$ is not well-defined and gets replaced by $\zm$-dependent quantity $\tilde{\eta}(\sk,\zm)$ containing a contribution $\asprop\ln(\zm\sk)$, as we shall explicitly confirm below. Since the scattering theory for Coulomb potentials is mathematically well controlled, we may trust that Eq.~\eqref{eq:tfv0} carries over to the border-line case when $\tilde{v}(z)=O(1/\zm)$, except for the mentioned replacement of the phase shift by a logarithmically divergent $\tilde{\eta}(\sk,\zm)$.
 
  \subsubsection{The case $v=v_\infty^+$ and $\tilde{v}=v^{\text{sg},+}$}
  
  For the choices $v(\zm)=v^+_\infty(\zm)-1$ and $\tilde{v}^{\text{sg},+}(\zm)$ the Schr\"odinger equations have no bound states. The spectra are continuous with $\varepsilon=\sk^2\in(0,\infty)$. The corresponding regular solutions $\tilde{\regsol}(\zm,\sk)$ are given by
\begin{equation}\label{eq:regsoltildeplus}
  \tilde{\regsol}(\zm,\sk;\kappa)=\frac{\rme^{\rmi\pi/4}}{\sqrt{2\sk}}\,M_{-\rmi\kappa,0}(-2\rmi\sk \zm) ,\;\;\kappa=\frac{2}{\pi^2\sk},
  \end{equation}
where $M_{-\rmi\kappa,0}(\zm)$ is a Whittaker-$M$ function \cite{AS72,NIST:DLMF,Olver:2010:NHMF}. Its asymptotic behaviors at small and large $\zm$ are \footnote{The linearly independent second solution of the Schr{\"o}dinger equation involving a Whittaker-$W$ function varies as $\zm^{1/2}\ln \zm$ for $\zm\to 0$, which is inconsistent with the boundary condition~\eqref{eq:regsolinfbc}.}
  \begin{equation}\label{eq:regsolcont}
  \tilde{\varphi}(\zm,\sk;\pm\kappa)\mathop{=}_{\zm\to 0}\sqrt{\zm}\left[1\pm\frac{4\zm}{\pi^2}+\left(\frac{4}{\pi^2}-\frac{\sk^2}{4}\right)\zm^2+O(\zm^3)\right]
  \end{equation}
and
 \begin{equation}
 \tilde{\regsol}(\zm,\sk;\pm\kappa)\mathop{\aseq}_{\zm\to\infty}\frac{\tilde{A}(\sk)}{\sk}\sin[\sk\zm+\tilde{\eta}(\sk,\zm)]
\end{equation}
with
\begin{equation}\label{eq:tildeA}
\tilde{A}(\sk)=\left\{\frac{\sk}{\pi}\left[1+\exp\left(\pm\frac{4}{\pi\sk}\right)\right]\right\}^{1/2}
\end{equation}
and
\begin{equation}\label{eq:tildeeta}
\tilde{\eta}(\sk,\zm)=\frac{\pi}{4}+\arg \Gamma\left(\frac{1}{2}\pm\frac{2\rmi}{\pi^2\sk}\right)\mp\frac{2}{\pi^2\sk}\ln(2\sk\zm).
\end{equation}
Here the upper signs apply to the case of the regular solutions specified in Eq.~\eqref{eq:regsoltildeplus}. The lower signs give the analogous results which hold when the sign of $\kappa$ is changed. They will be needed in our analysis of the case $v=v_\infty^-$  and $\tilde{v}=v^{\text{sg},-}$ presented below. Note that the results given in Eq.~\eqref{eq:tildeeta} exhibit the above-mentioned logarithmic $\zm$-dependence of $\tilde{\eta}(\sk,\zm)$. 
 
We insert these results along with the scattering amplitude $A_+(\sk)$ given in Eq.~\eqref{eq:Asol} into the trace formula~\eqref{eq:tfv0}. This gives
 \begin{equation}\label{eq:v0+calc}
 v^+_0-1=+\frac{4}{\pi}\int_0^\infty\rmd{\sk}\,\sk^2\bigg[\frac{\pi/\sk}{1+\exp[4/(\pi\sk)]}-\frac{\arctan\sk}{\sk}\bigg].
 \end{equation}
 Upon making a change of variable  $\sk\to q=1/\sk$, the integral on the right-hand side can be decomposed into a sum of two finite integrals $J_1$ and $J_2$. The first  can be rewritten as an integral of a series, which can be integrated termwise and subsequently summed to obtain
 \begin{eqnarray}
J_1&=&\frac{4}{\pi} \int_0^\infty\frac{\rmd{q}}{q^3}\left[\frac{\pi}{1+\rme^{4q/\pi}}-\frac{\pi}{2}+q\right]\nonumber\\
&=&\frac{32}{\pi^3}\int_0^\infty\rmd{q}\sum_{j=1}^\infty\frac{(2j-1)^{-2}}{q^2+\pi^4(2j-1)^2/16}=\frac{56\zeta(3)}{\pi^4}.\nonumber\\
\end{eqnarray}
The second integral is										
\begin{equation}
J_2=\frac{4}{\pi}\int_0^\infty\frac{\rmd{q}}{q^3}\left[\frac{\pi}{2}-q-\text{arccot}\,q \right]=-1.
\end{equation}
Its contribution to $v_0^+$ is compensated by the $-1$ on the left-hand side of Eq.\eqref{eq:v0+calc}. Hence we arrive at the value of $v_0^+$ given in Eq.~\eqref{eq:v0res}.
 
\subsubsection{The case $v=v^-$ and $\tilde{v}=v^{\text{sg},-}$}

For the potential $v(\zm)=v_\infty^-(\zm)$, no bound states exists. The spectrum is continuous with $\varepsilon=\sk^2>0$. However, in the case of the potential $\tilde{v}(\zm)=\tilde{v}_\infty^-(\zm)$ infinitely many bound states exist. To see this, note that the regular solutions satisfying the boundary condition~\eqref{eq:regsolinfbc} are again given by Eq.~\eqref{eq:regsoltildeplus} except that $\kappa$ must be replaced by  $-\kappa=-2\pi^{-2}\sk^{-1}$. For $\sk/\rmi =\sk''>0$, the requirement that $\regsol(\zm,\rmi\sk'';-\kappa)$ decays exponentially as $\zm\to\infty$ yields the discrete values and associated eigenenergies $\varepsilon_\nu=-(\sk''_\nu)^2$, namely
\begin{equation}
\sk''_\nu=\frac{4}{\pi^2(2\nu-1)},\;\;\varepsilon_\nu=\frac{-16}{\pi^4\,(2\nu-1)^2},\;\;\nu=1,2,\dotsc,\infty.
\end{equation}
The associated eigenstates simplify to
\begin{eqnarray}
\tilde{\regsol}_\nu(\zm)&=&\tilde{\regsol}(\zm,\rmi \sk'';-\kappa)\nonumber\\ &=&\sqrt{z}\exp\left[-\frac{4\zm}{\pi^2(2\nu-1)}\right]\,L_{\nu-1}\bigg(\frac{8z}{\pi^2(2\nu-1)}\bigg),\nonumber\\
\end{eqnarray}
where the $L_{\nu-1}(z)$ are Laguerre polynomials. 

Using these results, one can compute the required norm parameters $\tilde{\normreg}_\nu$. One obtains
\begin{equation}
\tilde{\normreg}_\nu=\frac{\pi^4}{64}\,(2\nu-1)^3,\;\,\nu=1,2,\dotsc,\infty.
\end{equation}

The regular solutions associated with the continuous part $\{{\varepsilon=\sk^2}|{0<\sk<\infty}\}$ of the spectrum are given by $\tilde{\regsol}(\zm,\sk;-\kappa)$ in terms of the functions defined in Eq.~\eqref{eq:regsoltildeplus}.

Insertion of the above results and Eq.~\eqref{eq:sigmsolf} for $\exp(-2\tilde{\sigma}(\sk))$ into the trace formula~\eqref{eq:tfv0} yields
\begin{eqnarray}\label{eq:v0mincalc}
v_0^- &=& \frac{4}{\pi}\int_0^\infty\rmd{\sk}\left[\frac{\pi\sk}{1+\exp[-4/(\pi\sk)]}-1-\frac{\pi\sk}{2}\right] \nonumber\\ &&\strut +\frac{128}{\pi^4}\sum_{\nu=1}^\infty\frac{1}{(2\nu-1)^3}\nonumber\\ &=&-\frac{56\,\zeta(3)}{\pi^4}+\frac{112\,\zeta(3)}{\pi^4}=\frac{56\,\zeta(3)}{\pi^4}.
\end{eqnarray}
To compute the integral in the first line, we again transformed to the variable $q=1/\sk$, represented the resulting integrand as the series 
\[
\sum_{\mu=1,3,\ldots}^\infty(-8\pi^{-2}\mu^{-2})[q^2+(\mu\pi^2/4)^2]^{-1}, 
\]
and integrated termwise to obtain the first term in the last line of Eq.~\eqref{eq:v0mincalc}. The series in the second line can be summed in a straightforward fashion. Hence we obtain the value of $v_0^-$ given in Eq.~\eqref{eq:v0res}.

 \subsection{The integrals $\alpha_\pm$ and the universal amplitude difference $\Delta A^{(\mathrm{s})}_0$}\label{sec:alphapm}
 
We next turn to the calculation of the integrals $\alpha_\pm$ defined in Eq.~\eqref{eq:alphapmdef} and the related universal amplitude difference $\Delta A^{(\mathrm{s})}_0$. To this end we use the following strategy. We determine the asymptotic large-$\sk$ behavior of the Jost function $F(\sk)$ by means of the semiclassical expansion. Details of this analysis can be found in Appendix~\ref{app:semiclass}. The asymptotic form of this function for the half-line case of a potential with a singular part $v^{\text{sg}}$ of the form specified in Eq.~\eqref{eq:vsg} involves $v_{-1}$-dependent analogs of the integrals $\alpha_\pm$.  The corresponding results for the  phase shifts read
\begin{equation}
\eta_\pm(\sk)=\frac{\pi}{4}-\frac{1}{2\sk}\left[ v_{-1}^\pm(\gammaE+\ln 8\sk)+\alpha_\pm\right]+O(\sk^{-3}).
\end{equation}
Upon inserting the known values~\eqref{eq:vm1} of $v_{-1}^\pm$, we can determine $\alpha_\pm$ by matching the derivatives $\eta'_\pm(\sk)$ of these results with the large-$\sk$ forms  given in Eq.~\eqref{eq:etaprimempas} or implied by Eq.~\eqref{eq:etamprimefres}. One thus obtains
\begin{equation}\label{eq:alphaplusres}
\alpha_+=J_+-\alpha_-
\end{equation}
and
\begin{equation}\label{eq:alphaminres}
\alpha_-=\frac{4}{\pi^2}[\gammaE-1+\ln(16/\pi)],
\end{equation}
where  $J_+$ was defined in Eq.~\eqref{eq:Jdef}.

The universal amplitude difference $\Delta A^{(\mathrm{s})}_0$ follows upon insertion of these expresssions for $\alpha_\pm$  into Eq.~\eqref{eq:DelA0pmres} and use of the numerical value  $J_+$ given in Eq.~\eqref{eq:Jplusnum}. One finds
\begin{align}\label{eq:DelA0anexp}
\Delta A^{(\mathrm{s})}_0&=\frac{\alpha_++\alpha_-}{16\pi}=\frac{J_+}{16 \pi}\\
&=0.00944132199\ldots.
 \end{align}
 This exact value of $\Delta A^{(\mathrm{s})}_0$ derived here has  recently been confirmed quite accurately in \cite{DGHHRS14} by numerical solutions of the self-consistency equations for two microscopically different $O(n)$ $\phi^4$ models representing the same universality class (called models A and B there).

\subsection{A surface excess quantity associated with the squared order parameter} \label{sec:surfexcq}

For local quantities of  semi-infinite systems whose $z$-dependent thermal average approaches the bulk value $Q_{\mathrm{b}}=Q(\infty)$ sufficiently fast, one conventionally defines the associated surface excess as \cite{Bin83,Die86a}
 \begin{equation}\label{eq:Qsdef}
 Q_{\mathrm{s}}=\int_0^\infty\rmd{z}\,[Q(z)-Q(\infty)].
 \end{equation}
 We have already seen in Sec.~\ref{sec:surfexcessen} that this definition entails UV singularities in the continuum limit $a\to 0$ in the case of the excess energy density because of the behavior $\asprop z^{-2}$ of $v_\infty$ in the regime ${a\ll z\ll |m|}$. These singularities required the subtractions made to define the integrals $\alpha_\pm$ in Eq.~\eqref{eq:alphapmdef}. 
 
In the case of the squared order-parameter density $\Phi^{(2)}(z)=\langle\Phi(\bm{y},z)\rangle^2$, no problems arise from the behavior in the near-boundary regime $z\lesssim |m|$ when $a=0$. As we know from Eq.~\eqref{eq:G1squscal}, the scaling function associated with the density $4\pi\,\Phi^{(2)}(z)$ is given by the square of the semi-bound state $\regsol_0(|m|z)$. This vanishes linearly in $z$ as $z\to 0$ and hence is integrable at the lower limit. However, problems arise at the upper integration limit $L=\infty$ when $m<0$ because the existence of Goldstone modes in the ordered phase implies the asymptotic behavior $\asprop z^{-(d-2)}$ at distances much larger than the Josephson coherence length $|m|^{-1}$ \cite{Die86a}. At $d=3$, this decay $\aseq z^{-1}$ is too slow to ensure convergence at the upper integration limit $\infty$. In order to define a finite surface excess $\Phi^{(2)}_{\mathrm{s}}$, the definition~\eqref{eq:Qsdef} must be appropriately modified. 
 
Let us begin by explicitly demonstrating the claimed slow decay, showing that the regular solution $\regsol_0(\zm)$ function behaves  asymptotically as 
\begin{equation}\label{eq:regsol0as}
\regsol_0(\zm)\mathop{\aseq}_{\zm\to\infty}1-\frac{1}{4\zm}+o(1/\zm).
\end{equation}
To determine this limiting behavior one can solve the ${m=-1}$ variant of the Schr\"odinger Eq.~\eqref{eq:SEhalfline} for $\zm\gg 1$. This requires knowledge of the asymptotic form of the potential $v_\infty^-(\zm)$. In the next section, we determine the behavior of the self-consistent potential ${v(z;L,m<0)=L^{-2}\,v(z/L;1,x)}$ in the regime ${1\ll z|m|\ll |x|}$. The result, given in Eq.~\eqref{eq:vxlargezLlarge}, means that
\begin{equation}\label{eq:vmininfas}
v_\infty^-(\zm)\mathop{\aseq}_{\zm\to\infty}-\frac{1}{2\zm^3}.
\end{equation}

Upon substituting this asymptotic form for the potential, one can determine the two linearly independent solutions \begin{eqnarray}\label{eq:regsol02sol}
\regsol_0(\zm\to\infty)&\aseq&(2\zm)^{1/2}\,J_1\Big(\sqrt{2/\zm}\Big),\nonumber\\
\psi_0(\zm\to\infty )&\aseq& -\pi  (\zm/2)^{1/2}\,Y_1\Big(\sqrt{2/\zm}\Big),
\end{eqnarray}
of Eq.~\eqref{eq:SEhalfline}, where $J_\nu(u)$ and $Y_\nu(u)$ are Bessel functions of the first and  second kind, respectively. We have normalized these solutions such that $\regsol_0(\infty)=1$ and the Wronskian satisfies $W[\regsol_0(\zm),\psi_0(\zm)]=1$. The first solution has the asymptotic behavior~\eqref{eq:regsol0as}; the second  behaves as
\begin{equation}\label{eq:psi0as}
\psi_0(\zm\to\infty)= \zm+\frac{\ln \zm}{2}+C_0+o(1/\zm)
\end{equation}
with $C_0=(1-2 \gammaE +\ln 2)/2$.

The result~\eqref{eq:regsol0as} tells us that a finite universal surface excess quantity associated with $4\pi\,\Phi^{(2)}(z)/|m|$ can be introduced as the integral
\begin{equation}\label{eq:Jvarphi}
J_\varphi=\int_0^\infty\rmd{\zm}\left[\regsol_0^2(\zm)-1+\frac{\Htheta(\zm-1)}{2\zm}\right].
\end{equation}

To compute it, we proceed as follows. We expand the regular solution $\regsol(z,\sk)$ of the ${m=-1}$ Schr\"odinger Eq.~\eqref{eq:SEhalfline} to $O(\sk^2)$, writing
\begin{equation}\label{eq:regsolkexp}
\regsol(\zm,\sk)=\regsol_0(\zm)+\sk^2\,\regsol_2(\zm)+O(\sk^4).
\end{equation}
The function $\varphi_2(\zm)$ satisfies the differential equation
\begin{equation}\label{eq:regsol2eq}
\mathcal{H}_{v}\regsol_2(\zm)=[-\partial_{\zm}^2+v(\zm)]\regsol_2(\zm)=\regsol_0(\zm).
\end{equation}
A special solution is
\begin{equation}\label{eq:solregsol2}
\regsol_2(\zm)=\regsol_0(\zm)\int_0^{\zm}\rmd{\zeta}\,\regsol_0(\zeta)\,\psi_0(\zeta)-\psi_0(\zm)\int_0^{\zm}\rmd{\zeta}\,\regsol_0^2(\zeta).
\end{equation}

The general solution is the sum of this solution and the general solution  $c_\regsol \regsol_0(\zm)+c_\psi \psi_0(\zm)$ of the homogeneous equation. However, since $\regsol_2(\zm)=\frac{1}{2}\partial_{\sk}^2\regsol(\zm,\sk=0)$, it must vanish faster than $\regsol(\zm,\sk)$ for $\zm\to0$, namely be $O(z^{5/2})$. Both solutions $\regsol_0$ and $\psi_0$ of the homogeneous solution violate this condition. Hence $c_\regsol=c_\psi =0$ and the correct solution must be given by Eq.~\eqref{eq:solregsol2}. 

Integrating the asymptotic form of $\regsol_0(\zm)\psi_0(\zm)$ implied by Eqs.~\eqref{eq:regsol0as} and \eqref{eq:psi0as}, one finds that the first integral on the right-hand side of   Eq.~\eqref{eq:solregsol2} behaves as
\begin{equation}\label{eq:varphi2dec}
\int_0^{\zm}\rmd{\zeta}\,\regsol_0(\zeta)\psi_0(\zeta)\mathop{=}_{\zm\to\infty}\frac{\zm^2}{2}
+\zm \bigg[\frac{\ln \zm}{2}+C_0-\frac{3}{4}\bigg]+O(\ln^2\zm).
\end{equation}
To determine the asymptotic behavior of the second integral, we split off $J_\varphi$  and then compute the integrals for large $\zm$ using again the limiting form of $\regsol_0$. This gives
\begin{eqnarray}
\int_0^{\zm}\rmd{\zeta}\,\regsol_0^2(\zeta)&=&J_\varphi-\int_{\zm}^\infty\rmd\zeta\left[\regsol_0^2(\zeta)-1+\frac{\Htheta(\zeta-1)}{2\zeta}\right]\nonumber\\ &&\strut +\int_0^{\zm}\rmd{\zm}\left[1-\frac{\Htheta(\zeta-1)}{2\zeta}\right]\nonumber\\ &\mathop{=}\limits_{\zm\to\infty}&\zm-\frac{1}{2}\ln\zm+J_\varphi+O(1/\zm).
\end{eqnarray}
Insertion of these results into Eq.~\eqref{eq:solregsol2} then leads us to
\begin{equation}\label{eq:regsol2as}
\regsol_2(\zm)\mathop{=}_{\zm\to\infty} -\frac{\zm^2}{2}+\frac{\zm}{2}\ln\zm-\Big(J_\varphi+\frac{7}{8}\Big)\zm+O(\ln^2\zm).
\end{equation}

Since the integral $J_\varphi$ appears in the asymptotic large-$\zm$ form of the regular solution's $\sk^2$-term, it should also appear in the corresponding $\sk^2$-term of the Jost function $F(\sk)$. Hence we should be able to determine $J_\varphi$ by computing this function and matching it with our exact scattering-data results of Sec.~\ref{sec:hssdmm} given in Eqs.~\eqref{eq:sigmsolf} and \eqref{eq:etamprimefres}. To obtain $F(\sk)$, we first calculate the Jost solution $f(\zm,\sk)$ for large $\zm$  and then use it  together with the above results for the regular solution to compute $F(\sk)$ as the Wronskian~\eqref{eq:FWronskian} of these functions.

The Jost solution $f(\zm,\sk)$ can be determined at large $\zm$ by means of perturbation theory.  We use the familiar equivalence of the differential Eq.~\eqref{eq:SEhalfline} with the integral equation \cite{CS89}
\begin{equation}
f(\zm,\sk)=\rme^{\rmi\sk\zm}+\int_{\zm}^\infty \rmd{\zeta}\,v_\infty^-(\zeta)\frac{\sin[\sk(\zeta-\zm)]}{\sk}\,f(\zeta,\sk),
\end{equation}
which we iterate to first order in $v_\infty^-$, replacing $f(\zeta,\sk)$ and $v_\infty^-$ in the integral on the right-hand side by $\rme^{\rmi\sk \zm}$ and the asymptotic form of $v_\infty^-$ given in Eq.~\eqref{eq:vmininfas}, respectively. We thus arrive at the expansion
\begin{equation}\label{eq:JostfOv}
f(\zm,\sk)=\bigg(1-\frac{1}{4\zm}\bigg)\rme^{\rmi\sk\zm}+\frac{\sk}{2}\rme^{-\rmi\sk\zm}[\pi+\rmi\, \text{Ei}(2 \rmi \sk \zm)]+\ldots,
\end{equation}
where $\text{Ei}(z)$ denotes a standard exponential-integral function.
For the contribution $f^{(j)}$ of $j$th order  in the potential, we have the bound (cf.\ p.~9 of \cite{CS89})
\begin{equation}\label{eq:Ovjext}
|f^{(j)}(\zm,k)|\le \text{const}\,\rme^{-\Im(\sk \zm)}\frac{1}{j!}\bigg[\int_{\zm}^\infty\rmd{\zeta}\,\frac{\zeta\,|v_\infty^-(\zeta)|}{1+|k|\zeta}\bigg]^j,
\end{equation}
where ``$\text{const}$'' is an appropriate (positive) numerical constant. Using once more the asymptotic form of $v_\infty^-$, we can determine the large-$\zm$ behavior of the integral. Since
\begin{eqnarray}
\int_{\zm}^\infty\rmd{\zeta}\,\frac{\zeta\,|v_\infty^-(\zeta)|}{1+|k|\zeta}&\mathop{=}\limits_{\zm\to\infty}&\frac{1}{2\zm}-|\sk|\, \text{arccoth}(1+2|\sk|\zm)\nonumber\\ 
&=&\frac{1}{4|\sk|\zm^2}+O(\zm^{-3}),
\end{eqnarray}
the contributions of second order in $v_\infty^-$ are smaller by a factor of $\zm^{-2}$ than the $O(v_\infty^-)$ term.  Note that the latter varies as
\begin{equation}
f^{(1)}(\zm,\sk>0)=-\rme^{\rmi \sk\zm}\bigg[\frac{1}{8\sk\zm^2}+O(\zm^{-3})\bigg]
\end{equation}
for large $\zm$, in accordance with the bound~\eqref{eq:Ovjext}. 

We now wish to compute the Wronskian $W[f(\zm,\sk),\regsol(\zm,k)]$,  expand it to $O(\sk^2)$, and take the limit $\zm\to\infty$. To this end it is sufficient to  retain in $f(\zm,\sk)$ only the contributions to  $O(v_\infty^-)$ explicitly given on the right-hand side of Eq.~\eqref{eq:JostfOv} because the foregoing considerations imply that the terms $f^{(j)}$ with $j\ge 2$ do not contribute to this limit. Proceeding in this manner, we find
\begin{align}\label{eq:Wronskexp}
&\lim_{\zm\to\infty}W[f(\zm,\sk),\regsol(\zm,k)]\nonumber\\ 
&=-\rmi \sk+\sk^2\bigg[\frac{\rmi \pi}{4} -J_\varphi-\frac{\gammaE}{2}-\frac{\ln(2\sk)}{2}\bigg]+o(\sk^2).
\end{align}
On the other hand, the exact Jost function $F({\sk>0})$ given by Eqs.~\eqref{eq:FJostk}, \eqref{eq:sigmsolf}, and \eqref{eq:etamfres} has the expansion
\begin{equation}\label{eq:Jostsmallk}
F(\sk)=-\rmi\sk+\sk^2\bigg[\frac{1}{2}+\rmi\frac{\pi}{4}-\frac{\ln(\pi\sk/2)}{2}\bigg]+o(\sk^2).
\end{equation}
Matching these results finally yields the analytically exact and numerical values
\begin{eqnarray}\label{eq:Jvarphires}
J_\varphi&=&-\frac{1}{2}[1+\gammaE+\ln(4/\pi)]\nonumber\\
&\etwa& -0.9093900700860\ldots.
\end{eqnarray}

 \section{Asymptotic behavior in the low-temperature scaling limit $x\to-\infty$}\label{sec:aslowt}
 
 \subsection{Asymptotic behavior of eigenvalues and eigenfunctions}\label{sec:asspecprop}

In \cite{DGHHRS12} and \cite{DGHHRS14} we showed that the eigenfunctions $\oneigfct_\nu$ and eigenvalues $E_\nu$ of the self-consistent Schr\"odinger problem defined by Eqs.~\eqref{eq:Hv},  \eqref{eq:StLc}, \eqref{eq:vsurfsing}, and \eqref{eq:vsc} approach in the limit $x\equiv mL\to-\infty$  those of a one-dimensional free massless bosonic field theory on the strip $(0,L)$ subject to Neumann boundary conditions. This was achieved by generalizing previous arguments for the semi-infinite case \cite{DN86} to map the low-temperature behavior of the model to a nonlinear sigma model (see Appendix B of \cite{DGHHRS14}) and confirming the results via numerical solutions of the self-consistency equations. The aim of this section is to  compute the leading corrections to the limiting Neumann eigenvalues  and eigenvectors for large negative values of $x$.

To this end it will be convenient to scale $L$ to unity and use the scaling properties of $E_\nu$ and $\oneigfct_\nu$ implied by dimensional arguments to express them as 
\begin{eqnarray}\label{eq:eigvalfctscal}
L^2\,E_\nu(L,m)&=&E_\nu(1,mL)\equiv E_\nu(x),\nonumber \\ \sqrt{L}\,\oneigfct_\nu(z;L,m)&=&\oneigfct_\nu(z/L;1,mL)\equiv \oneigfct_\nu(\zL,x),
\end{eqnarray}
in terms of the dimensionless variables $x$ and $\zL=z/L$.  Their limiting large-$x$ behavior stated above is
\begin{eqnarray}
\lim_{x\to-\infty}E_\nu(x)&=&E_\nu(-\infty)=\mathring{E}_\nu\,,\nonumber\\
 \lim_{x\to-\infty}\oneigfct_\nu(\zL,x)&=&\oneigfct_\nu(\zL,-\infty)=\mathring{\oneigfct}_\nu(\zL)\,,
\end{eqnarray}
where 
\begin{align}\label{eq:spectrummxinfty}
&\mathring{E}_\nu=\sko_\nu^2\equiv[\pi(\nu-1)]^2,\;\;\nu=1,2,\dotsc,\infty,\nonumber \\
&\mathring{\oneigfct}_1(\zL)=1,\;\;\mathring{\oneigfct}_{\nu\ge 2}=\sqrt{2}\cos(k_\nu \zL),
\end{align}
denote the familiar Neumann eigenvalues and eigenfunctions.

Perturbation theory about the ${x=-\infty}$~limit is not standard because it involves several unusual, though interesting and challenging, features. First, since $\mathring{E}_1=0$ and $E_1$ vanishes $\asprop \rme^{-|x|}$ as $x\to-\infty$, we must deal with an asymptotic zero-energy mode. Second, perturbative corrections to the potential must be determined self-consistently along with those of the eigenvalues and eigenvectors. Third, the perturbation expansions of $\zL$-dependent quantities differ, depending on whether $\zL$ belongs to the boundary regions of thickness $\asprop |x|^{-1}$ near the two surface planes at $\zL=0$ and $\zL=1$ or else to the remaining inner region. Thus the problem involves aspects of singular perturbation theory and bears similarities to boundary layer theory in fluid dynamics \cite{KC96}.

To set  up perturbation theory, we start from the scaled analog of Eq.~\eqref{eq:vsc} corresponding to the replacements $z\to \zL=z/L$, $m\to x=tL$, and $L\to 1$, respectively, and choose $\zL\in(0,1)$ so that the $\delta$-distributions do not contribute. This gives 
\begin{equation}\label{eq:scaledsce}
x=\int_{-\infty}^0\rmd{E}\bigg[\mathcal{G}(\zL,\zL;E)+\
\frac{1}{2\sqrt{-E}}\bigg],\quad\zL\in (0,1).
\end{equation}
Since the left-hand side diverges $\asprop x$ as $x\to-\infty$, so must the right-hand side. The latter divergence arises due to the behavior $\ln E_1\asprop -|x|$. We therefore subtract from  both sides of the equation the term
\begin{align}
&\oneigfct_1^2(\zL,x)\lim_{\rho\to 0+}\bigg\{\int_{-\infty}^{-\rho}\rmd{E}\bigg[\frac{1}{E-E_1(x)}-\frac{1}{E}\bigg]+\ln\rho\bigg\}\nonumber\\ &=\oneigfct_1^2(\zL,x)\ln E_1(x),
 \end{align}
 obtaining
 \begin{equation}\label{eq:logE1Peq}
x-\oneigfct_1^2(\zL,x)\ln E_1(x)= P(\zL,x)
 \end{equation}
 with
 \begin{eqnarray}\label{eq:Pdef}
 \lefteqn{P(\zL,x)}&&\nonumber\\ &=&\lim_{\rho\to 0+}\bigg\{\int_{-\infty}^{-\rho}\rmd{E}\left[\tilde{\mathcal{G}}(\zL,\zL;E)+\frac{1}{2\sqrt{-E}}\right]-\oneigfct_1^2(\zL,x)\ln\rho\bigg\},\nonumber\\
 \end{eqnarray}
where
\begin{equation}\label{eq:Gtilde}
\tilde{\mathcal{G}}(\zL,\zL';E)=\frac{\oneigfct_1(\zL)\,\oneigfct_1(\zL')}{E}+\sum_{\nu=2}^\infty\frac{\oneigfct_\nu(\zL)\,\oneigfct_\nu(\zL)}{E-E_\nu}.
\end{equation}

As $x\to-\infty$, the function $\oneigfct_1(\zL,x)\to 1$ while $\tilde{\mathcal{G}}$ approaches the Green's function  $\mathcal{G}_{\text{NN}}$ for Neumann boundary conditions at $\zL=0$ and $\zL=1$. The latter function is given by
\begin{equation}
\mathcal{G}_{\text{NN}}(\zL,\zL';E=-p^2)=-\frac{\cosh(p\zL_<)\cosh[p(1-\zL_>)]}{p\sinh p}.
\end{equation} 
It can be determined by means of standard image methods, which yield it as a sum of contributions from infinitely many image charges.  Using this representation, one finds that the $x\to-\infty$ limit of the integrand on the right-hand side of Eq.~\eqref{eq:Pdef} becomes
  \begin{align}
&\mathcal{G}_{\text{NN}}(\zL,\zL;E)+\frac{1}{2\sqrt{-E}}=-\bigg\{\frac{\rme^{-2p\zL}}{2p}\nonumber\\ &\strut +\frac{1}{2p} \sum_{j=1}^\infty\Big[2\,\rme^{-2pj}
+\rme^{-2p(j-\zL)}+\rme^{-2p(j+\zL)}\Big]\bigg\}_{p=\sqrt{-E}}.
 \end{align}
 
 Upon substitution of this expression into Eq.~\eqref{eq:Pdef}, the integral and  limit $\rho\to 0+$ that $P(\zL,x)$ involves can be computed in a straightforward fashion using {\sc Mathematica} \cite{Mathematica10}. One obtains
 \begin{equation}\label{eq:Pinftyres}
 P(\zL,-\infty)=\gammaE+2\ln 2+\frac{1}{2}\,\psi(\zL)+\frac{1}{2}\,\psi(1-\zL),
 \end{equation}
 where $\psi(\zL)=\rmd\ln\Gamma(\zL)/\rmd\zL$ is the digamma function. Hence we have shown that  $\ln E_1\mathop{=}\limits_{x\to-\infty}-|x|[1+o(1)]$.
 
Second, we can solve Eq.~\eqref{eq:logE1Peq} for $\oneigfct_1(\zL)$ and then use Eq.~\eqref{eq:Pinftyres} to conclude that
 \begin{equation}
 \oneigfct_1(\zL,x)=\sqrt{\frac{x-P(\zL,x)}{\ln E_1(x)}}=\sqrt{\frac{x}{\ln E_1(x)}}\,\Psi(\zL,x)
 \end{equation}
 with
\begin{equation}\label{eq:Psixexp}
\Psi(\zL,x)=1+\frac{\psi(\zL)+\psi(1-\zL)+2\gammaE+4\ln 2}{4|x|}+o\big(|x|^{-1}\big).
\end{equation}

The result tells us that for arbitrary $\zL\in(0,1)$, the eigenfunction $\oneigfct_1$ becomes a constant in the limit $x\to-\infty$, so that $\oneigfct_1(\zL,-\infty)\equiv 1$ indeed, as stated in Eq.~\eqref{eq:spectrummxinfty}.

In order for this perturbation expansion to be valid, the $O(1/|x|)$ must be small compared to the leading $O(1)$ contribution. Expanding in $\zL$ gives
\begin{equation}\label{eq:PsismallzL}
\Psi(\zL,x)=1-\frac{1}{4|x|}\left[\frac{1}{\zL}-4\ln 2+O(\zL^2)\right]+o(1/|x|).
\end{equation}

Owing to the symmetry $\zL\to 1-\zL$, a corresponding expansion holds about $\zL=1$. Hence, for the perturbation expansion~\eqref{eq:Psixexp} to hold, $\zL$ must not belong to  boundary layers of thickness $\asprop 1/|x|$ near either one of the two surface planes at $\zL=0$ and $\zL=1$. In fact, very close to the surfaces we know from Eq.~\eqref{eq:noneigfctnorm} that $\oneigfct_1(\zL,x)\propto \noneigfct_1(\zL,x)$ vary $\asprop \sqrt{\zL}$ and $\asprop\sqrt{1-\zL}$, respectively. 

The crossover between the behavior in the boundary layers and the remaining ``inner region'' is expected to occur at a distance $b/|x|$ away from the surface planes, where $b$ should be of order unity. In terms of the unscaled variable $z$ this means that for given $m$ and $L$ with $b/|m|\ll L/2$, the crossover from the near-boundary behavior to the behavior in the inner region occurs at  $z\etwa b/|m|$ and $1-z\etwa b/|m|$. The length $b/|m|$ agrees up to a factor of order $1$ with the Josephson coherence length \cite{Jos66}. 

The perturbative result for $\oneigfct_1$ given by Eqs.~\eqref{eq:Psixexp} and \eqref{eq:PsismallzL} can be used to obtain the expansion of the potential in the inner region. To this end, we solve the differential equation for $\oneigfct_1$, obtaining
\begin{equation}
v(\zL,x)=\frac{\partial_{\zL}^2\oneigfct_1(\zL,x)}{\oneigfct_1(\zL,x)}+E_1(x),
\end{equation}
and then substitute the above result for $\oneigfct_1$. This gives
\begin{equation}\label{eq:vinner}
v(\zL,x)=\frac{\psi''(\zL)+\psi''(1-\zL)}{4|x|}+o(1/|x|)
\end{equation}
for $b/|x|\lesssim \zL\lesssim 1-b/|x|$ (inner region) and
\begin{equation}\label{eq:vxlargezLlarge}
v(\zL,x)=-\frac{1}{|x|}\left[\frac{1}{2\zL^3}+\zeta(3)+O(\zL^2)\right]+o(1/|x|)
\end{equation}
for $b/|x|\ll \zL\ll 1/2$. By contrast, in the near-boundary regime $\zm=z|m|\lesssim 1$, the asymptotic behavior of the  potential $v(\zL,x)=v(1-\zL,x)$ must be in conformity with that of the semi-infinite case, i.e., with Eqs.~\eqref{eq:vnearboundary}--\eqref{eq:v0res}. 

We next turn to the determination of the asymptotic behavior of the lowest eigenvalue $E_1(x)$ for $x\to-\infty$. To this end we write the orthonormality relation for $\oneigfct_1(\zm,x)$ as
\begin{equation}\label{eq:onf1}
1=
2\int_{0}^{1/2}\rmd{\zL}\,\oneigfct_1^2(\zL,x)=I_<(\tilde{b},x)+I_>(\tilde{b},x)
\end{equation}
with
\begin{equation}\label{eq:I<}
I_<(\tilde{b},x)=2\int_0^{\tilde{b}/|x|}\rmd{\zL}\,\oneigfct_1^2(\zL,x)
\end{equation}
and
\begin{equation}\label{eq:I>}
I_>(\tilde{b},x)=2\int_{\tilde{b}/|x|}^{1/2}\rmd{\zL}\,\oneigfct_1^2(\zL,x),
\end{equation}
where $\tilde{b}\gg 1$ but $\tilde{b}/|x|<1/2$. To compute $I_>$ we can substitute our perturbation results, Eqs.~\eqref{eq:Psixexp} and \eqref{eq:PsismallzL}, for $\oneigfct_1$. This yields
\begin{eqnarray}
I_>(\tilde{b},x)&=&\frac{2x}{\ln E_1(x)}\int_{\tilde{b}/|x|}^{1/2}\rmd{\zL}\bigg[1\nonumber\\ &&\strut +\frac{\psi(\zL)+\psi(1-\zL)+2\gammaE+4\ln 2}{2|x|}\bigg]+o(1/|x|)\nonumber\\&=&\frac{-1}{\ln E_1(x)}\left[|x|+\gammaE-2\tilde{b}+\ln\frac{4\tilde{b}}{|x|}+o\big(|x|^0\big)\right].\nonumber\\
\end{eqnarray}

In order to evaluate $I_<(\tilde{b},x)$ for large negative $x$, note that according to
Eq.~\eqref{eq:varphipsi1} $\oneigfct_1(z/L;m L)=\sqrt{|m|L}\,\oneigfct_1(|m|z;|m|L,-1)$ approaches the half-bound state $\varphi_0(|m|z)$ in the limit $L\to\infty$. 
Hence  $\oneigfct_1(\zL,x)$ must behave as
\begin{equation}
\oneigfct_1(\zL,x)\mathop{=}_{x\to-\infty}\varphi_0(\zm)+o(|x|^0)
\end{equation}
and $I_<(\tilde{b},x)$ becomes
\begin{equation}
I_<(\tilde{b},x)=\frac{2}{|x|}\int_0^{\tilde{b}}\rmd{\zm}\,\regsol_0^2(\zm)+o(1/|x|).
\end{equation}
The integral on the right-hand side has been considered in Eq.~\eqref{eq:varphi2dec}. Its behavior for large $\tilde{b}$ is given by the last line of this equation with the replacement $\zm\to\tilde{b}$:
\begin{equation}
\int_0^{\tilde{b}}\rmd{\zm}\,\regsol_0^2(\zm)=J_\varphi+\tilde{b}-\frac{1}{2}\ln\tilde{b}+o(1/\tilde{b}).
\end{equation}

Upon inserting the above results for $I_>(\tilde{b},x)$ and $I_<(\tilde{b},x)$ into Eq.~\eqref{eq:onf1}, one sees that the $O(1/|x|)$ terms that diverge linearly or logarithmically as $\tilde{b}\to\infty$ cancel. Hence we can take the limit $\tilde{b}\to\infty$ \footnote{The reader might wonder whether our taking of the limit $\tilde{b}\to\infty$ is in conflict with the original requirement that $\tilde{b}/|x|<1/2$. To see that this is not the case, note that both the terms of order $1/|x|$ and $(\ln|x|)/|x|$ are ${x=-\infty}$ properties. Thus, we can first take the limit $x\to-\infty$ and subsequently let $\tilde{b}\to\infty$.}. Matching the terms of order $|x|$, $\ln|x|$ and $|x|^0$ then yields
\begin{equation}
\ln E_1(x)\mathop{=}_{x\to-\infty}-|x|+\ln|x|-\gammaE-2J_\varphi- \ln 4 +o(1).
\end{equation}
With the aid of Eq.~\eqref{eq:Jvarphires} our result for $E_1(x)$ thus becomes
\begin{align}\label{eq:E1asres}
\ln E_1(x)&\mathop{=}_{x\to-\infty}-|x|+\ln|x|+\ln\frac{\rme}{\pi}+o(1/|x|^0),\nonumber\\
E_1(x)&\mathop{=}_{x\to-\infty}\frac{\rme}{\pi}\,|x|\,\rme^{-|x|+o(1/|x|^0)}.
\end{align}

In order to determine the leading shifts $\Delta E_\nu$ of the eigenvalues $E_{\nu>1}(x)$ from their ${x=-\infty}$~values $\mathring{E}_\nu$ given in Eq.~\eqref{eq:spectrummxinfty}, it is natural to use Rayleigh-Schr\"odinger perturbation theory in $v(\zL,x)$. Since we know from Eq.~\eqref{eq:E1asres} that $E_1(x)$ remains exponentially small at large negative $x$, the  shift
\begin{equation}
\Delta E_\nu=-2\int_0^{1/2}\rmd{\zL}\,v(\zL,x)\big[\mathring{\oneigfct}_\nu(\zL)\big]^2
\end{equation}
must vanish when $\nu=1$. Upon subtracting $0=2\Delta E_1$  from $\Delta E_{\nu>1}$ and remembering that $\mathring{\oneigfct}_1\equiv 1$, we obtain
\begin{eqnarray}
\lefteqn{\Delta E_{\nu>1}(x)=-2\int_0^{1/2}\rmd{\zL}\,v(\zL,x)\left\{2-\big[\mathring{\oneigfct}_\nu(\zL)\big]^2\right\}}&& \nonumber\\
&=&\strut-4\int_{b/|x|}^{1/2}\rmd{\zL}\,v(\zL,x)\sin^2[\pi(\nu-1)\zL]+O(1/|x|)\nonumber \\
&=&2[\pi(\nu-1)]^2\,\frac{\ln |x|}{|x|}+O(1/|x|),
\end{eqnarray}
where we used the fact that the contribution from the omitted integral $\int_0^{b/|x|}\rmd{\zL}$ is $O(1/|x|)$.
To compute the remaining integral, we inserted Eq.~\eqref{eq:vxlargezLlarge} for $v(\zL,x)$ and kept only the terms $\asprop |x|^{-1} \ln |x|$. The result given in the last line means that the eigenvalues $E_{\nu>1}(x)$ behave in the  limit ${x\to-\infty}$ as 
\footnote{Following steps analogous to those taken in Appendices~\ref{app:semiclass} and \ref{app:semclchi}, one can extend this result to order $1/|x|$. A lengthy and cumbersome calculation yields $E_{\nu>1}/[\pi^2(\nu-1)^2]=1+2|x|^{-1}\ln[2|x|/[\pi(\nu-1)]+o(1/|x|)$.}
\begin{equation}\label{eq:Enugtr1}
E_{\nu>1}(x)=\pi^2(\nu-1)^2\bigg[1+\frac{2\ln |x|}{|x|}\bigg]+O(1/|x|).
\end{equation}

The result has a simple interpretation: the eigenvalues $E_{\nu>1}$ agree to the given order with those of the free $(v=0)$ system with Neumann boundary conditions up to a reduction  
\begin{eqnarray}\label{eq:Lred}
L&\to& L\big[1+2|x|^{-1}\ln|x|+O(1/|x|)\big]^{-1/2}\nonumber\\
&=&L\left[1-|x|^{-1}\ln |x|+O(1/|x|)\right]
\end{eqnarray}
of the thickness of the slab.

\subsection{Free-energy functional for the slab in the  continuum limit}\label{sec:freeenergyfunc}

In Sec.~\ref{sec:sclsce} we considered  the continuum limit of the self-consistency equation for the potential in detail.  In order to investigate the scaling function  $\Theta(x)$ of the residual free energy, it is useful to integrate the corresponding self-consistency Eq.~\eqref{eq:vsc} to obtain a free-energy functional whose extremum yields the residual free energy.

Since the bulk free energy and the surface free energy involve UV singularities, appropriate subtractions must be made to eliminate these singularities. In the translation-invariant case of pbc, this can be done by subtracting the Taylor expansion in the mass parameter $m^2$ to the lowest   order required to yield UV finite expressions \cite{DDG06}. In our case of fbc this procedure is impractical because the self-consistent potential has a nontrivial dependence on the temperature variable $m$. This entails a nontrivial $m$-dependence of the  eigenenergies $E_\nu$, which is not known in analytic closed form. We therefore use an alternative, more feasible approach.

Starting from the self-consistency equation for the continuum theory, Eq.~\eqref{eq:vsc}, we will construct a free-energy functional whose value at the maximizing potential $v_*(z)$ yields the free energy per area of the slab up to regular background terms. To achieve this, we multiply Eq.~\eqref{eq:vsc} by a ``test function'' $\delta v(z)$ with support on $[0,L]$, integrate $z$ over this interval, and interchange the order of integrations. We thus arrive at
\begin{equation}\label{eq:intscv}
mL\,\overline{\delta v}+\frac{\delta v(0)+\delta v(L)}{2}=\int_0^\infty\rmd{\omega}\left[\frac{L\,\overline{\delta v}}{2\sqrt{\omega}}+\delta\chi(\omega)\right],
\end{equation}
where
\begin{equation}\label{eq:deltachi}
\delta\chi(\omega)=\int_0^L\rmd{z}\,\delta v(z)\,\mathcal{G}(z,z;-\omega),
\end{equation}
while
$\overline{\delta v}$ denotes the average
\begin{equation}\label{eq:avdelv}
\overline{\delta v}=\int_0^L\frac{\rmd{z}}{L}\,\delta v(z).
\end{equation}
At this stage, it is sufficient to presume continuity of  $\delta v(z)$ on the closed interval $[0,L]$. Since this implies boundedness and integrability, the average~\eqref{eq:avdelv} exists. 

Let us show that $\delta\chi(y)$ is indeed the variation of a function $\chi(\omega)$ caused by a modification $\delta v(z)$ of the potential. Note, first, that the Green's function $\mathcal{G}(z,z';E)$ can be written in terms of the left and right regular solutions $\regsol_{l,r}$ of the Schr\"odinger equation 
\begin{equation}\label{eq:SElr}
-\partial_z^2\regsol_{l,r}(z;E)+[v(z)-E]\regsol_{l,r}(z,E)=0
\end{equation}
on $[0,L]$ which are fixed by the boundary conditions
\begin{eqnarray}\label{eq:bclr}
\regsol_l(z;E)&\mathop{=}\limits_{z\to 0+}&\sqrt{z}\,[1+O(z)],\nonumber\\
\regsol_l(z;E)&\mathop{=}\limits_{z\to L-}&\sqrt{L-z}\,[1+O(L-z)].
\end{eqnarray}

The Wronskian of these functions,
\begin{equation}
\Delta(E)\equiv W[\regsol_r(z;E),\regsol_l(z;E)],
\end{equation}
is called characteristic function of the Sturm-Liouville problem defined by Eqs.~\eqref{eq:SElr} and \eqref{eq:bclr}. It  is an entire function of the complex variable $E$ with simple zeros at the eigenvalues $\{E_\nu\}_{\nu=1}^\infty$ of this problem. Further, for  any complex $E$ not in the spectrum $\{E_\nu\}$ one has
\begin{equation}\label{eq:Gintermsoflr}
\mathcal{G}(z,z';E)=-\frac{\regsol_l(z_<;E)\,\regsol_r(z_>;E)}{\Delta(E)}.
\end{equation}
This should be obvious because the right-hand side of Eq.~\eqref{eq:Gintermsoflr} is a solution to Eq.~\eqref{eq:SElr} for $z\ne z'$ and the jump condition $\partial_z G(z,z';E)|_{z=z'-0}^{z=z'+0}=1$ as well as the boundary conditions~\eqref{eq:bclr} are satisfied.

The above-mentioned properties of  $\Delta(E)$ imply that
\begin{equation}\label{eq:DeltaElnG}
\ln[\text{const}\, \Delta(E)]=\Tr \ln (E-\mathcal{H}_v)=\Tr\ln\mathcal{G}^{-1}(E).
\end{equation}
Noting that $\delta \mathcal{G}^{-1}=-\delta\mathcal{H}=-\delta v$, we see that the variation of the left-hand side yields the negative of the expression on the right-hand side of Eq.~\eqref{eq:deltachi}. Equation~\eqref{eq:deltachi} can therefore we interpreted as the variation of the function
\begin{equation}\label{eq:DeltaEdef}
\chi({\omega})\equiv -\ln[\pi\Delta(-\omega)],
\end{equation}
where the prefactor $\pi$ was chosen to simplify subsequent relations.

We wish to consider potentials $v(z)$ that are even with respect to reflections at the midplane $z=L/2$ and have a Laurent expansion about $z=0$ with a fixed pole part. Accordingly, we require that
\begin{equation}\label{eq:vsymcond}
 v(z)= v(L-z)
\end{equation}
and that $v(z)$ can be expanded about $z=0$ as 
\begin{equation}\label{eq:vLaurent}
v(z)=-\frac{1}{4z^2}+\frac{4 m}{\pi^2z}+u_0+u_1z+\ldots,
\end{equation}
where the principal part of the Laurent series (pole terms) is fixed while the remaining nonsingular expansion terms including $u_0$ may vary. Note that we use the notation $u_j$ for the coefficients of the varying Taylor series part of the potentials considered to avoid confusion with coefficients such as  $v_0$ of the self-consistent potentials for $m=\pm 1$.  Variations $\delta v(z)$ therefore satisfy the relations
\begin{equation}\label{eq:delv0cond}
\delta v(L)=\delta v(0)=\delta u_0
\end{equation}
and do not involve pole terms. Furthermore, the function $\Delta(E)$ simplifies to
\begin{equation}\label{eq:Deltaregsol}
\Delta (E)=\partial_z \regsol^2_l(\zm;E)\big|_{z=L/2}.
\end{equation}

In order to integrate $\chi(\omega)$ with respect to $\omega$ to determine a free-energy functional that yields Eq.~\eqref{eq:intscv} as stationarity condition, we must know the behavior of $\chi(\omega)$ at large and small $\omega$. The leading asymptotic behavior of $\chi(\omega)$ for large $\omega$ is a bulk property and easily determined. Setting $m=0$ and subtracting the bulk term at $\omega=0$ to eliminate UV singularities, we obtain $\chi({\omega\to\infty})\aseq-(L/\pi)\int_0^\infty\rmd{k}\ln[(k^2+\omega)/k^2]=-L\sqrt{\omega}$. Since $\int\rmd{z}\,v(z)$ has momentum dimension $1$, we can conclude that the next-to-leading terms are of order $\omega^{-1/2}\mod \ln\omega$. Furthermore, $\chi(\omega)$ can be expressed in terms of its derivative as
\begin{equation}\label{eq:chiitrho}
\chi(\omega)=-L\sqrt{\omega}-\int_\omega^\infty\rmd{\omega'}\left[\chi'(\omega')+\frac{L}{2\sqrt{\omega'}}\right].
\end{equation}

In Appendix~\ref{app:semclchi} we extend the above asymptotic expansion  of $\chi(\omega)$  by computing the two next-to-leading terms by means of the semiclassical expansion, obtaining
\begin{eqnarray}\label{eq:chilargeom}
\chi(\omega)&\mathop{=}\limits_{\omega\to\infty}&-L\sqrt{\omega}-\frac{1}{2\sqrt{\omega}}\bigg[L\mathcal{R}([v];L,m)+\frac{4m}{\pi^2}\ln(\omega L^2)\bigg]\nonumber\\ &&\strut +\frac{2m^2}{\pi^2\omega}+O(\omega^{-3/2}),
\end{eqnarray}
where $\mathcal{R}([v];L,m)$ denotes the functional
\begin{equation}\label{eq:Rdef}
\mathcal{R}([v];L,m)\equiv \frac{1}{L^2}+\frac{8m}{\pi^2L}\,(\gammaE+\ln 4)+\overline{v^{\text{ns}}},
\end{equation}
while
\begin{equation}\label{eq:vnsav}
\overline{v^{\text{ns}}}\equiv 
\frac{2}{L}\int_0^{L/2}\rmd{z}\bigg[v(z)+\frac{1}{4z^2}-\frac{4m}{\pi^2z}\bigg]
\end{equation}
is the average of the nonsingular part of the potential.

In the limit $\omega\to 0$, the function $\chi(\omega)$ approaches a constant $\chi(0)$, whose bulk contribution $-Lm$ follows from a calculation analogous to the one performed above. The value $\chi_*(0)$ at  the extremal point $v_*(z;L,m)$ that maximizes the functional $\mathcal{F}_\mu$ can be computed in a straightforward manner. To this end, we integrate the self-consistency equation~\eqref{eq:vsc} over the interval $(0,L)$, interchange the integrations over $z$ and $\omega$, and use
\begin{equation}\label{eq:chiprime}
\chi'(\omega)=\int_0^L\rmd{z}\,\mathcal{G}(z,z;-\omega)
\end{equation}
to conclude that
\begin{equation}\label{eq:chistar0}
mL+1=\int_0^\infty\rmd{\omega}\left[\frac{L}{2\sqrt{\omega}}+\chi'_*(\omega)\right]=-\chi_*(0).
\end{equation}
Here the contribution from the upper integration limit vanishes by Eq.~\eqref{eq:chiitrho}, and the asterisk indicates evaluation at the maximizing potential $v_*(z;L,m)$.

We are now ready to introduce a free-energy functional whose variation subject to the constraints specified in Eqs.~\eqref{eq:vsymcond}--\eqref{eq:delv0cond} is in conformity with the stationarity condition~\eqref{eq:vsc} and gives back the implied Eq.~\eqref{eq:intscv}. To integrate $\chi(\omega)$, we subtract from it a contribution with the same large-$\omega$ behavior as specified in  Eq.~\eqref{eq:chilargeom}, so that the integral of the difference $\int_0^\infty\rmd{\omega}$ converges at the upper integration limit. To avoid convergence problems at the lower integration limit, we make the replacement $\omega\to\mu^2+\omega$ in the last one of the terms explicitly displayed in Eq.~\eqref{eq:chilargeom}, where $\mu>0$ is an arbitrary momentum scale, and introduce the free-energy functional
\begin{subequations}\label{eq:fmudef}
\begin{equation}\label{eq:fmusumdef}
\mathcal{F}_\mu([v];L,m)=\mathcal{F}^{(1)}([v];L,m)+\mathcal{F}^{(2)}_\mu([v];L,m)
\end{equation}
with
\begin{equation}\label{eq:fmu1def}
\mathcal{F}^{(1)}([v];L,m)=\frac{mL\,\mathcal{R}([v];L,m)+u_0}{8\pi}
\end{equation}
and
\begin{eqnarray}\label{eq:fmu2def}
\mathcal{F}^{(2)}_\mu([v];L,m)&=& - \int_0^\infty\frac{\rmd{\omega}}{8\pi}\bigg[\chi([v],\omega)+L\sqrt{\omega}\nonumber \\ 
&&\strut  +\frac{2m\ln(\omega L^2)}{\pi^2\sqrt{\omega}}+\frac{L}{2\sqrt{\omega}}\,\mathcal{R}([v];L,m)\nonumber\\ &&\strut  -\frac{2m^2}{\pi^2(\mu^2+\omega)}\bigg],
\end{eqnarray}
\end{subequations}
where we have explicitly indicated the functional dependence of $\chi$ on the potential $v$.

Since $\delta\mathcal{R}([v];L,m)=\overline{\delta v}$, the stationarity condition $\delta\mathcal{F}_\mu([v],L,m)|_{v=v_*}=0$ is satisfied as a consequence of Eq.~\eqref{eq:intscv}, where it should be remembered that only the nonsingular part of $v(z)$ 
(including the parameter $u_0$) 
is to be varied while the singular one is fixed at $v_*^{\text{sg}}(z)$, as indicated in Eq.~\eqref{eq:vLaurent}.

The logarithmic derivative of $\mathcal{F}_\mu$ with respect to $\mu$ at fixed $v,L$, and $m$ is given by
\begin{equation}
\mu\partial_\mu\big|_{v,L,m}\mathcal{F}_\mu([v];L,m)=-\frac{m^2}{2\pi^3}.
\end{equation}
Under a change $\mu\to\mu \ell$, it transforms as
\begin{align}\label{eq:Fell}
\mathcal{F}_{\mu\ell}([v];L,m)&=\mathcal{F}_{\mu}([v];L,m)-\frac{m^2}{2\pi^3}\ln\ell\nonumber\\
&=\mu^2\ell^2\mathcal{F}_1([\mu^{-2}\ell^{-2}v];\mu \ell L,m\mu^{-1}\ell^{-1}),
\end{align}
where the last line follows from dimensional considerations. We denote the value of this functional at the maximizing potential $v_*$ as
\begin{equation}\label{eq:Fvstar}
f_\mu(L,m)\equiv \max_{v^{\text{ns}}=v- v_*^{\text{sg}}}\mathcal{F}_\mu([v];L,m)=\mathcal{F}_\mu([v_*];L,m).
\end{equation}
Choosing $\ell=(\mu L)^{-1}$ in Eq.~\eqref{eq:Fell} then gives us the relation
\begin{equation}\label{eq:fscalform}
f_\mu(L,m)=L^{-2}\,Y(mL)-\frac{m^2}{2\pi^3}\ln(\mu L)
\end{equation}
with
\begin{equation}\label{eq:Ydef}
Y(x)\equiv f_1(1,x).
\end{equation}

To relate the scaling function  $\Theta(x)$ of the residual free energy to $f_\mu(L,m)$, we must determine the bulk and surface free energy contributions $L \fb(m)$ and $2 \fs(m;\mu)$ contained in this free energy. The former is readily determined using $\mathcal{R}([v_*];L,m)=m^2\Htheta(m)+o(1/L)$ and $\chi(\omega)=-L\sqrt{\omega+m^2\Htheta(m)}+o(L)$. One obtains
\begin{eqnarray}
\lefteqn{\fb(m)=\lim_{L\to\infty}L^{-1}f(L,m)}&& \nonumber\\ &=&\Htheta(m)\bigg[\frac{m^3}{8\pi}+\int_0^\infty\frac{\rmd{\omega}}{8\pi}\left(\sqrt{\omega+m^2}-\sqrt{\omega}-\frac{m^2}{2\sqrt{\omega}}\right)\bigg]\nonumber\\
&=& \frac{m^3}{24\pi}\,\Htheta(m)
\end{eqnarray}
in accordance with Eq.~\eqref{eq:fbsing}.

To determine the surface contribution to $\mathcal{F}_\mu([v_*];L,m)$ we must compute the limit 
\begin{equation}
2\fs(m;\mu)=\lim_{L\to\infty}\big[\mathcal{F}_\mu([v_*];L,m)-L\fb(m)+\frac{m^2}{\pi^3}\ln(\mu L)\big].
\end{equation}
In Appendix~\ref{app:fs} we show that
\begin{align}\label{eq:fmutoooflnL}
\mathcal{F}_\mu([v_*];L,m)=&\Htheta(m)\bigg[L\frac{m^3}{24\pi}+2m^2\Delta A^{(\mathrm{s})}_0\bigg]\nonumber\\ &
\strut+\frac{m^2}{\pi^3}\bigg[\frac{3}{4}+\frac{7\zeta(3)}{\pi^2}\bigg] -\frac{m^2}{2\pi^3}\ln\frac{2L|m|}{\pi}\nonumber\\ &\strut-\frac{m^2}{2\pi^3}\ln(\mu L)
+O(1/L).
\end{align}
This yields
\begin{align}\label{eq:fsres}
\fs(m;\mu)=&\frac{m^2}{2\pi^3}\bigg[\frac{3}{4}+\frac{7\zeta(3)}{\pi^2}-\frac{1}{2}\ln\frac{2|m|}{\pi\mu}\bigg]\nonumber\\ &\strut+m^2\Delta A^{(\mathrm{s})}_0\,\Htheta(m).
\end{align}

The above results enable us to identify those parts of the scaling function $Y(x)$ that correspond to bulk and surface contributions to the free energy per area. 
It follows that the scaling function $\Theta(x)$ of the residual free energy is related to $Y(x)$ via
\begin{align}\label{eq:YitofTheta}
Y(x)=&\bigg[\frac{x}{48\pi}+\Delta A_0^{(\mathrm{s})}\bigg]2x^2\Htheta(x)+\frac{x^2}{\pi^3}\bigg[\frac{3}{4}+\frac{7\zeta(3)}{\pi^2}\nonumber\\ &\strut -\frac{1}{2}\ln\frac{2|x|}{\pi} \bigg]+\Theta(x).
\end{align}

We are now ready to turn to a discussion of exact properties of the scaling functions $\Theta(x)$ and $\vartheta(x)$ of the residual free energy and the Casimir force. In the next subsection, we recapitulate the qualitatively distinct asymptotic behaviors of these functions in the high-temperature, low-temperature and critical-temperature limits $x\to+ \infty$, $x\to -\infty$, and $x\to 0\pm$, recall their known exact properties, and then focus on the derivation of their limiting forms for $x\to-\infty$. For the latter purpose we shall need the following property of the free energy introduced in Eq.~\eqref{eq:Fvstar},
\begin{equation}\label{eq:tracerelcont}
\frac{\partial f_\mu(L,m)}{\partial m}=\frac{L\mathcal{R}([v_*];L,m)+2v_0m}{8\pi}-\frac{m}{\pi^3}\ln(\mu L).
\end{equation}
This relation may be viewed as the continuum analog of Eq.~\eqref{eq:dfNdt}. Since the term $L\mathcal{R}([v_*];L,m)$ agrees with $L\overline{v^{\text{ns}}}$, the ``trace'' of $v^{\text{ns}}$, up to an $L$-dependent constant, its form~\eqref{eq:tracerelcont} is plausible. However, its derivation is not entirely trivial because of the singular parts of the full self-consistent potential $v_*$. From Eq.~\eqref{eq:fscalform} we see that what needs to be proven to establish Eq.~\eqref{eq:tracerelcont} is the implied relation for the scaling function $Y(x)$, namely
\begin{equation}\label{eq:Yprimerel}
Y'(x)=\frac{1}{8\pi}\mathcal{R}([v_*];1,x)+C_1x
\end{equation}
with
\begin{equation}\label{eq:Cex}
C_1=\frac{v_0}{4\pi}=\frac{14\zeta(3)}{\pi^5}=0.054992529830367\ldots.
\end{equation}

In Appendix~\ref{app:derivrel} we give a direct derivation of Eq.~\eqref{eq:Yprimerel} from the free-energy functional~\eqref{eq:fmudef}. For the constant $C_1$ we obtain the result
\begin{align}\label{eq:Cintexpr}
C_1=&\frac{\gammaE+\ln 8}{\pi^3}+\frac{4}{\pi^5}\int_0^\infty\rmd{z} \bigg\{\frac{1}{\sqrt{2z}} K_0(z)\,\frac{\partial M_{\kappa,0}(2z)}{\partial \kappa}\Big|_{\kappa=0}\nonumber\\ 
&\strut+I_0(z)\bigg[K_0(z)\,(\gammaE+\ln 4)\nonumber\\ &\strut
+\sqrt{\frac{\pi}{2z}}\,\frac{\partial W_{\kappa,0}(2z)}{\partial \kappa}\Big|_{\kappa=0}\bigg]\bigg\}\ln z,
\end{align}
where $\partial_\kappa M_{\kappa,0}(z)$ and $\partial_\kappa W_{\kappa,0}(z)$ are derivatives of Whittaker $M$ and $W$ functions.

We have not been able to evaluate the integral on the right-hand side analytically. However, upon computing the corresponding expression for $C_1$  by numerical integration via {\sc Mathematica} \cite{Mathematica10}, we verified that the analytical value given in Eq.~\eqref{eq:Cex} is reproduced to 13 digits. Moreover, we have  shown by independent arguments that $C_1$ has the exact analytical value given in Eq.~\eqref{eq:Cex}: We have determined the asymptotic $x\to-\infty$ behavior of $Y(x)$ in two different ways, one that uses Eq.~\eqref{eq:Yprimerel} with an unspecified constant $C_1$, and a second one based on inverse scattering results, which involves the potential parameter $v_0$ (see Appendix~\ref{app:scfctxminfty}). Matching the corresponding results yields the analytical value $C_1=v_0/4\pi$. 

Let us also note the following useful relation between the scaling function $Y(x)$ and the value of the functional $\mathcal{F}^{(2)}_1([v];1,x)$ at the self-consistent potential $v_*(z;1,x)$:
\begin{equation}\label{eq:LTr}
x\frac{\rmd Y(x)}{\rmd x}-Y(x)=-\mathcal{F}^{(2)}_1([v_*];1,x)+\frac{7\zeta(3) x^2}{\pi^5},
\end{equation}
which follows immediately  from Eqs.~\eqref{eq:fmudef}, \eqref{eq:Ydef}, \eqref{eq:Yprimerel}, and  \eqref{eq:Cex}.

\subsection{Asymptotic behaviors of the scaling functions $\Theta(x)$ and $\vartheta(x)$}\label{sec:asbehscalf}

The scaling functions $\Theta(x)$ and $\vartheta(x)$ are known in numerical form from the results of \cite{DGHHRS12} and \cite{DGHHRS14}. They exhibit qualitatively distinct asymptotic behaviors in the high-temperature, low-temperature, and critical limits $x\to+\infty$, $x\to-\infty$ and $x\to 0\pm$, respectively. Although some discussion of these issue can be found in \cite{DGHHRS12}, \cite{DGHHRS14}, \cite{DR14}, and elsewhere (e.g.,  in \cite{KD91,GD08,DS11} for general values of d), it will be helpful to recall the necessary background and some known results. 

For $x>0$, we are dealing with a massive theory confined to a film with free surfaces. Therefore, both scaling functions must vanish exponentially; one has
 \begin{eqnarray}\label{eq:Thetaaslargex}
\begin{array}[c]{l}
\Theta(x\to\infty) \\
\vartheta(x\to\infty)
\end{array}\bigg\}\asprop\rme^{-2x+O(\ln x)},
\end{eqnarray}
where the factor $2$ in the exponential is specific to free boundary conditions \footnote{Roughly speaking, perturbations of a local quantity at position $\bm{x}=(\bm{y},z)$ due to the presence of surfaces involves squares and higher powers of correlators between $\bm{x}$ and the surface points. The associated minimal decay length is twice the true bulk correlation length. In the case of periodic boundary conditions \cite{Dan96},  the decay  is  $\asprop \rme^{-x}$ instead.}. 

In the critical limit $x\to 0$, $\Theta(x)$ approaches the Casimir amplitude $\D{}$ [cf.\ Eq.~\eqref{eq:DeltaCdef}], whose exact analytical value is not known but for which the  numerically precise result
\begin{equation}\label{eq:CAnum}
\D{}=-0.01077340685024782(1)
\end{equation}
was obtained in \cite{DGHHRS14}.

Since the free energy must be regular in $m$ when $L<\infty$, the singular bulk and surface free energy contributions must be canceled by corresponding contributions $\asprop x^3$, $\asprop x^2$, and $\asprop x^2\ln|x|$ of $\Theta(x)$. As discussed in \cite{DR14}, this implies the limiting behavior
\begin{align}\label{eq:Thetacrit}
\Theta(x)\mathop{=}_{x\to 0\pm}& \D{}-\left[\Delta A_0^{(\rm{s})}+\frac{x}{48\pi}\right]2x^2\,\Htheta(x)\nonumber\\&\strut +\frac{1}{2\pi^3}\,x^2\ln|x|+\Theta_1 x+\Theta_2 x^2+\Theta_3x^3\nonumber\\ &\strut+o(x^3),
\end{align}
where $\Delta A_0^{(\rm{s})}$ is the universal amplitude difference~\eqref{eq:DelA0anexp}. Here $\sum_{j=1}^3\Theta_j x^j$ are regular contributions and the $o(x^3)$ terms involve regular contributions of orders $x^4$ and higher. Setting $d=3$ in Eq.~\eqref{eq:relvarthetaTheta} to determine the consequences for $\vartheta(x)$ yields
\begin{align}\label{eq:varthetacrit}
\vartheta(x)\mathop{=}_{x\to 0\pm}& 2\D{}+\Theta_1x-\frac{x^2}{2\pi^3}+\frac{x^3}{24\pi}\Htheta(x)\nonumber\\ &\strut -\Theta_3 x^3+o(x^3),
\end{align}
from which one recovers the previously mentioned result \cite{DR14}
\begin{equation}\label{eq:vartheta2der}
\vartheta''(0)=-1/\pi^3.
\end{equation}

In the low-temperature limit $x\to-\infty$, $\Theta(x)$ approaches the exactly known Casimir amplitude $-\zeta(3)/(16\pi)$ of a free field theory subject to Neumann boundary conditions \cite{KD91}, as follows from the mapping to the nonlinear sigma model \cite{DGHHRS12,DGHHRS14,DN86}.

The results given in Eqs.~\eqref{eq:Enugtr1} and \eqref{eq:E1asres} can be combined in a straightforward fashion with the known result $-\zeta(3)/(16\pi L^2)$ for the residual free energy of a massless free field theory subject to Neumann boundary conditions to gain information about the asymptotic behavior of the scaling function $\Theta(x)$ for $x\to-\infty$. The reduction of $L$ specified in Eq.~\eqref{eq:Lred} gives us the leading correction $\asprop |x|^{-1}\ln|x|$ to the limiting value $\Theta(-\infty)$. Upon including also the next-to-leading term $\asprop d_1/|x|$, we arrive at the limiting form
\begin{equation}\label{eq:Thetalargemx}
\Theta(x)\mathop{=}_{x\to-\infty}-\frac{\zeta(3)}{16\pi}\left[1+\frac{d_1+2\ln |x|}{|x|}+o\big(1/|x|\big)\right].
\end{equation}

This limiting behavior has previously been stated in \cite{DGHHRS12} and \cite{DGHHRS14}. The exact value of $2$ of the coefficient in front of the $|x|^{-1}\ln|x|$ --- here obtained as a consequence of Eqs.~\eqref{eq:Enugtr1} and \eqref{eq:Lred} --- was derived  there by matching the low-temperature behavior of our model to that of a nonlinear sigma model. Unfortunately, the latter analysis left the exact value of the universal coefficient $d_1$ undetermined. 

We now turn to the direct derivation of the asymptotic form~\eqref{eq:Thetalargemx} within the framework of the ${n\to\infty}$ analysis of our $O(n)$ $\phi^4$ model. For this purpose  we must extend our results 
 for the free energy and its scaling function $Y(x)$ given in Eqs.~\eqref{eq:fmutoooflnL} and \eqref{eq:YitofTheta}  by working out their asymptotic contributions in the limit ${x\to-\infty}$ up to $O(1/|x|)$. Let us start by determining those of these contributions that originate from the free-energy term $\mathcal{F}_1^{(1)}([v_*];1,x)$. We need
the asymptotic ${x\to-\infty}$~behavior of $\mathcal{R}_*(1,x)$ to $o(1/|x|)$. This can most easily be computed via Eq.~\eqref{eq:Yprimerel}. Upon inserting the $x\to-\infty$ form~\eqref{eq:Thetalargemx} of $\Theta(x)$ into Eq.~\eqref{eq:YitofTheta}, we can compute $Y'(x)$ and solve for $\mathcal{R}_*(1,x)$. Taking into account Eq.~\eqref{eq:Cex}, we find
\begin{align}\label{eq:Rstartoxm2}
\mathcal{R}_*(1,x)=&-\frac{8|x|}{\pi^2}[1-\ln(2|x|/\pi)]\nonumber\\ &\strut+\zeta(3)\,\frac{1-d_1/2-\ln|x|}{|x|^2}+o(1/|x|^2),
\end{align}
i.e., no terms of order $1/|x|$ appear in the asymptotic $x\to-\infty$ form of this quantity.

The calculation of $\mathcal{F}_1^{(2)}([v_*];1,x)$ to $o(1/|x|)$ is more involved and relegated to Appendix~\ref{app:scfctxminfty}. It confirms the asymptotic form of $\Theta(x)$ given in Eq.~\eqref{eq:Thetacrit} and 
yields for the coefficient $d_1$ the result
\begin{eqnarray}
\label{eq:d1res}
d_1&=&2\bigg[\gammaE+\ln\frac{4}{\pi}\bigg]-1-2\frac{\zeta'(3)}{\zeta(3)}\nonumber\\
&=&0.96720564466060109\ldots .
\end{eqnarray}

This value is  consistent with the result $d_1= 1.0(1)$ obtained in  \cite{DGHHRS14} by  numerical solutions of the self-consistency equations.

The $x\to-\infty$ behavior of the Casimir force scaling function implied by Eq.~\eqref{eq:relvarthetaTheta} is, of course, the same as given in \cite{DGHHRS14}, namely
\begin{equation}\label{eq:varthetalargemx}
\vartheta(x)\mathop{=}_{x\to-\infty}-\frac{\zeta(3)}{8\pi}\bigg[1+\frac{3d_1/2-1+3\ln |x|}{|x|}+o(1/|x|)\bigg],
\end{equation}
except that we now have the exact analytical value~\eqref{eq:d1res} for $d_1$.

\section{Summary and conclusions}\label{sec:SumConcl}

In this paper we explored the potential of inverse scattering theory to  gain exact information about the $n=\infty$ solution of the $O(n)$ $\phi^4$ on a three-dimensional film of size $\infty^2\times L$ bounded by a pair of free surfaces. A wealth of exact results could be determined. Since their derivation required a combination of various tools and lengthy calculations (most of which are presented in the appendixes), it will be helpful to briefly summarize them.

The essence of our strategy was to eliminate the potential $v(z)$ of the self-consistent Schr\"odinger problem that the  exact $n=\infty$ solution involves in favor of scattering data. This enabled us to reformulate the self-consistency equation in terms of scattering data.

For the semi-infinite case $L=\infty$ we succeeded in the exact determination of the scattering data pertaining to the self-consistent potential $v(z;{L=\infty},m)$ for all values of the temperature variable $m$ at, above and below criticality  ($m=0$). The corresponding results are presented in Secs.~\ref{sec:hssdmp} and  \ref{sec:hssdmm}; see Eqs.~\eqref{eq:scatdatm0}, \eqref{eq:Asol} and \eqref{eq:etapfres}, and \eqref{eq:sigmsolf} and \eqref{eq:etamfres} for the cases $m=0$, $m>0$, and $m<0$, respectively. The obtained phase shifts $\eta_0(k)$ and $\eta_\pm(\sk)$ are depicted in Fig.~\ref{fig:eta}.

The knowledge of these scattering data enabled us to obtain exact results for the two-point boundary correlation function for all $m\gtreqless 0$. The results are described in Secs.~\ref{sec:crit2ptfct}--\ref{sec:2ptord} and summarized in Table~\ref{tab:Gcomp}.

We then exploited these scattering data in conjunction with information about the potential obtained via boundary-operator expansions, on the one hand,  and various semiclassical expansions and a trace formula, on the other hand, to derive a variety of exact results for the case of a film of finite thickness $L$. In Sec.~\ref{eq:potrelq} we computed several quantites related to the self-consistent potential, among them the potential coefficient $v_0=m^{-2}\lim_{z\to 0}v^{\text{ns}}(z;\infty,m)$, the universal amplitude difference $\Delta A^{(\mathrm{s})}_0$, and an excess quantity associated with the squared order-parameter density. The results are given in Eqs.~\eqref{eq:v0res}, \eqref{eq:DelA0anexp}, and \eqref{eq:Jvarphires}, respectively.

A major issue to which we then turned was the asymptotic behavior of the spectrum and the scaling functions of the residual free energy and the Casimir force in the low-temperature scaling limit $x\to-\infty$. Our findings for the asymptotic behavior of the  spectrum in this limit are presented in Sec.~\ref{sec:asspecprop}. Our result for the exponential vanishing of the lowest eigenvalue $E_1$ is given in Eq.~\eqref{eq:E1asres}. The remaining energy levels $E_{\nu>1}$ approach their limiting values associated with a free massless theory subject to Neumann boundary conditions algebraically modulo logarithmic corrections; see Eq.~\eqref{eq:Enugtr1} and \cite{Note16}. We also obtained the asymptotic behavior of the self-consistent potential in the inner region of the film; our results for the scaled potential $v(z/L;1,x<0)$ are presented in Eqs.~\eqref{eq:vinner} and \eqref{eq:vxlargezLlarge}.

To derive the asymptotic behavior of the scaling functions $\Theta(x)$ and $\vartheta(x)$, we extended our analysis by constructing a free-energy functional (cf.\ Sec.~\ref{sec:freeenergyfunc}) whose stationarity condition yields the self-consistent Schr\"odinger equation involving the potential (see Sec.~\ref{sec:freeenergyfunc}). This enabled us to obtain an integral representation for the scaling function $Y(x)$ of the free-energy density in terms of the Green's function of the continuum theory and the self-consistent  potential.

Using semiclassical expansions in conjunction with the $L=\infty$ scattering data and our other results, we then determined the asymptotic $x\to-\infty$ behavior of this function $Y(x)$, from which those of $\Theta(x)$ and $\vartheta(x)$ follow directly. Our  results for the latter are given in Eqs.~\eqref{eq:Thetalargemx} and \eqref{eq:varthetalargemx}. They agree with the asymptotic forms obtained in \cite{DGHHRS12} and \cite{DGHHRS14} and corroborated there by numerical results.

The analytic results of the present work for the asymptotic $x\to-\infty$ behavior of the scaling functions go in several ways beyond those of \cite{DGHHRS12} and \cite{DGHHRS14}: In the latter two papers, both the asymptotic forms~\eqref{eq:Thetalargemx} and \eqref{eq:varthetalargemx}  and the exact values of the amplitudes of the logarithmic anomalies $\asprop |x|^{-1} \ln|x|$ of $\Theta(x)$ and $\vartheta(x)$ were obtained via the mapping to a nonlinear sigma model. Here, we derived these results directly from our  $O({n\to\infty})$ $\phi^4$ model in the scaling limit. In addition, we calculated the exact analytical value of the coefficient $d_1$ which governs the amplitudes of the ${x\to-\infty}$ contributions $\propto |x|^{-1}$ of $\Theta(x)$ and $\vartheta(x)$. This value, given in Eq.~\eqref{eq:d1res}, is in conformity with the numerical estimate of \cite{DGHHRS14}.

Note that in our analysis in this paper we started from a lattice model to avoid UV singularities, but then focused quickly on its continuum scaling limit (specified at the beginning of Sec.~\ref{sec:sclsce}). The rationale of this approach is easily understood: Of primary interest are the universal features encoded in the scaling functions $\Theta(x)$ and $\vartheta(x)$. Focusing on the scaling limit is a convenient way to eliminate the less important microscopic details associated with the behavior on microscopic scales. If one determines the self-consistent potential and the free energy of the $n\to\infty$ solution of a given $O(n)$  lattice model  representing the universality class of our model by numerical means, as was done in \cite{DGHHRS12} and \cite{DGHHRS14} for two families of such models, 
care must be exercised when extracting the universal scaling functions $\Theta(x)$ and $\vartheta(x)$ from the numerical data. The latter include the full information about the nonuniversal behavior of the model on microscopic scales. The determination of the scaling functions requires the elimination of such nonuniversal small-scale features.  To this end, the appropriate scaling limit $m\to0$ and $L\to\infty$ at fixed $x=mL$ must be investigated, where corrections to scaling should be properly taken into account. As $x$ takes on large negative values, it becomes increasingly more difficult to determine reliable results from the numerical data. Nevertheless, sufficiently large values of $-x$ could be reached in \cite{DGHHRS14} to see the asymptotic $x\to-\infty$ behavior. Moreover, the numerical results of  \cite{DGHHRS14}  for the scaling functions $\Theta(x)$ and $\vartheta(x)$ are consistent with all exactly known results, including the value of the coefficient $d_1$ obtained here. The precision with which the latter has been numerically checked is  admittedly still modest. However, other exactly known properties such as $\vartheta''(0)$, the universal amplitude difference $\Delta A^{(\mathrm{s})}_0$, or the amplitude of the $|x|^{-1}\ln|x|$ contribution to the asymptotic $x\to\infty$ behavior of $\Theta(x)$ have been confirmed with much greater numerical accuracy.

In a recent paper \cite{DBR14}, Dantchev \emph{et al} (DBR) studied the exact solution of a mean  spherical lattice model on a film of size $\infty^2\times L$ bounded by two free surfaces with separate constraints imposed on the average of the squares on the spins in each layer $z$. This model is known to describe the $n\to\infty$ limit  of a fixed-length spin $O(n)$ spin model \cite{Kno73}. DBR investigated, on the one hand, its exact solution by numerical means and derived,  on the other hand, a low-temperature expansion for finite $L$. A discussion of how their work relates to the one of Diehl \emph{et al} (DGHHRS),  presented first in the earlier letter \cite{DGHHRS12} and subsequently in more details in \cite{DGHHRS14}, has already been given in \cite{DGHHRS15}. Hence we restrict ourselves here to a few clarifying remarks. 

DBR overlooked the fact that  $O(n)$ $\phi^4$ models with quadratic and quartic interaction constants $\tau$ and $g$ turn into fixed spin-length models in the limit $g\to\infty$ at fixed $\tau/g$. Therefore they did not realize that DGHHRS's work includes the exact solution of their mean spherical model as special case and that the numerical data DGHHRS presented for the $g\to\infty$ limit of DGHHRS's $\phi^4$ model  B directly apply to this mean spherical model.

DBR's numerical data appear to be in conformity with those of DGHHRS. However, differences occur in the analysis of the numerical data. In the near-critical regime (small $m$) DBR  recovered the scaling behavior reported by DGHHRS. On the other hand, they had problems to observe scaling for larger values of $-x=-mL$. As is discussed in \cite{DGHHRS15}, this may be attributed to their inclusion of data well outside the scaling regime in attempted scaling plots and an insufficient study of the scaling limit. In their low-temperature analysis, DBR focused on the behavior for low temperatures at fixed $L$ rather than on a proper determination of the scaling functions for large values of $-x$. They found that their low-temperature data at fixed $L$ are well described by the analytical low-temperature expansion they determined. 

Note that the asymptotic $x\to-\infty$ behavior of the scaling functions $\Theta(x)$ must of course comply with the low-temperature expansion. Inspection of this expansion by DBR shows indeed that the limiting value $\Theta(-\infty)$ and the coefficient of the $|x|^{-1}\ln |x|$ contribution [cf.\ Eq.~\eqref{eq:Thetalargemx} and the related one for $\vartheta(x)$, Eq.~~\eqref{eq:varthetalargemx}] may be recovered from it. On the other hand, the coefficient $d_1$ of the correction term $\propto |x|^{-1}$ cannot be derived from it. Unfortunately, the information that can be extracted from the low-temperature expansion is rather limited. The reasons are twofold. First of all, the expansion is not in inverse powers of the scaling variable $x$ (modulo logarithms). Second, nonuniversal microscopic contributions are still contained in this expansion.

As can be seen from the lengthy and, in part, tedious calculations we had to perform, the precise  determination of the asymptotic $x\to-\infty$ forms of the scaling functions to the given order in $1/|x|$ is a quite demanding task. 
By taking the continuum limit, we could eliminate the nonuniversal behavior at microscopic distances from the boundaries. For local densities such as the energy  and magnetization densities distinct nontrivial behaviors in the near-boundary regimes of distances  $L-z$ and $z\lesssim 1/|m|\ll L$ from the surfaces and the inner regime $1/|m|\ll z, L-z\lesssim L/2$ occur, both of which generally contribute to quantities involving integrals over the film such as the free energy or the Casimir force.

Although we have not been able to determine the desired scaling functions $\Theta(x)$ and $\vartheta(x)$ exactly in closed analytic form, the wealth of  exact analytic results we have been able to obtain by means of inverse scattering theory indicates that the strategy we have pursuit --- eliminating the potential from the self-consistency equation in favor of scattering data --- has considerable potential for  successful applications to analogous self-consistency problems. Interesting examples of such problems where similar strategies might prove fruitful are  studies of stochastic dynamic models and quantum quenches in the large-$n$ limit.

\begin{acknowledgments}
We have enjoyed fruitful interactions and discussions with Martin Hasenbusch, Fred Hucht, and Felix Schmidt during the course of this work.  In particular, we are grateful to Fred Hucht for pointing out that our original result for $d_1$ can be rewritten as in Eq.~\eqref{eq:AHd1}. We gratefully acknowledge partial support by DFG through Grant No.\ Ru 1506/1.
\end{acknowledgments}

\appendix
\section{Properties of the functions $W_2$ and $U_2$}\label{app:W2U2}
In this appendix we give the explicit analytic expressions for the functions $W_2(\lambda)$ and $U_2(\lambda)$ in terms of special functions and briefly discuss some of their properties.

The function $W_2(\lambda)$ can be expressed exactly as
\begin{equation}\label{eq:W2lambda}
W_2(\lambda)=\frac{2 }{\pi  (\lambda+4)}\,{\ellipK}\bigg(\frac{4}{\lambda+4}\bigg).
\end{equation}
in terms of the complete elliptic integral of the first kind
\begin{align}\label{eq:Klambdadef}
\ellipK(\lambda)&=\int_0^{1}\frac{\rmd x}{\sqrt{(1-x^2)(1-\lambda^2\,x^2)}}\nonumber\\
&=\frac{\pi}{2}\,{}_2F_1\!\left(\frac{1}{2},\frac{1}{2};1;\lambda^2\right),
\end{align} 
where ${}_pF_q$ denotes the generalized hypergeometric function. 

Its antiderivative $U_2(\lambda)$ defined in Eq.~\eqref{eq:U2def} can be determined in closed form by integrating Eq.~\eqref{eq:Klambdadef}.  This gives  (cf.\ Eq.~(48) of \cite{HGS11}, and \cite{Gut10})
\begin{align}\label{eq:U2lambda}
U_2(\lambda)={}&{}\frac{-2}{(\lambda +4)^2}\,
{}_4F_3\bigg[1,1,\frac{3}{2},\frac{3}{2};2,2,2;\Big(\frac{4}{\lambda+4}\Big)^2\bigg]\nonumber\\
& +\ln (\lambda +4).
\end{align}

From the definition of $W_2(\lambda)$ in Eq.~\eqref{eq:Wddef} one can read off that  it can be analytically continued into the complex $\lambda$-plane except for the branch cut $[-8,0]$. To determine the form of  $W_2(\lambda)$ for small $\lambda$, we use the expansion 
\begin{equation}
\ellipK(k)=\sum_{j=0}^\infty\frac{\overline{k}^{2 j}}{(j!)^2}\left[\Big(\frac{1}{2}\Big)_j\right]^2 \left[\ln\left(1/\overline{k}\right)+d(j)\right]
\end{equation}
of $\ellipK(k)$ near the singularity at $k=1$  (see Eq.~19.12.1 in \cite{NIST:DLMF,Olver:2010:NHMF}),
where
\begin{equation}
\overline{k}=\sqrt{1-k^2}
\end{equation}
is the complementary modulus, while
\begin{equation}\label{eq:Pochh}
(a)_j=a(a-1)(a-2)\dotsm(a-j+1)
\end{equation}
denotes the Pochhammer symbol and $d(j)$ can be expressed in terms of the digamma function $\psi(z)$ as
\begin{equation}
d(j)=\psi(1+j)-\psi(\tfrac{1}{2}+j).
\end{equation}
This expansion converges for ${0<|\overline{k}|<1}$ \cite{NIST:DLMF,Olver:2010:NHMF}.
Upon substituting it into Eq.~\eqref{eq:W2lambda} and expanding in $\lambda$, we conclude that $W_2(\lambda)$ can be written as 
\begin{equation}\label{eq:W2anstruc}
W_2(\lambda)=-w_2(\lambda)\ln\lambda+R_2(\lambda),
\end{equation}
where $w_2(\lambda)$ and $R_2(\lambda)$ are both analytic for sufficiently small $|\lambda|$. The former function, which for $\lambda\in(-8,0)$ measures the discontinuity across the branch cut, is the spectral function introduced in Eq.~\eqref{eq:w2def}. It can be  computed  for real  values of $\lambda$ in a straightforward fashion either from Eq.~\eqref{eq:W2lambda} or the integral representation for $w_{2}$ implied by Eq.~\eqref{eq:Wddef}, namely
\begin{eqnarray}\label{eq:w2intrep}
w_2(\lambda\in\mathbb{R})&=&
\int_{-\pi}^{\pi}\frac{\rmd p_1}{2\pi}\int_{-\pi}^{\pi}\frac{\rmd p_2}{2\pi}\,\delta[\lambda+\nonumber\\ &&\strut +
4\sin^2\left(p_1/2\right)+4\sin^2\left(p_2/2\right)].
\end{eqnarray}
 One finds that
\begin{equation}\label{eq:w2} 
w_2(\lambda\in\mathbb{R})=\frac{1}{2\pi^2}\ellipK\left(\sqrt{-\lambda(\lambda+8)/16}\right)\,\Htheta(8+\lambda)\,\Htheta(-\lambda),
\end{equation}
where $\ellipK(k)$ is the elliptic function~\eqref{eq:Klambdadef}.

Integration of Eq.~\eqref{eq:W2anstruc} implies that $U_2(\lambda)$ can be written as
\begin{equation}
U_2(\lambda)=\frac{\lambda(1-\ln\lambda)}{4\pi}\left[1+\lambda\,A(\lambda)\right]+B(\lambda),
\end{equation}
where $A(\lambda)$ and $B(\lambda)$ are analytic for sufficiently small $|\lambda |$ and related to the functions $w_2(\lambda)$ and $R_2(\lambda)$ via
\begin{equation}\label{eq:w2A}
w_2(\lambda)=\frac{1}{4\pi}[1+2\lambda\,A(\lambda)+\lambda^2 A'(\lambda)]
\end{equation}
and
\begin{equation}
R_2(\lambda)=\frac{\lambda}{4\pi}[A(\lambda)+\lambda\,A'(\lambda)]+B'(\lambda).
\end{equation}
Their power series expansions
 \begin{equation}\label{eq:Aexp}
 A(\lambda)=-\frac{1}{16} +\frac{5 \lambda
   }{768}-\frac{7 \lambda
   ^2}{8192}+\frac{169 \lambda
   ^3}{1310720}
   +O(\lambda^4)
 \end{equation}
 and
 \begin{equation}\label{eq:Bexp}
B(\lambda)= \frac{4 \catalan}{\pi}+\frac{5 \lambda  \ln 2}{4 \pi }+\frac{\lambda ^2 (3-10 \ln 2)}{128\pi }+O(\lambda ^3)
\end{equation}
can be determined in a straightforward fashion. Here 
\begin{equation}
\catalan\equiv\sum_{k=0}^\infty\frac{(-1)^k}{(2k+1)^2}= 0.9159655942\ldots
\end{equation}
 is Catalan's constant. The value $4 \catalan/\pi$ of $B(0)=U_2(0)$ follows directly from Eq.~\eqref{eq:U2lambda} by evaluating the right-hand side of this equation at $\lambda=0$.

\section{Derivation of equations for the scattering phase $\eta(E)$}\label{app:scatphase}
We start by giving a direct proof of Eq.~\eqref{eq:etaE} for the derivative $\rmd{\eta_E}/\rmd{E}$ of the scattering phase.

Let $\mathcal{G}_\pm(\zm,\zm';E)=\langle\zm|(E-\mathcal{H}_{v_\pm})^{-1}|\zm^\prime\rangle$ be the Green's function introduced in Eq.~\eqref{eq:Gpm} for the cases $m=\pm 1$. It is defined for all complex values of $E$ that do not belong to the spectrum $\spek(\mathcal{H}_{v_\pm})=[\delta_{\pm 1,1},\infty)$ and has a branch cut along the positive real axis starting at $\delta_{\pm,1}$. For energies $E+\rmi 0=\delta_{\pm 1,1}+\sk^2+\rmi 0$ on the upper rim of the branch cut, it can be represented as
\begin{equation}\label{eq:GEimE0}
\mathcal{G}_\pm(\zm,\zm^\prime;E+\rmi 0)=-\frac{\varphi(\zm,\sk)f(\zm^\prime,\sk)}{F(\sk)}\;\; \text{ for }\zm^\prime \ge \zm,
\end{equation}
where $\varphi(\zm,\sk)$ and $f(\zm,\sk)$  denote respectively the regular and Jost solutions of the Schr\"odinger Eq.~\eqref{eq:SEhalfline}, while $F(\sk)$ is the Jost function introduced in Eqs.~\eqref{eq:regsolJost} and \eqref{eq:FJostk}. Note that Eq.~\eqref{eq:FWronskian} for $F(\sk)$ has been used.

To relate the numerator of the expression for the diagonal element of the Green's function given on the right-hand  side of Eq.~\eqref{eq:GEimE0} to a Wronskian, we consider these regular and Jost solutions for two different positive values $\sk$ and $\sk^\prime$.  We multiply the differential Eq.~\eqref{eq:SEhalfline} for $\varphi(\zm,\sk^\prime)$ by $f(\zm,\sk)$, that for  $f(\zm,\sk)$ by $\varphi(\zm,\sk^\prime)$, and then subtract the results from each other, obtaining
\begin{equation}\label{eq:Wronskpartialz}
(\sk^{\prime 2}-\sk^2)\varphi(\zm,\sk')f(\zm,\sk)=\partial_{\zm}W[\varphi(\zm,\sk^\prime),f(\zm,\sk)].
\end{equation}
This can be integrated to yield
\begin{equation}\label{eq:intvarphif}
\int_0^{\zm_u}\rmd{\zm}\,\varphi(\zm,\sk')f(\zm,\sk)=\frac{W[\varphi(\zm,\sk^\prime),f(\zm,\sk)]\big|_0^{\zm_u}}{\sk^{\prime 2}-\sk^2},
\end{equation}
where  $\zm_u>0$ is a finite upper integration limit whose limit  $\zm_u\to\infty$ we intend to take.

We first evaluate the Wronskian at the lower limit $0+$. In addition to the regular solution $\regsol(\zm,\sk)$, the Schr\"odinger Eq.~\eqref{eq:SEhalfline} has a linearly independent second solution, which we denote as $\varrho(\zm,\sk)$. Unlike the small-$\zm$ behavior~\eqref{eq:varphinormal} of $\regsol(\zm,\sk)$, the latter function varies as $\varrho(\zm,\sk)\asprop \zm^{1/2}\ln\zm$ for $\zm\to 0$. More precisely, insertion of the ansatz
\begin{equation}
\varrho(\zm,\sk)=\regsol(\zm,\sk)\ln\zm+\rho(\zm,\sk)
\end{equation}
into the differential Eq.~\eqref{eq:SEhalfline} shows that
\begin{equation}
[-\partial_{\zm}^2+v_\pm-\sk^2]\rho(\zm,\sk)=\bigg[\frac{2}{\zm}\partial_{\zm}-\frac{1}{\zm^2}\bigg]\regsol(\zm,\sk)=O(\zm^{-1/2}),
\end{equation}
which in turn implies that
\begin{equation}\label{eq:varrhosmallzm}
\varrho(\zm,\sk)\mathop{=}_{\zm\to 0}\regsol(\zm,\sk)\ln\zm+O(\zm^{3/2}).
\end{equation}

We can now expand the Jost solution as
\begin{equation}\label{eq:Jostsolsmallzm}
f(\zm,\sk)=A_\sk\, \regsol(\zm,\sk)+B_\sk\, \varrho(\zm,\sk)
\end{equation}
and compute the limit 
\begin{align}\label{eq:Wzmzero}
&\lim_{\zm\to 0+}W[\regsol(\zm,\sk^\prime),f(\zm,\sk)]=B_\sk \lim_{\zm\to 0+}W[\regsol(\zm,\sk^\prime),\varrho(\zm,\sk)]\nonumber\\ &=B_\sk \lim_{\zm\to 0+}W[\sqrt{\zm},\sqrt{\zm}\ln\zm]=B_\sk=-F(\sk),
\end{align}
where the last equality follows from the fact that the result is independent of $\sk^\prime$ and Eq.~\eqref{eq:FWronskian}.

To compute the Wronskian for large $\zm_u$, we use Eq.~\eqref{eq:regsolJost} for $\regsol(\zm,\sk^\prime)$, substitute the Jost solutions $f(\zm,\pm \sk^\prime)$ and $f(\zm,\sk)$ by their asymptotic forms~\eqref{eq:flargez}, insert the resulting value of the Wronskian at $\zm_u$  together with Eq.~\eqref{eq:Wzmzero} into the right-hand side of Eq.~\eqref{eq:intvarphif}, and then take the limit $\sk^\prime\to\sk$. We thus arrive at
\begin{eqnarray}
\int_0^{\zm_u}\rmd{\zm}\,\mathcal{G}_\pm(\zm,\zm;E+\rmi 0)&=&\frac{1}{4\sk^2}\bigg[ \frac{2\sk\,F'(\sk)+F(-\sk)\,\rme^{2\rmi \sk\zm_u}}{F(\sk)}\nonumber\\
&&\strut-1-2\rmi \sk \zm_u)\bigg].
\end{eqnarray}

The (oscillating) term $\propto \rme^{2\rmi\sk\zm_u}$ vanishes as $\zm_u\to\infty$ owing to the positive imaginary part of $\sk$ implied by the $\rmi 0$ of the energy. 
Proceeding to the limit  $\zm_u\to\infty$ yields:
\begin{equation}\label{eq:GreenF}
\int_0^{\infty}\rmd{\zm}\,\left[\mathcal{G}_\pm(\zm,\zm;E+\rmi 0)+\frac{\rmi}{2 \sk}\right]=
\frac{1}{2\sk}\frac{\rmd}{\rmd \sk}\bigg[\frac{\ln F(\sk)}{\sk^{1/2}} \bigg].
\end{equation}
For real positive $\sk$, the term  $\propto (\ln F)'(\sk)$ becomes $[\sigma'(\sk)-\rmi\eta'(\sk)]/(2\sk)$. Taking the  imaginary part of the above relation finally gives
\begin{equation}
\Im\int_0^\infty\rmd{\zm}\,\bigg[\mathcal{G}_\pm(\zm,\zm;E+\rmi 0)+\frac{1}{2\sk}\bigg]=-\frac{\eta'(\sk)}{2\sk},
\end{equation}
which is Eq.~\eqref{eq:etaE}.

We next turn to the proof of Eq.~\eqref{eq:deletav}. Let us consider changes of the potential $v_\pm(\zm)\to v_\pm(\zm)+\delta v(\zm)$ in Eq.~\eqref{eq:etaE}, where the variation $\delta v(\zm)$ has the properties (i)--(iv) specified in Eq.~\eqref{eq:delvprop}. For the implied linear variation $\delta\eta_E$ of the scattering phase we obtain from Eq.~\eqref{eq:deletav}:
\begin{align}\label{eq:delG}
\frac{\rmd\delta\eta_E}{\rmd E}&=-\int_0^\infty\rmd{\zm}\,\Im\,\delta\mathcal{G}(\zm,\zm;E+\rmi 0)\nonumber \\& =-\Im\Tr\!\left[\big(E+\rmi 0-\mathcal{H}_{v_\pm}\big)^{-1}\delta v\,\big(E+\rmi 0-\mathcal{H}_{v_\pm}\big)^{-1}\right]\nonumber\\ &=\frac{\rmd}{\rmd{E}}\Im\Tr\left[\delta v\,\big(E+\rmi 0-\mathcal{H}_{v_\pm}\big)^{-1}\right].
\end{align}
This can be integrated with respect to $E$ from $0$ to $E$ to obtain Eq.~\eqref{eq:deletav}. The integration constant $\delta\eta|_{E=0}$ is zero because condition (iv) of Eq.~\eqref{eq:delvprop} ensures  that the values $\eta_E|_{E=0}=0$ and $\eta_E|_{E=0+}=\pi/2$ for the cases $m=1$ and $m=-1$ are not changed by $\delta v(\zm)$.

Note that the above analysis can be extended to all complex $\sk$ in the upper half-plane. In particular, Eq.~\eqref{eq:GreenF} holds for all complex $\sk$ with $0<\arg \sk <\pi$.

\section{Semiclassical expansions}\label{app:semiclass}
\subsection{Half-line case}
Here we use semiclassical expansions to gain information about the asymptotic $\sk\to\infty$ behaviors of the Jost solution, the regular solution, and the Jost function of the Schr\"odinger problem on the half-line specified in Theorem~\ref{thm:traceform}.

Following a standard procedure (described on pages 35ff of \cite{NMPZ84}), we make  the ans{\"a}tze
\begin{eqnarray}\label{eq:semclansaetze}
f(\zm,\sk)&=&\rme^{\rmi \sk\zm+\rmi\mathcal{Y}(\zm,\sk)},\nonumber\\
\regsol(\zm,\sk)&=&\mathcal{A}(\sk)\,\rme^{\rmi \sk\zm+\rmi\mathcal{Y}(\zm,\sk)}+\text{cc}
\end{eqnarray}
for the Jost and regular solutions, where ``$\text{cc}$'' means complex conjugate. They lead to a Riccati equation for $\rmi\partial_{\zm}\mathcal{Y}(\zm,\sk)$, namely,
\begin{equation}\label{eq:Riccati}
2\sk\partial_\zm\mathcal{Y}(\zm,\sk)+[\partial_\zm\mathcal{Y}(\zm,\sk)[^2-\rmi\partial_\zm^2\mathcal{Y}(\zm,\sk)+v(\zm)=0.
\end{equation}
The function $\mathcal{Y}(\zm,\sk)$ has for $\sk\to \infty$ the asymptotic expansion
\begin{equation}\label{eq:Ylargekas}
\mathcal{Y}(\zm,\sk)=\sum_{j=1}^\infty\mathcal{Y}_j(\zm)\,\sk^{-j}.
\end{equation}
The coefficients $\mathcal{Y}_j(\zm)$ follow from the equations that result upon iteration of Eq.~\eqref{eq:Riccati}. For $\mathcal{Y}_1$ and $\mathcal{Y}_2$ one finds the differential equations $\mathcal{Y}_1'(\zm)=-v(\zm)/2$ and $\mathcal{Y}_2'(\zm)=\rmi\mathcal{Y}''_1(\zm)/2$, respectively. Solving these with the boundary conditions $\mathcal{Y}_1(\infty)=\mathcal{Y}_2(\infty)=0$ gives
\begin{eqnarray}
\mathcal{Y}_1(\zm)&=&\frac{1}{2}\int_{\zm}^\infty\rmd{s}\,v(s),\label{eq:Y1}\\
\mathcal{Y}_2(\zm)&=&-\frac{\rmi\,v(\zm)}{4}.\label{eq:Y2}
\end{eqnarray}

The large-$\sk$ expansion~\eqref{eq:Ylargekas} holds in the regime $\zm\gg 1/\sk$, but cannot be used in the narrow  boundary layer $0<\zm\lesssim 1/\sk$. For the latter, we will give an alternative large-$\sk$ expansion below. Before we turn to it, let us first investigate the behavior of $\mathcal{Y}_1(\zm)$ for $\zm\ll1$.

Owing to the singular near-boundary behavior of the potential [see Eqs.~\eqref{eq:traceform} and \eqref{eq:vsg}],  divergences linear in $1/\zm$ and logarithmic divergences  appear in the asymptotic form of $\mathcal{Y}_1(\zm)$ for $\zm\to 0$. To uncover these divergences, we add and subtract appropriate contributions to $\mathcal{Y}_1(\zm)$, rewriting it as 
\begin{eqnarray}\label{eq:Y1exp}
\mathcal{Y}_1(\zm)&=&\frac{\alpha}{2}-\frac{1}{2}\int_0^{\zm}\rmd{\zeta}\Big[v(\zeta)+\frac{1}{4\zeta^2}-\frac{v_{-1}}{\zeta}\Big]\nonumber\\ &&\strut -\frac{1}{8\zm}-\frac{v_{-1}}{2}\ln\zm\nonumber\\
&=&\frac{-1}{8\zm}-\frac{1}{2}v_{-1}\ln \zm+\frac{\alpha}{2}-\frac{v_0\zm}{2}+O(\zm^2)
\end{eqnarray}
with
\begin{equation}
\alpha=\int_0^\infty\rmd{z}\left[v(\zm)+\frac{1}{4\zm^2}-\frac{v_{-1}}{\zm}\,\Htheta(1-\zm)\right].
\end{equation}
We thus arrive at the following two-parameter asymptotic expansion
\begin{eqnarray}\label{eq:asJsol}
\ln f(\zm,\sk)&=&\rmi\sk\zm-\frac{\rmi }{2\sk}\left[
\frac{1}{4\zm}+v_{-1}\,\ln \zm-\alpha+v_0\zm+O(\zm^2)
\right]\nonumber\\
&&\strut-\frac{1}{4\sk^2}\bigg[
\frac{1}{4\zm^2}-\frac{v_{-1}}{z}-v_0+O(\zm)
\bigg]+O(\sk^{-3}), \nonumber\\
\end{eqnarray}
which holds for $\sk\to\infty$ with $1/\sk\ll\zm\ll 1$.

In order to develop an alternative large-$\sk$ expansion that holds in the near-surface regime, we express the regular solution in terms of the variable $\varsigma=\zm\sk$, defining
\begin{equation}
\tilde{\regsol}(\varsigma,\sk)=\sqrt{\sk}\,\regsol(\varsigma /\sk,\sk),
\end{equation}
and expand $v(\varsigma/\sk)$ in inverse powers of $\sk$. We thus arrive at the differential equation
\begin{equation}
\left[-\partial_{\varsigma}^2-\frac{1}{4\varsigma^2}-1+\frac{v_{-1}}{\varsigma\sk}+\frac{v_0}{\sk^2}+O(\sk^{-3})\right]\tilde{\regsol}(\varsigma,\sk)=0,
\end{equation}
which is to  be solved subject to the boundary condition
\begin{equation}
\tilde{\regsol}(\varsigma\to 0,\sk)=\sqrt{\varsigma}\,[1+O(\varsigma)].\nonumber\\
\end{equation}

By analogy with Eq.~\eqref{eq:regsoltildeplus}, the solution is again given by a Whittaker $M$-function; one has
\begin{equation}
\tilde{\regsol}(\varsigma,\sk)=\frac{\exp(\rmi\pi/4)}{\sqrt{2 r_{\sk}}}\,M_{-\rmi\mu_k,0}(-2\rmi r_{\sk}\varsigma)[1+O(\sk^{-3})],
\end{equation}
with
\begin{equation}
r_{\sk}=\sqrt{1-v_0/\sk^2}
\end{equation}
and
\begin{equation}
\mu_k=\frac{v_{-1}}{2\sk r_{\sk}}.
\end{equation}

Using the asymptotic expansion of the Whittaker $M$-function given in Eq.~(13.19.2) of \cite{NIST:DLMF} yields 
\begin{equation}
\tilde{\regsol}(\varsigma,\sk)=\frac{1}{\sqrt{2\pi}}\,\rme^{\tilde{W}(\varsigma,\sk)}+\text{cc},
\end{equation}
where $\tilde{W}(\varsigma,\sk)$ behaves as
\begin{eqnarray}\label{eq:largevarzetaexp}
\tilde{W}(\varsigma,\sk)&=&-\frac{\ln r_{\sk}}{2}-\ln \frac{\Gamma\big[\frac{1}{2}-\rmi \mu_k\big]}{\Gamma(1/2)}-\rmi\mu_k\ln\varsigma +\frac{\rmi\varsigma}{2}\nonumber\\ &&-\frac{\rmi\pi}{4}+\frac{\pi \mu_k}{2} -\frac{1}{4}\Big(1+\rmi \,2\mu_k\Big)^2\bigg[\frac{\rmi}{\varsigma}  \nonumber\\ &&\strut+\Big(1+\rmi \mu_k\Big)\frac{1}{\varsigma^2}+O(\varsigma^{-3})\bigg]
\end{eqnarray}
for $2r_{\sk}\varsigma\gg1$. The result can be expanded in inverse powers of $\sk$. This gives
\begin{eqnarray}
\tilde{W}(\varsigma,\sk)&=&\rmi\varsigma-\frac{\rmi \pi}{4}-\frac{\rmi}{8\varsigma}-\frac{1}{16\varsigma^2}+O(\varsigma^{-3})+\frac{v_{-1}}{\sk}\bigg[\frac{\pi}{4}\nonumber\\
&&\strut -\frac{\rmi\gammaE}{2}-\frac{\rmi\ln(8\varsigma)}{2}+\frac{1}{4\varsigma}-\frac{5\rmi}{32\varsigma^2}+O(\varsigma^{-3})\bigg]
\nonumber\\
&&\strut+\frac{1}{\sk^2}\bigg[-\frac{\rmi v_0}{2}\,\varsigma+\frac{v_0}{4}+\frac{\pi^2v_{-1}^2}{16}\nonumber\\
&&\strut +\frac{\rmi}{16\varsigma}\,(2v_{-1}^2-v_0)+\frac{1}{\varsigma^2}\Big(\frac{1}{8}v_{-1}^2-\frac{v_0}{16}\Big)\nonumber\\&&\strut +O(\varsigma^{-3})\bigg]+O(\sk^{-3}).
\end{eqnarray}
Returning to the variable $\zm=\varsigma \sk$ and the function $\regsol(\zm,\sk)$, we arrive at the result
\begin{equation}\label{eq:regsolitoW}
\regsol(\zm,\sk)=\frac{1}{\sqrt{\pi\sk/2}}\,\Re\, \rme^{W(\zm,\sk)}
\end{equation}
with
\begin{eqnarray}\label{eq:Wres}
W(\zm,\sk)&\equiv&\tilde{W}(\zm \sk,\sk)\nonumber\\ 
&=&\rmi\sk\zm-\frac{\rmi\pi}{4}+\frac{1}{\sk}\bigg\{\frac{-\rmi}{8\zm}-v_{-1}\bigg[\frac{\rmi\gammaE}{2}-\frac{\pi}{4} \nonumber\\
&&\strut +\frac{\rmi\ln(8\zm\sk)}{2}\bigg]-\frac{\rmi\zm v_0}{2}\bigg\}+\frac{1}{4\sk^2}\bigg\{\frac{-1}{4\zm^2}
+\frac{v_{-1}}{\zm}\nonumber\\ 
&&\strut +v_0+\frac{\pi^2 v_{-1}^2}{4}\bigg\}+O(\sk^{-3}\ln\sk).
\end{eqnarray}

In the regime $1/\sk\ll\zm\ll 1$ (i.e., $1\ll \varsigma\ll \sk$) both this expansion  as well as the one implied by Eqs.~\eqref{eq:asJsol} and \eqref{eq:semclansaetze} for the regular solution hold. Hence we can match them using Eq.~\eqref{eq:regsolJost}. Upon substituting our result~\eqref{eq:asJsol} for the Jost solution into this equation and equating it to Eq.~\eqref{eq:regsolitoW} with $W(\zm,\sk)$ given by Eq.~\eqref{eq:Wres}, one can determine the functions $F(\sk)$, $\sigma(\sk) $, and the phase shift $\eta(\sk)$ in a straightforward fashion. One obtains
\begin{eqnarray}\label{eq:Fplargek}
F(\sk)&=&\sqrt{\frac{2\sk}{\pi}}\exp\bigg\{-\frac{\rmi\pi}{4}+\frac{\rmi v_{-1}}{2\sk}[\gammaE+\ln(8\sk)]\nonumber\\ &&
 \strut +\frac{\rmi\alpha}{2\sk}+\frac{v_{-1}\pi}{4\sk}+\frac{v_{-1}^2\pi^2}{16\sk^2}+O(\sk^{-3})\bigg\},
\end{eqnarray}

\begin{equation}
\sigma(\sk)=\frac{1}{2}\ln(2\sk/\pi)+\frac{v_{-1}\pi}{4\sk}+\frac{v_{-1}^2\pi^2}{16\sk^2}+O(\sk^{-3}),
\end{equation}
\begin{equation}
\eta(\sk)=\frac{\pi}{4}-\frac{\alpha}{2\sk}-\frac{v_{-1}}{2\sk}[\gammaE+\ln(8\sk)]+O(\sk^{-3}),
\end{equation}
and
\begin{equation}\label{eq:emin2siglargek}
\rme^{-2\sigma(\sk)}=\frac{\pi}{2\sk}-\frac{\pi^2v_{-1}}{4\sk^2}+O(\sk^{-4}).
\end{equation}
In the ${m=-1}$~case, where $\sigma(\sk)=\sigma_{-}(\sk)$ and $v_{-1}=v_{-1}^-$, the last equation becomes Eq.~\eqref{eq:emin2sigminlargek}.

\subsection{The finite interval case}

An analogous calculation can be made for the singular Sturm-Liouville problem 
\[
-\partial_\zm^2 \varphi(\zm,\sk_\nu)+v(\zm)\varphi(\zm,\sk_\nu)=\sk_\nu^2 \,\varphi(\zm,\sk_\nu), \quad \nu\in\mathbb{N},
\]
on the  finite interval $0<\zm<|m|L$ with an even potential 
$v(\zm)=v(L|m|-\zm)$ that has a Laurent expansion of the form 
\[
v(\zm)=-\frac{1}{4\zm^2}+\frac{v_{-1}}{\zm}+v_0+\ldots, 
\]
near the boundary plane $\zm=0$. The usual boundary conditions for the eigenfunctions $\varphi(\zm,\sk_\nu)$
are implied:
\begin{eqnarray}\label{eq:BC1}
&&\varphi(\zm,\sk_\nu)=\\
&&\begin{cases}
\sqrt{\zm}\,[1+O(\zm)]& \text {for }  z\to 0,\\
(-1)^{\nu+1}\sqrt{L|m|-\zm}\,[1+O(L|m|-\zm)]& \text {for }  z\to L|m|.
\end{cases}\nonumber
\end{eqnarray}
While the result~\eqref{eq:Y2} for $\mathcal{Y}_2(\zm)$ remains valid except that $\zm$ now is restricted to $[0,L|m|]$,  the analog of Eqs.~\eqref{eq:Y1} and \eqref{eq:Y1exp} becomes
\begin{eqnarray}\label{eq:Y1L}
\mathcal{Y}_1(\zm)&=&\frac{1}{2}\int_{\zm}^{L|m|/2}\rmd{\zm}\,v(\zm)\nonumber\\
&=&\frac{L|m|}{4}\overline{v^{\text{ns}}}+\frac{1}{4L|m|}-\frac{1}{8\zm}-\frac{v_{-1}}{2}\ln\frac{2\zm}{L|m|}-\frac{v_0\zm}{2}\nonumber\\ &&\strut+O(\zm^2),
\end{eqnarray}
where 
\[
\overline{v^{\text{ns}}}=\frac{2}{L|m|}\int_0^{L|m|/2}\rmd{\zm}\,\left[v(\zm)+\frac{1}{4\zm^2}-\frac{v_{-1}}{\zm}\right].
\]
Instead of Eq.{\eqref{eq:asJsol}, we now have
\begin{eqnarray}\label{eq:asJsolN}
\ln f(\zm,\sk)&=&\rmi\sk\zm+ \frac{\rmi L|m|}{4\sk}\,\overline{v^{\text{ns}}}-\frac{\rmi }{2\sk}\bigg[
\frac{1}{4\zm}
\nonumber\\
&&\strut
-\frac{1}{2L|m|}+v_{-1}\,\ln\frac{ 2\zm}{L|m|}+v_0\zm+O(\zm^2)\bigg]
\nonumber\\&&\strut
-\frac{1}{4\sk^2}\bigg[
\frac{1}{4\zm^2}-\frac{v_{-1}}{\zm}-v_0+O(\zm)
\bigg]+O(\sk^{-3}) \nonumber\\
\end{eqnarray}
for the logarithm of the Jost solution. 

Proceeding as in the previous subsection, one finds that the result for the regular solution in the region $1/\sk\ll\zm\le L|m|/2$ can be written as
\begin{equation}
\regsol(\zm,\sk)=\sqrt{\frac{2}{\pi\sk}}\,\exp[P(\zm,\sk)] 
\cos[Q(\zm,\sk)]
\end{equation}
with
\begin{eqnarray}
P(\zm,\sk)&=&\frac{v(\zm)}{4\sk^2}+\frac{\pi v_{-1}}{4\sk}+\frac{\pi^2 v_{-1}^2}{16\sk^2}+O(\sk^{-3}),\\
Q(\zm,\sk)&=&\sk \zm-\frac{\pi}{4}-\frac{1}{4\sk}\left(\frac{1}{L|m|}+L|m|\,\overline{v^{\text{ns}}}\right)\nonumber\\&&\strut -\frac{v_{-1}}{2\sk}[\gammaE+\ln(4\sk L|m|)]\nonumber\\&&\strut +\frac{1}{2\sk}\int_{\zm}^{L|m|/2}\rmd{\zeta}\,v(\zeta)+O(\sk^{-3}\ln\sk).\quad
\end{eqnarray}
The discrete values of $\sk_\nu$ corresponding to the eigenenergies $\varepsilon_\nu=\sk_\nu^2$ follow from the Neumann and Dirichlet boundary conditions that hold at the midplane $\zm=L|m|/2$ for odd and even values of $\nu=1,2,\ldots,\infty$, respectively. Noting that $v'(L|m|/2)=0$, one concludes that the $\sk_\nu$ for all $\nu\in\mathbb{N}$ are given by the solutions to the equation
\begin{equation}\label{eq:knucond}
Q(L|m|/2,\sk_\nu)= \pi(\nu-1)/2.
\end{equation}
Solving this in an iterative manner yields
\begin{eqnarray}
\sk_\nu&=&\sko_\nu+\frac{1}{2\sko_\nu}\bigg\{\frac{1}{(L|m|)^2}+\overline{v^{\text{ns}}}+\frac{2v_{-1}}{L|m|}[\ln(\sko_\nu L|m|)\nonumber\\
&&\strut +\gammaE+\ln4]\bigg\} +O(\sko_\nu^{-3}\ln\sko_\nu),
\end{eqnarray}
where $\sko_\nu=\pi(\nu-1/2)/(L|m|)$.

\section{Asymptotic behavior of $\chi(\omega)$}\label{app:semclchi}

In order to derive the asymptotic large-$\omega$ behavior of $\chi(\omega)$ asserted in Eq.~\eqref{eq:chilargeom}, we use again the semiclassical expansion, proceeding along lines analogous to those followed in Appendix~\ref{app:semiclass}. Since we are now dealing with the case of negative energy $E$, we write 
$q\equiv\sqrt{-E}$ and make the ansatz 
\begin{equation}\label{eq:regsolleftansatz}
\regsol_l(z,q)=A_q\,\rme^{qz+\mathcal{Z}(z,q)}
\end{equation}
for the left regular solution, analogous to Eq.~\eqref{eq:semclansaetze}. The implied Riccati equation for $\partial_z\mathcal{Z}(z,q)$ becomes
\begin{equation}
-2q\partial_z\mathcal{Z}(z,q)-[\partial_z\mathcal{Z}(z,q)]^2-\partial_z^2\mathcal{Z}(z,q)+v(z)=0.
\end{equation}
We expand $\mathcal{Z}(z,q)$ in inverse powers of $q$ and  separate the contributions from even and odd powers,  writing
\begin{equation}\label{eq:Zdecoddeven}
\mathcal{Z}(z,q)=\mathcal{Z}_{\text{o}}(z,q)+\mathcal{Z}_{\text{e}}(z,q)
\end{equation}
with
\begin{eqnarray}\label{eq:Zexp}
\mathcal{Z}_{\text{o}}(z,q)&=&\sum_{j=0}^\infty b_{2j+1}(z)\,q^{-(2j+1)},\nonumber\\
\mathcal{Z}_{\text{e}}(z,q)&=&\sum_{j=1}^\infty b_{2j}(z)\,q^{-2j}.
\end{eqnarray}
For the two lowest-order coefficients one finds
\begin{equation} \label{eq:b1ofz}
b_1(z)=-\frac{1}{2}\int_z^{L/2}\rmd{z}'\,v(z')
\end{equation}
and
\begin{equation}
 \label{eq:b2ofz}
b_2(z)=-v(z)/4.
\end{equation}
The function $\mathcal{Z}_{\text{e}}(z,q)$ must satisfy the equation
\begin{equation}
2\partial_z\mathcal{Z}_{\text{e}}(z,q)[q+\partial_z\mathcal{Z}_{\text{o}}(z,q)]=-\partial_z^2\mathcal{Z}_{\text{o}}(z,q).
\end{equation}
This can be integrated in a straightforward fashion to obtain 
\begin{equation}\label{eq:ZeZo}
\mathcal{Z}_{\text{e}}(z,q)=-\frac{1}{2}{\ln}\big[1+\partial_z\mathcal{Z}_{\text{o}}(z,q)/q\big],
\end{equation}
where the integration constant has been chosen such that  the consistency of the result with the expansions~\eqref{eq:Zexp} and Eqs.~\eqref{eq:b1ofz} and \eqref{eq:b2ofz} is ensured.

Inserting Eq.~\eqref{eq:ZeZo} into Eq.~\eqref{eq:Zdecoddeven} and the resulting expression in turn into the ansatz~\eqref{eq:regsolleftansatz} then gives
\begin{equation}\label{eq:regsollsemcl}
\regsol_l(z,q)=A_q\,q^{1/2}[q+\partial_z\mathcal{Z}_{\text{o}}(z,q)]^{-1/2}\,\rme^{qz+\mathcal{Z}_{\text{o}}(z,q)}.
\end{equation}
To determine the consequences for the functions $\Delta({E=-q^2})$ and $\chi(q^2)$, we use Eq.~\eqref{eq:Deltaregsol}, taking into account that 
$\mathcal{Z}_1(L/2,q)$ and $\partial^2_z\mathcal{Z}_1(z,q)|_{z=L/2}$ vanish since $\mathcal{Z}_1(z,q)$ is an odd function with respect to reflections about the midpoint $L/2$, i.e., $\mathcal{Z}_1(z,q)=-\mathcal{Z}_1(L-z,q)$. We thus arrive at
\begin{eqnarray}
\Delta(E)&=&2qA_q^2\,\rme^{qL}|_{E=-q^2},\label{eq:DeltaEAq}\\
\chi(q^2)&=&-qL-\ln[2\pi qA_q^2].\label{eq:chiomegaAq}
\end{eqnarray}
Hence the problem is reduced to the calculation of the large-$q$ asymptotics of the amplitude $A_q$, which is fixed by the boundary condition~\eqref{eq:bclr} at $z=0$ for the left regular solution $\regsol_l$.

As before, the above semiclassical expansion cannot be used in the near-boundary region with $0<z\lesssim 1/q$ because of the singularity of the potential at $z=0$.  By analogy with the analysis in Appendix~\ref{app:semiclass}, the remedy is to go over to the rescaled variable $\varsigma=zq$, to introduce
\begin{equation}
\tilde{\regsol}_l(\varsigma,q)=\sqrt{q}\,\regsol_l(z,q),
\end{equation}
expand $v(\varsigma/q)$ in inverse powers of $q$, and solve the resulting differential equation
\begin{equation}
\bigg[-\partial_\varsigma^2-\frac{1}{4\varsigma^2}+1+\frac{4m}{\pi^2 q\varsigma}+\frac{v_0\,m^2}{q^2}+O\big(q^{-3}\big)\bigg]\tilde{\regsol}_l(\varsigma,q)=0
\end{equation}
subject to the boundary condition
\begin{equation}
\tilde{\regsol}_l(\varsigma,q)\mathop{=}\limits_{\varsigma\to 0}\sqrt{\varsigma}[1+O(\varsigma)].
\end{equation}
The solution is
\begin{equation}
\tilde{\regsol}_l(\varsigma,q)=2^{-1/2}\,M_{-2m\pi^{-2}\kappa^{-1}_q,0}(2\varsigma\kappa_q/q)\big[1+O(q^{-3})\big]
\end{equation}
with
\begin{equation}
 \label{eq:kappaq}
\kappa_q=\sqrt{q^2+v_0\,m^2}.
\end{equation}
Using the asymptotic expansion (Eq.~13.9.1 of \cite{NIST:DLMF}) 
\begin{align}
M_{\kappa,0}(x)\mathop{\aseq}\limits_{x\to\infty}\frac{\rme^{-x}x^{-\kappa}}{\Gamma(1/2-\kappa)}\sum_{j=0}^\infty\frac{(1/2-\kappa)_j(1/2+\kappa)_j}{j!\,x^j}
\end{align}
of the Whittaker function, where $(a)_j$ is the Pochhammer symbol introduced in Eq.~\eqref{eq:Pochh}, and expanding subsequently $\kappa=-2m\pi^{-2}/\kappa_q$ in powers of $1/q$, one is led to the following two-parameter expansion for large $\varsigma$ and $q$:
\begin{align}\label{eq:varphitlres}
\ln[\tilde{\regsol}_l(\varsigma,q)]=&-\frac{1}{2}\ln(2\pi)+\varsigma+\frac{1}{8\varsigma}+\frac{1}{16\varsigma^2}+O\big(\varsigma^{-3}\big)\nonumber \\
&\strut +\frac{2m}{\pi^2 q}\bigg[\ln(2\varsigma)+\gammaE+\ln 4-\frac{1}{2\varsigma}+O\big(\varsigma^{-2}\big)\bigg]\nonumber\\ &\strut +\frac{1}{q^2}\bigg[\frac{v_0\,m^2(2\varsigma-1)}{4}-\frac{m^2}{\pi^2}+O(1/\varsigma)\bigg]\nonumber\\&\strut +O(q^{-3}).
\end{align}

In the crossover regime $1/q\ll z\ll 1$ both Eq.~\eqref{eq:varphitlres} and the original semiclassical expansion are valid. In this regime, the coefficient $b_1(z)$ is given by
\begin{eqnarray}
b_1(z)&=&\frac{1}{8z}-\frac{1}{4L}+\frac{2m}{\pi^2}\ln(2z/L)-\frac{1}{2}\int_0^{L/2}\rmd{z}'\,v^{\text{ns}}(z') \nonumber\\
&&\strut +\frac{1}{2}\int_0^z\rmd{z}'\,v^{\text{ns}}(z') \nonumber\\
&=&\frac{v_0\,m^2z}{2}-\frac{1}{2}\int_0^{L/2}\rmd{z}'\,v^{\text{ns}}(z')+\frac{1}{8z}-\frac{1}{4L}\nonumber\\
&&\strut +\frac{2m}{\pi^2}\ln(2z/L)+O(z^2).
\end{eqnarray}
Upon substituting this small-$z$ expansion along with that of $b_2(z)$ into Eq.~\eqref{eq:varphitlres}, we arrive at the expansion
\begin{eqnarray}\label{eq:varphivarphit}
\lefteqn{\ln[\regsol_l(\varsigma/q,q)]= \ln\left[q^{1/2}\tilde{\regsol}_l(\varsigma,q)\right]=}&&\nonumber\\ &=&
\ln(A_q)+\bigg[\varsigma+\frac{1}{8\varsigma}+\frac{1}{16\varsigma^2}+O(\varsigma^{-3})\bigg]\nonumber\\ &&\strut +\frac{1}{2q}\bigg[-\frac{1}{2L}-\frac{2m}{\pi^2\varsigma}+\frac{4m}{\pi^2}\ln\frac{2\varsigma}{qL}-\int_0^{L/2}\rmd{z'}v^{\text{ns}}(z')\nonumber\\ &&\strut +O(\varsigma^{-2})\bigg]+\frac{1}{q^2}\bigg[\frac{v_0\,m^2(2\zeta-1)}{4}+O(\varsigma^{-1})\bigg]\nonumber\\ &&\strut +O(q^{-3})
\end{eqnarray}
in the crossover regime. This can be matched with the  two-parameter expansion~\eqref{eq:varphitlres} to conclude that
\begin{align}
\ln A_q=&-\frac{1}{2}\ln(2\pi q)+\frac{1}{4q}[L\mathcal{R}([v_*];L,m)+\frac{8m}{\pi^2}\ln(qL)]\nonumber\\ &\strut -\frac{m^2}{\pi^2q^2}+O(q^{-3}),
\end{align}
where $\mathcal{R}([v_*];L,m)$ is given by Eq.~\eqref{eq:Rdef}. Substitution of this result into Eq.~\eqref{eq:chiomegaAq} finally yields the asymptotic form stated in Eq.~\eqref{eq:chilargeom}.

\section{Calculation of $\fs(m;\mu)$}\label{app:fs}

In this appendix we compute the surface free energy $\fs(m;\mu)$ from the free-energy functional~\eqref{eq:fmudef}. We begin by computing the contribution $\mathcal{F}^{(1)}([v_*];L,m)$ defined in Eq.~\eqref{eq:fmu1def}. Upon adding and subtracting to the nonsingular part $v_*^{\text{ns}}$ of the self-consistent potential $v_*$ the bulk potential $v_{\mathrm{b}}(m)=m^2\Htheta(m)$ and inserting $u_{0*}(L,m)=v_0\,m^2$, we arrive at
\begin{align}
\mathcal{F}^{(1)}([v_*];L,m)=&\frac{mL}{8\pi}\,v_{\mathrm{b}}(m)+\frac{m^2}{\pi^3}(\gammaE+\ln 4)\nonumber\\ &\strut +\frac{mL}{8\pi}\,\overline{v_*^{\text{ns}}-v_{\mathrm{b}}}+\frac{v_0\,m^2}{8\pi}+O(1/L).
\end{align}

The integral $L\overline{v_*^{\text{ns}}-v_{\mathrm{b}}}$ can be decomposed as
\begin{align}
L\,\overline{v_*^{\text{ns}}-v_{\mathrm{b}}}=&2|m|\alpha_\pm -2|m|\int_1^{|m|L/2}\rmd{\zm}\frac{4\sgn(m)}{\pi^2\zm}\nonumber\\ &\strut
-2|m|\int_{|m|L/2}^\infty \rmd{\zm}\bigg[v(\zm;\infty,\sgn(m))\nonumber\\ &\strut
-v_{\text{b}}(\sgn(m))+\frac{1}{4\zm^2}\bigg]+O(1/L).
\end{align}

In the second integral, we can replace $v(\zm;\infty,\sgn(m))-v_{\text{b}}(\sgn m)$ by the limiting forms $\asprop \rme^{-2\zm}$ and $-(2\zm^3)^{-1}$ for $m>0$ and $m<0$, respectively,  to see that it yields contributions $\asprop \rme^{-mL/2}$ and $\asprop L^{-2}$. Dropping these, we arrive at
\begin{equation}\label{eq:RstarooflogL}
L\,\overline{v_*^{\text{ns}}-v_{\mathrm{b}}}=2|m|[\alpha_\pm\mp4\pi^{-2}\ln(|m|L/2)]+O(1/L),
\end{equation}
 which upon addition of the remaining contributions to $\mathcal{F}^{(1)}([v_*];L,m)$ and insertion of our results for  $\alpha_\pm$ given in Eqs.~\eqref{eq:alphaplusres}, and \eqref{eq:alphaminres} yields
\begin{align}\label{eq:F1largeL}
\mathcal{F}^{(1)}([v_*];L,m)=& \theta(m)\,m^2\bigg[L\frac{m}{8\pi}+4\Delta A_0^{(\mathrm{s})}\bigg]+\frac{v_0\,m^2}{8\pi}\nonumber\\ &\strut
+\frac{m^2}{\pi^3}\left[1-\ln\frac{2L|m|}{\pi}\right]
+O(1/L).
\end{align}

Turning to the calculation of $\mathcal{F}_\mu^{(2)}([v_*];L,m)$, we first consider the case $m<0$. In this case 
one has $v_{\text{b}}=0$, together with the relation
\begin{equation}\label{eq:LR}
L\mathcal{R}_*(L,m)=\frac{8m}{\pi^2}\left(
1-\ln \frac{2|m| L}{\pi}
\right)+O(1/L),
\end{equation}
following directly from the above analysis. After substitution of \eqref{eq:LR} into   \eqref {eq:fmu2def} and transformation of the integration variable $\omega=m^2 y$,
one obtains
\begin{eqnarray}\label{eq:F2largeL}
\mathcal{F}_\mu^{(2)}([v_*];L,m)=\frac{-m^2}{8\pi}\int_0^\infty\rmd{y}\,I(y;mL)\\
-\frac{m^2}{2\pi^3}\ln\frac{\mu}{|m|}+O(1/L),\nonumber
\end{eqnarray}
where
\begin{eqnarray}\label{eq:Iyx}
I(y;x)=\chi (y;L=|x|,m=-1)+|x|\sqrt{y}\\
-\frac{4}{\pi^2 \sqrt{y}}\left(1+\ln \frac{\pi\sqrt{y}}{2}\right)
-\frac{2}{\pi^2(y+1)}.\nonumber
\end{eqnarray}
To obtain the  contribution of $\chi (y;|x|,-1)=\chi(\omega;L,m)$ to $I(y;x)$, we have set $L=\infty$ in the regular solution $\regsol(z|m|,k/|m|;L,-1)$ and used its relation
\begin{equation}
\regsol(\zm,\sk;\infty,-1)\mathop{=}\limits_{\zm\gg 1}\frac{F_-({k=\rmi\sqrt{y}})}{2\sqrt{y}}\,\rme^{\zm\sqrt{y}}[1+O(1/\zm)]
\end{equation}
to the Jost function for large $\zm$, obtaining
 \begin{eqnarray}\label{eq:chilargexm}
\chi(y;|x|,-1)
\mathop{=}\limits_{x\to-\infty}-|x|\sqrt{y}+\ln(2\sqrt{y}/\pi)\nonumber\\ 
\strut-2\ln F_-(\rmi\sqrt{y})+o(|x|^0).
\end{eqnarray}

To compute the integral over $I(y;x)$ in \eqref{eq:F2largeL}, one can integrate by parts, using the fact that the boundary terms vanish since $I(y;x)=O(y^{-3/2}\ln y)$ for $y\to\infty$. For the required derivative of the analytically continued logarithm of the Jost function one finds
\begin{equation}\label{eq:dlnFres}
\frac{\rmd\ln F(\rmi\sqrt{y})}{\rmd y}=\frac{1}{2y}+\frac{1}{2(4+\pi^2y)}\bigg[\frac{\ln(\pi^2y/4)}{\sqrt{y}}-\frac{\pi^2}{2}\bigg].
\end{equation}

Substituting this, one can compute the integral in a straightforward manner using {\sc Mathematica} \cite{Mathematica10}. One obtains
\begin{eqnarray}\label{eq:Qpartintegral}
\int_0^\infty\rmd{y}\,I(y;x)&=&-\int_0^\infty\rmd{y}\,y\,\frac{\partial}{\partial y}I(y;x)\nonumber\\
&=&2\pi^{-2}[1-2\ln(2/\pi)],
\end{eqnarray}
which yields
\begin{equation}\label{eq:F2}
\mathcal{F}_\mu^{(2)}([v_*];L,m)=\frac{m^2}{4\pi^3}\left(-1+2\ln\frac{2 |m|}{\pi\mu} \right)+O(1/L).
\end{equation}

The right-hand sides of Eqs.~\eqref{eq:F1largeL} and \eqref{eq:F2} can now be added and inserted into Eq.~\eqref{eq:fmusumdef} to arrive at
\begin{eqnarray}\nonumber
f_\mu(L,m)&\equiv&\mathcal{F}_\mu([v_*];L,m)=\frac{3m^2}{4\pi^3}-\frac{m^2}{2\pi^3}\ln\frac{2 |m|L}{\pi}\\
&&\strut +\frac{m^2 v_0}{8\pi}-\frac{m^2}{2\pi^3}\ln (\mu L)+
O(1/L).\label{eq:Fmin}
\end{eqnarray}
As a consequence of Eq.~\eqref{eq:v0mincalc}, the result~\eqref{eq:Fmin} is identical with Eq.~\eqref{eq:fmutoooflnL} when $m<0$.

At the end of  Subsection~\ref{sec:freeenergyfunc}, we had announced a check of the asymptotic behavior of the scaling function $Y(x)$ for $x\to-\infty$ by an independent calculation. The foregoing result provides such a check. It enables us to prove that the coefficient $C_1$ has the analytic value given in Eq.~\eqref{eq:Cex}. In fact, 
straightforward differentiation of Eq.~\eqref{eq:Fmin} gives
\begin{eqnarray}\nonumber
\lefteqn{\frac{\partial f_\mu(L,m)}{\partial m}}&&\\ &=&\frac{m}{\pi^3}-\frac{m}{\pi^3}\ln\frac{2 |m|L}{\pi}+\frac{m v_0}{4\pi}
-\frac{m\ln(\mu L)}{\pi^3}+O(1/L)\nonumber\\
&=&\frac{L\mathcal{R}([v_*];L,m)}{8\pi}+\frac{m v_0}{4\pi}-\frac{m\ln(\mu L)}{\pi^3}+O(1/L),\nonumber\\\label{eq:dfdm}
\end{eqnarray}
where the second equality follows via Eq.~\eqref{eq:LR}. Setting $\mu=1$ in the latter and comparing it with Eq.~\eqref{eq:Yprimerel} yields the analytic value of $C_1$ given in Eq.~\eqref{eq:Cex}.

To obtain the result \eqref{eq:fmutoooflnL} for $m>0$, one must simply add the bulk contribution $\propto m^3\,\Htheta(m)$ along with the term $m^2\Htheta(m)$ involving the universal amplitude difference. Note that this {$m>0$}~result means that the scaling function $Y(x)$ for $x>0$ can be written as
\begin{eqnarray}\label{eq:Ygtrx}
Y(x>0)&=&\frac{x^3}{24\pi}+\bigg[\frac{7\zeta(3)}{\pi^5}+\frac{3-2\ln(2x/\pi)}{4\pi^3}+\frac{J_+}{8\pi}\bigg]x^2\nonumber\\&&\strut +\Theta(x),
\end{eqnarray}
where $\Theta(x)$ vanishes $\asprop \rme^{-2x}$ for $x\to\infty$ [cf.\ Eq.~\eqref{eq:Thetaaslargex}].

We checked that the two terms in the first line of Eq.~\eqref{eq:Ygtrx} can be derived directly from the free-energy functional~\eqref{eq:fmu1def} by showing that $Y(x)$ agrees asymptotically for $x\to+\infty$ with these terms to $O(x)$. In this calculation, we used the representation of the logarithm $\ln F_+(\sk)$ of the Jost function given in Eq.~\eqref{eq:logFplus}. As a by-product, we obtained the following alternative representation of the integral defined in Eq.~\eqref{eq:Jdef}:
\begin{equation}
J_+=-\frac{4\ln(\pi/2)}{\pi^2}+\int_0^\infty\frac{\rmd{t}}{\pi}\,\frac{1}{1+t}\bigg[\frac{1}{\arctan\sqrt{t}}-\frac{2}{\pi}\bigg].
\end{equation}
We also verified by numerical integrations  that the integrals on the right-hand side of this equation and in the original definition~\eqref{eq:Jdef} yield the same values up to an error $\lesssim 10^{-22}$, even though we did not work out an analytical proof of their equality. 

\section{Derivation of the relation~\eqref{eq:Yprimerel}}\label{app:derivrel}

To prove the relation~\eqref{eq:Yprimerel}} from which Eq.~\eqref{eq:tracerelcont} follows by means of Eq.~\eqref{eq:fscalform}, we set $\mu=L=1$ and $m=x$, and consider the derivative $\rmd/\rmd{x}$ of the free-energy functional 
$\mathcal{F}_1([v_*];1,x)$, 
\begin{eqnarray}\label{eq:F1x}
\mathcal{F}_1([v_*];1,x)=\frac{x\,\mathcal{R}([v_*];1,x)}{8\pi}+\frac{u_0}{8\pi}\\\nonumber
-\int_0^\infty \frac{\rmd \omega}{8\pi}\Big[
\chi_*(\omega;1,x)+\sqrt{\omega}+\frac{2 x \ln \omega}{\pi^2\sqrt{\omega}}\\\nonumber
+\frac{\mathcal{R}([v_*];1,x)}{2\sqrt{\omega}}-\frac{2 x^2}{\pi^2(1+\omega)}
\Big].
\end{eqnarray}

 The latter has an explicit dependence on $x$ and  an implicit one through the $x$-dependence of the maximizing potential $v_*(z)\equiv v_*(z;1,x)$. Hence
\begin{eqnarray}\label{eq:dFdx}
\frac{\rmd}{\rmd{x}}\mathcal{F}_1([v_*];1,x)&=& 2\int_0^{1/2}\rmd{z}\,\frac{\delta\mathcal{F}_1([v_*];1,x)}{\delta v_*(z)}\frac{\partial v_*(z;1,x)}{\partial x}\nonumber\\ && \strut+\frac{\partial \mathcal{F}_1([v_*];1,x)}{\partial x}.
\end{eqnarray}
Since $[\partial v_*^{\text{ns}}/\partial x] \rmd{x}$ is a variation $\delta v$ that does not modify the singular part of $v_*$  and we have shown that $\delta\mathcal{F}_1([v_*];1,x)=0$ for such variations, the nonsingular part $v_*^{\text{ns}}$ of $v_*$ does not contribute to the first term on the right-hand side. The only contribution to this term originates from the $x$-dependent singular part  $4x\pi^{-2}z^{-1}$ of $v_*$. Moreover, we can decompose this part into a sum of a  component  that is singular at $z=0$, namely $4x\pi^{-2}z^{-1}\Htheta(z_0-z)$, and the nonsingular one $4x\pi^{-2}z^{-1}\Htheta(z-z_0)$, where $z_0\in(0,1/2)$ can be taken to be arbitrarily small. Since the latter corresponds to the derivative of a nonsingular variation $\delta v$, it also does not contribute to the first term on the right-hand side of Eq.~\eqref{eq:dFdx}. This decompostion of the term $\propto z^{-1}$ of $v$ must be consistently made also in other $v$-dependent terms such as $\mathcal{R}([v_*];1,x)$ to determine those of their contributions to  $\rmd \mathcal{F}_1([v_*];1,x)/\rmd x$ that must be kept. In the case of $\mathcal{R}([v_*];1,x)$, we may replace $v_*$ by $4x\pi^{-2}z^{-1}\theta(z_0-z)$. In conjunction with the subtracted $1/z$ term in $v_*^{\text{ns}}$, a contribution $4x\pi^{-2}z^{-1}\Htheta(z-z_0)$ results whose implied contribution to  $\rmd\mathcal{R}([v_*];1,x)/\rmd x$ must be retained. One thus obtains
\begin{align}
 \label{eq:dF1dx}
\frac{\rmd}{\rmd x}\mathcal{F}_1([v_*];1,x)=
\frac{\mathcal{R}_*}{8\pi}+\frac{x(\gammaE+\ln4)}{\pi^3}\nonumber\\ 
-\frac{ x}{\pi^3}\int_{z_0}^{1/2}\frac{\rmd z}{z}-\frac{1}{8\pi}\lim_{\Omega\to\infty}I_\Omega,
\end{align}
where $\mathcal{R}_*\equiv \mathcal{R}([v_*];1,x)$ while  $I_\Omega$ denotes the integral
\begin{align}\label{eq:IOmega}
I_\Omega=&\int_0^\Omega\rmd{\omega}\bigg[\int_0^{z_0}\rmd{z}\bigg(\frac{\delta\chi_*}{\delta v_*(z)}\,\frac{8}{\pi^2 z}\bigg)\nonumber\\ &\strut +\frac{4}{\pi^2\sqrt{\omega}}\bigg(\gammaE+\ln 4-\int_{z_0}^{1/2}\frac{\rmd z}{z}\bigg)+\frac{2\ln\omega}{\pi^2\sqrt{\omega}}\nonumber\\ &\strut -\frac{4x}{\pi^2(1+\omega)}\bigg]
\end{align}
with $\chi_*\equiv\chi([v_*];\omega)$.
From Eqs.~\eqref{eq:DeltaElnG} and \eqref{eq:vsymcond} one sees that the functional derivative $\delta\chi_*/\delta v_*(z)$ yields the Green's function $\mathcal{G}(z,z;-\omega)$, i.e.,  
\begin{equation}
\frac{\delta\chi_*}{\delta v_*(z)}=\mathcal{G}(z,z;-\omega;1,x)=-\langle z|(\omega+\mathcal{H}_{v_*})^{-1}|z\rangle.
\end{equation}
Since $z_0$ is arbitrarily small, we need the behavior of this function asymptotically close to the boundary plane $z=0$. For such values of $z$, we can safely replace the Green's function $\mathcal{G}(z,z;-\omega;1,x)$
by its semi-infinite ($L=\infty$) analog $\mathcal{G}(z,z;-\omega;\infty,x)$
and omit $o(z)$ contributions to $v_*$. 
This gives
\begin{eqnarray}\label{eq:delchidelv}
\frac{\delta\chi_*}{\delta v_*(z)}\mathop{\aseq}\limits_{z\ll 1}-\langle z|(\omega+\mathcal{H}_{v_*^{\text{sg}}+v_0x^2})^{-1}|z\rangle_{{L=\infty}}\nonumber\\
=-\varphi(z,p,\varkappa)\,\mu(z,p,\varkappa),
\end{eqnarray}
where $p=\sqrt{\omega+v_0x^2}$, $\varkappa=4 x/\pi^2 $, and 
$\varphi(z,p,\varkappa)$ and $\mu(z,p,\varkappa)$ denote the two linearly independent solutions 
of the  differential equation 
\begin{equation}
-\partial_z^2 \psi(z) +\left(-\frac{1}{4 z^2}+\frac{\varkappa}{ z}+p^2
\right)\psi(z)=0
\end{equation}
on the half-line $0<z<\infty$ that are fixed by the boundary conditions
\begin{eqnarray}
\varphi(z\to0+,p,\varkappa)&=&\sqrt{z}\,[1+O(z)],\nonumber\\
\mu(z\to 0+,p,\varkappa)&=&-\sqrt{z}\,[\ln z+O(1)],\nonumber\\
\mu(z\to\infty,p,\varkappa)&\to&0.
\end{eqnarray}
Explicitly, these two solutions read 
\begin{eqnarray}
\varphi(z,p,\varkappa)&=&\frac{1}{\sqrt{2 p}}\, M_{- \varkappa/(2p),0}(2 p z),\\
\mu(z,p,\varkappa)&=&\frac{\Gamma\left(\frac{1}{2}+\frac{\varkappa}{2 p}\right)}{\sqrt{2 p}}\, W_{- \varkappa/(2p),0}(2 p z),
\end{eqnarray}
where $M_{\kappa,\nu}(x)$ and $W_{\kappa,\nu}(x)$ denote Whittaker functions 
\cite{AS72,NIST:DLMF,Olver:2010:NHMF}.
We expand these solutions at fixed $pz$ to first order in $\varkappa$, 
\begin{align}\nonumber
\varphi(z,p,\varkappa )=\sqrt{z}\,\left[ I_0(p z)+\varkappa  z \,h_1 (p z)+O(\varkappa^2 z^2) \right],\\
\mu(z,p,\varkappa )=\sqrt{z}\,\left[ K_0(p z)+\varkappa  z \,h_2 (p z)+O(\varkappa^2 z^2) \right],\label{eq:h1h2}
\end{align}
and introduce the two functions 
\begin{equation}
h_1(z)=-(2z)^{-3/2}\partial_\kappa M_{\kappa,0}(2z)\big|_{\kappa=0},
\end{equation}
and 
\begin{eqnarray}
h_2(z)&=&-K_0(z)\frac{\gammaE+\ln 4}{2z}\nonumber\\&&\strut -(2z)^{-3/2}\sqrt{\pi}\, \partial_\kappa W_{\kappa,0}(2z)\Big|_{\kappa=0}.
\end{eqnarray}
Substitution of  the expansions \eqref{eq:h1h2}  into \eqref{eq:delchidelv} then yields
  \begin{eqnarray}\label{eq:delchi1}
\frac{\delta\chi_*}{\delta v_*(z)}&\mathop{\aseq}\limits_{z\ll 1}&-zI_0(pz)K_0(pz)-\frac{4x}{\pi^2}
z^2\big[K_0(pz)\,h_1(pz) \nonumber \\
&&\strut+I_0(pz)\,h_2(pz)\big]+\ldots.
\end{eqnarray}

We can now compute the contribution of the first term on the right-hand side  of Eq.~\eqref{eq:delchi1}  to the first term on the right-hand side of Eq.~\eqref{eq:IOmega} for $\Omega\to\infty$ and small $z_0>0$:
\begin{eqnarray}
\lefteqn{-\frac{8}{\pi^2}\int_0^\Omega\rmd{\omega}\int_0^{z_0}\frac{\rmd{z}}{z}z\,I_0(pz)K_0(pz)}&&\nonumber\\ &=&\frac{8}{\pi^2}\Omega^{1/2}[1-\gammaE-\ln(8 z_0)-\frac{1}{2}\ln\Omega]+O(\Omega^{-1/2})\nonumber
\\ &=&
-\int_0^\Omega\rmd{\omega}\bigg[\frac{4}{\pi^2\sqrt{\omega}}
\bigg(\gammaE+\ln 4-\int_{z_0}^{1/2}\frac{\rmd z}{z}\bigg)+\frac{2\ln\omega}{\pi^2\sqrt{\omega}}\bigg]\nonumber\\ &&\strut +O(\Omega^{-1/2}).
\end{eqnarray}
The result shows that this contribution is canceled by the sum of terms in the second line  of Eq.~\eqref{eq:IOmega}. It follows that
\begin{eqnarray}
\lim_{\Omega\to\infty}I_\Omega&=&\frac{-4x}{\pi^2}\lim_{\Omega\to \infty}\int_0^\Omega\rmd{\omega}\bigg\{\frac{1}{1+\omega}\nonumber\\ 
&&\strut +\frac{8}{\pi^2} \int_0^{z_0} \rmd z \,z[K_0(pz)h_1(pz)+I_0(pz)h_2(pz)]\bigg\}\nonumber\\
&=&\frac{-4x}{\pi^2}\lim_{\Omega\to \infty}\bigg\{\int_0^{z_0\sqrt{\Omega}}\rmd{u}\,\frac{16u}{\pi^2}\big[K_0(u)h_1(u)\nonumber\\
&&\strut+I_0(u)h_2(u)\big]\ln\big(z_0\sqrt{\Omega}/u\big)+\ln\Omega\bigg\}+O(z_0).\nonumber\\
\end{eqnarray}

One of the required integrals can be computed exactly, namely
\begin{equation}
\int_0^\infty\rmd{u}\,u[K_0(u)h_1(u)+I_0(u)h_2(u)]=-\frac{\pi^2}{8}.
\end{equation}
One way to obtain this result is to express the integral in terms of the hypergeometric function ${}_2F_{3}$ and the Meijer $G$-function $G^{4,0}_{2,4}$, thereby showing that it agrees with the negative of the integrand of the integral denoted $H_3(0)$ in \cite{DR14}, whose value $\pi^2/8$ was computed there. 

Inserting the above results into Eq.~\eqref{eq:dF1dx} finally yields the desired relation~\eqref{eq:Yprimerel} with $C_1$ given by Eq.~\eqref{eq:Cintexpr}.

 \section{Asymptotic behavior of $\mathcal{F}_1^{(2)}([v]_*;1,x)$ for $x\to-\infty$ and computation of $d_1$ }\label{app:scfctxminfty}
The results for the free-energy functional $\mathcal{F}_1([v]_*;1,x)$ derived in Sec.~\ref{sec:freeenergyfunc} and Appendix~\ref{app:fs} yield the contributions to the scaling function $Y(x)$ that do not vanish as $|x|\to\infty$. Upon exploiting the relationship of $\mathcal{F}^{(1)}([v_*];1,x)$ to $Y'(x)$ along with the asymptotic ${x\to-\infty}$ form~\eqref{eq:Thetalargemx}, we have also been able to obtain the asymptotic behavior of $\mathcal{F}^{(1)}([v_*];1,x)$ for $x\to-\infty$ to order $1/|x|$, expressed in terms of the still unknown coefficient $d_1$. What remains to be done is the calculation of $\mathcal{F}_1^{(2)}([v_*];1,x)$ for $x\to-\infty$ to $O(1/|x|)$.

The contributions of order $|x|^{-1}\ln|x|$ and $1/|x|$ arise from two sources. First, the self-consistent potential $v_*(\zm;|x|,-1)$ differs from the asymptotic one $v_*(\zm;\infty,-1)$ by a distant-wall correction. As will be become clear below, the latter affects the Jost function in a nontrivial fashion. Second, the limiting $x\to-\infty$ form given in Eq.~\eqref{eq:chilargexm} cannot be used for values $y\lesssim |x|^{-1}$ because contributions from the second Jost solution $\asprop \rme^{-\zm \sqrt{y}}$ are not negligible.  

In order to separate these two types of corrections, we  split the integral $\int_0^\infty\rmd{\omega}$ on the right-hand side of Eq.~\eqref{eq:fmu2def} as $\int_0^\Omega\rmd{\omega}+\int_\Omega^\infty\rmd{\omega}$, where $\Omega$ is an arbitrary number such that 
\begin{equation}
1\ll x^2(2\pi/\tilde{b})^2\ll\Omega\ll x^2.
\end{equation}
Thus  $\mathcal{F}_1^{(2)}([v_*];1,x)$ becomes
\begin{equation}\label{eq:F12intermsofJs}
\mathcal{F}_1^{(2)}([v_*];1,x)=J_{<,1}(x;\Omega)+J_{<,2}(x;\Omega)+J_>(x;\Omega),
\end{equation}
where the first two integrals are given by
\begin{equation}\label{eq:Jsmaller1}
J_{<,1}(x;\Omega)=-\int_0^{\Omega}\frac{\rmd{\omega}}{8\pi}\,\big[\chi(\omega;1,x)+\sqrt{\omega}\big] 
\end{equation}
and
\begin{eqnarray}\label{eq:Jsmaller2}
J_{<,2}(x;\Omega)&=&-\int_0^{\Omega}\frac{\rmd{\omega}}{8\pi}\bigg[\frac{\mathcal{R}_*(1,x)}{2\sqrt{\omega}}+\frac{2x\ln\omega}{\pi^2\sqrt{\omega}}\nonumber\\ &&\strut -\frac{2x^2}{\pi^2(1+\omega)}\bigg],
\end{eqnarray}
respectively.
The third integral, $J_>(x;\Omega)$, stands for $-\int_\Omega^\infty\frac{\rmd{\omega}}{8\pi}\ldots$ of the sum of the integrands of $J_{<,1}(x;\Omega)$ and $J_{<,2}(x;\Omega)$. 

The calculation of  $J_{<,2}$ to the required order is easy. Using Eq.~\eqref{eq:Rstartoxm2}, one can conclude that it is sufficient to substitute the contributions $\asprop |x|$ for $\mathcal{R}_*(1,x)$ since the remaining terms yield contributions of order $O(x^{-2}\ln|x|)$. The integration can be done in a straightforward fashion, giving
\begin{eqnarray}\label{eq:Jsma2res}
J_{<,2}(x;\Omega)&=&\frac{x^2}{4\pi^3}\ln(1+\Omega)+\frac{|x|\sqrt{\Omega}}{\pi^3}
\ln\frac{\pi\sqrt{\Omega}}{2|x|}\nonumber\\ &&\strut+O(x^{-2}\ln|x|).
\end{eqnarray}

We next turn to the calculation of $J_{<,1}(x;\Omega)$. To determine its $x\to-\infty$ asymptotics, we need the regular solution $\regsol(\zL,\kL;1,x)$ of the differential equation
\begin{equation}\label{eq:dglregsolomega}
(-\partial_{\zL}^2+v(\zL;1,x)+\omega)\regsol(\zL,\kL;1,x)=0,\quad\omega=\kL^2,
\end{equation}
both in the inner region $\tilde{b}/|x|\lesssim \zL<1/2$ and the boundary region $0<\zL\ll \tilde{b}/|x|$ (cf.\ Sec.~\ref{sec:asspecprop}).

In the inner region, $v_*$ is small $\asprop 1/|x|$ and according to Eq.~\eqref{eq:vinner}, can be written as 
\begin{equation}\label{eq:vstar}
v_*(\zL;1,x)=\frac{w'''(\zL)}{4|x|}+O(|x|^{-2}),
\end{equation}
where $w(\zL)$ is defined as
\begin{equation}\label{eq:wdef}
w(\zL)\equiv \ln \Gamma(\zL)-\ln\Gamma(1-\zL)-\psi(1/2)(2\zL-1).
\end{equation}
We treat $v_*$ as a perturbation and subtract from the integrand of the resulting integral its Taylor expansion to $O(\zL^{\prime 2})$ about $\zL'=0$  to ensure its convergence at the lower  integration limit $\zL'=0$. We thus obtain
\begin{eqnarray}\label{eq:regsolomega}
\lefteqn{\regsol(\zL,\kL;1,x)}&&\nonumber\\ &=&\frac{\cosh(\zL \kL)}{|x|^{1/2}}+\frac{1}{4|x|^{3/2}}\bigg\{\int_{1/2}^{\zL}\rmd{\zL'}\,w'''(\zL')P_2(\zL',\zL,\kL)
\nonumber\\ && \strut+\int_0^{\zL}\rmd{\zL'}w'''(\zL')\bigg[\frac{\sin[(\zL-\zL')\kL]\cosh(\zL'\kL)}{\kL}\nonumber\\&&\strut -P_2(\zL',\zL,\kL)\bigg]+4A_1\cosh(\zL \kL)\nonumber\\ &&\strut+4B_1\sinh(\zL \kL)\bigg\}+o\big(|x|^{-3/2}\big)
\end{eqnarray}
with
\begin{equation}
P_2(\zL',\zL,\kL)=\kL^{-1}\sinh(\zL \kL)-\zL'\cosh(\zL \kL)+\kL\,\zL^{\prime 2}\sinh(\zL \kL),
\end{equation}
where the coefficients $A_1$ and $B_1$ still need to be determined. 

Before doing this, we first integrate by parts to rewrite Eq.~\eqref{eq:regsolomega} as
\begin{eqnarray}\label{eq:regsolinnreg}
\regsol(\zL,\kL ;1,x)&=&\frac{\cosh(\zL \kL)}{|x|^{1/2}}+\frac{1}{4|x|^{3/2}}\bigg\{w'(\zL)\cosh(\zL \kL)\nonumber\\ &&\strut -2w(\zL)\sinh(\zL \kL)\,\kL\nonumber\\&& \strut+4\omega \int_0^{\zL}\rmd{\zL'}\,w(\zL')\cosh[(\zL-2\zL')\kL]\nonumber\\ && \strut+4A_1\cosh(\zL   \kL)+4B_1\sinh(\zL \kL)\bigg\}\nonumber\\&& \strut +o\big(|x|^{-3/2}\big).
\end{eqnarray}
This can be inserted into Eq.~\eqref{eq:Deltaregsol} to compute $\Delta(-\omega;1,x)$. From the latter, $\chi(\omega;1,x)$ follows via Eq.~\eqref{eq:DeltaEdef}. We thus arrive at
\begin{eqnarray}\label{eq:chiomega1xres}
\lefteqn{\chi(\kL^2 ;1,x)}&&\nonumber\\
&=&- \kL-\ln\frac{\pi \kL}{2|x|}-\ln\big(1-\rme^{-2 \kL}\big)
 -2\,\frac{A_1+B_1\coth \kL}{|x|} \nonumber\\&&\strut
 -\frac{2\kL^2}{|x|}\int_0^{1/2}\rmd{\zL'}\,w(\zL')\big[\rme^{-2\zL' \kL}
+(1-\coth \kL)\sinh(2\zL' \kL)\big]
\nonumber\\&&\strut
+O\big(|x|^{-2}\big).
\end{eqnarray}

In order to determine $A_1$ and $B_1$, we  compute the regular solution in the boundary region and match the result with the one for the inner region given in Eq.~\eqref{eq:regsolinnreg} at the matching point (mp) $\zL_{\text{mp}}=\tilde{b}/|x|$. For $\zL\etwa z_{\text{mp}}\ll 1$, Eq.~\eqref{eq:regsolinnreg} simplies to
\begin{eqnarray}\label{eq:regszmp}
\lefteqn{|x|^{1/2}\regsol(\zL,\kL;1,x)}&\nonumber\\ &=&1+\frac{\kL^2\zL^2}{2}-\frac{1}{4|x|\zL}+\frac{A_1+\ln 2}{|x|}+\frac{\zL \kL}{|x|}\bigg\{B_1\nonumber\\ 
&&\strut +\frac{\kL}{2}\bigg[\frac{7}{4}-\gammaE-\ln(4\zL)\bigg]\bigg\}\nonumber\\&& \strut +\frac{\zL^2}{2|x|}[(A_1+\ln2)\kL^2-\zeta(3)]+O(\zL^3)+O\big(x^{-2}\big).\nonumber\\
\end{eqnarray}

In the boundary region, the potential $v_*$ cannot be treated by perturbation theory. To determine the regular solution in this region, we use the scaling properties  $v_*(\zm;|x|,-1)=v_*(\zL;1,x)/x^2$ and
\begin{equation}\label{eq:rescalingregsol}
\regsol(\zm,{\sk=\kL/|x|};|x|,-1)=\sqrt{|x|}\,\regsol(\zL,\kL;1,x),
\end{equation}
where  $\zm=|x|\zL$, to draw from Eqs.~\eqref{eq:vxlargezLlarge}, \eqref{eq:Bminlargezm}, and \eqref{eq:vstar} the following conclusions:
\begin{enumerate}
\item
the potential $v_*(\zm;|x|,-1)$ can be expanded about $x=-\infty$ as
\begin{equation}\label{eq:vstarlargex}
v_*(\zm;|x|,-1)=v_*(\zm;\infty,-1) +\frac{u_3(\zm)}{|x|^3}+O(|x|^{-4}),
\end{equation}
where $u_3(\zm)$  behaves asymptotically as    
\begin{eqnarray}\label{eq:u3smallzm}
u_3(0)&\mathop{=}\limits_{\zm\to 0}&0,\\
 \label{eq:u3largezm}
u_3(\zm)&\mathop{=}\limits_{\zm\to\infty}&-\zeta(3)-\beta_2\,\zm^{-1}+O(1/\zm^2),
\end{eqnarray}
by consistency with Eqs.~\eqref{eq:Aminlargezm} and \eqref{eq:Bminlargezm}.

Since this remains to be proven, we will not make use of it here, taking $\beta_2$ to be an unknown number. 
\item
the function 
$\regsol(\zm,\sk;|x|,-1)$ is the solution to
\begin{align}\label{eq:SEwithzeta3}
&[-\partial_{\zm}^2+v_*(\zm;\infty,-1)]\regsol(\zm,\sk;|x|,-1)\nonumber\\ &=-\bigg[\sk^2+
\frac{u_3(\zm)}{|x|^3}+O(x^{-4})\bigg]\regsol(\zm,\sk;|x|,-1)
\end{align} 
subject to the boundary condition~\eqref{eq:regsolinfbc}. 
\end{enumerate}

The contributions to the potential describing the deviation from $v_*(\zm;\infty,-1)$, which are given on the right-hand side, can be treated perturbatively because $\sk^2=O(x^{-2})$. The term $u_3(\zm)/|x|^3$ can be replaced by its value at $\zm=\infty$, namely $-\zeta(3)/|x|^3$. The reason is that the difference $u_3(\zm)-u_3(\infty)$ should be expandable in inverse powers of $|x|$ (modulo eventual logarithms) and hence vary for  $\zm\to\infty$   as $ \zm^{-\kappa}$ with $\kappa\ge1$. In fact, $\kappa$ is exactly unity since we shall show below that the potential coefficient $\beta_2$ is given by Eq.~\eqref{eq:beta2res} and hence does not vanish.

Once $u_3(\zm)$ has been replaced by its limiting value, it can be absorbed by introducing the shifted wavenumber $\sk_x$ through
\begin{equation}\label{eq:kxdef}
\sk_x^2\equiv \sk^2-\zeta(3)/|x|^3.
\end{equation}

To determine the coefficients $A_1$ and $B_1$, we will match the perturbative result for $\regsol(\zm,\sk;|x|,-1)$ obtained from Eq.~\eqref{eq:SEwithzeta3} at the right edge of the boundary region,  namely for $\zm=|x|\zL\etwa \tilde{b}\gg 1$, with the result given in Eq.~\eqref{eq:regszmp}. For such values of $\zm$ the term $\propto u_3(\zm)$ is smaller than the second one by a factor of $\tilde{b}^{-\kappa}$. It does not contribute at the order of the terms involving the coefficients $A_1$ and $B_1$, vanishes in the limit $\tilde{b}\to\infty$, and hence may be ignored here.

The associated perturbative solution of Eq.~\eqref{eq:SEwithzeta3} can be read off from previous ones derived  in Sec.~\ref{sec:surfexcq}. Noting that the quantity $\sk^2$ in Eq.~\eqref{eq:regsolkexp} corresponds here to $\sk^2-\zeta(3)|x|^{-3}$, we can conclude from 
Eqs.~\eqref{eq:regsol0as} and \eqref{eq:regsol2as} that the regular solution at $\zm=|x|\zL\etwa \tilde{b}\gg 1$ can be written as
\begin{eqnarray}
\regsol(\zm,\sk;|x|,-1)&\etwa&1-\frac{1}{4\zm}+\bigg[\sk^2-\frac{\zeta(3)}{|x|^3}\bigg]\bigg[\frac{\zm^2}{2}\nonumber\\ &&\strut +\frac{\zm}{2}\bigg(\frac{3}{4}-\gammaE-\ln\frac{4\zm}{\pi}\bigg)\bigg].
\end{eqnarray}
Matching this with Eq.~\eqref{eq:regszmp} at $\zm=|x|\zL$ then shows that the coefficients $A_1$ and $B_1$ are given by
\begin{eqnarray}
A_1&=&-\ln2,\nonumber\\
B_1&=&-\frac{1}{2}[1+\ln(|x|/\pi)]\sqrt{\omega}.
\end{eqnarray}

We can now insert Eq.~\eqref{eq:chiomega1xres} into Eq.~\eqref{eq:Jsmaller1}. For those parts of the integrand of the integral $\int_0^\Omega\rmd{\omega}$ that are exponentially decaying,  such as the ones proportional to $\ln\big(1-\rme^{-2\sqrt{\omega}}\big)$, $1-\coth\sqrt{\omega}$, and $\rme^{-2\sqrt{\omega}}$, we can safely replace the upper integration limit  by $\infty$. Using
\begin{eqnarray}
-\zeta(3)&=&2\int_0^\infty\rmd{\omega}\ln\big(1-\rme^{-2\sqrt{\omega}}\big)\nonumber\\ &=&\int_0^\infty\rmd{\omega}\sqrt{\omega}\,(1-\coth\sqrt{\omega}),
\end{eqnarray}
one arrives at
\begin{eqnarray}\label{eq:Jgrt1first}
J_{<,1}(x;\Omega)&=&\frac{\Omega}{8\pi}\bigg(-\frac{1}{2}+\ln\frac{\pi\sqrt{\Omega}}{2|x|}\bigg)-\frac{\Omega^{3/2}}{12\pi|x|}\bigg(1+\ln\frac{|x|}{\pi}\bigg)\nonumber\\ 
&&\strut-\frac{1}{4\pi|x|}\,\Omega\ln2-\frac{\zeta(3)}{8\pi}\bigg[\frac{1}{2}+\frac{1+\ln(|x|/\pi)}{|x|}\bigg]\nonumber\\&&\strut +J_{<,3}(x;\Omega)+J_{<,4}(x)+O\big(x^{-2}\big)\nonumber\\
\end{eqnarray}
with
\begin{eqnarray}
J_{<,3}(x;\Omega)&=&\frac{1}{4\pi|x|}\int_0^\Omega\rmd{\omega}\,\omega\int_0^{1/2}\rmd{z}\,w(z)\,\rme^{-2z\sqrt{\omega}}\nonumber\\
&=& \frac{1}{2\pi|x|}\int_0^{1/2}\rmd{z}\,w(z)\,R(z;\Omega)
\end{eqnarray}
and
\begin{eqnarray}
J_{<,4}(x)&=&\frac{1}{4\pi|x|}\int_0^{1/2}\rmd{z}\,w(z)\int_0^\infty\rmd{\omega}\,\omega\,\big[1\nonumber \\ &&\strut -\coth(\sqrt{\omega})\big]\sinh(2z\sqrt{\omega})\nonumber\\
&=&-\frac{d_{1,1}}{8\pi|x|}.
\end{eqnarray}
Here
\begin{equation}
R(z;\Omega)=\frac{3-\rme^{-2z\Omega^{1/2}}(3+6z\Omega^{1/2}+6z^2\Omega+4z^3\Omega^{3/2})}{8z^4}
\end{equation}
and
\begin{equation}\label{eq:d11def}
d_{1,1}=\frac{3}{2}\int_0^{1/2}\rmd{z}\,w(z)\big[z^{-4}-w^{(4)}(z)/6\big],
\end{equation}
where  $w(z)$ denotes the function 
\begin{equation}
w(z)\equiv \ln\Gamma(z)-\ln\Gamma(1-z) -(2z-1) \psi(1/2),
\end{equation}
which varies as
\begin{equation}
w(z)\mathop{=}_{z\to 0}w_{\text{as}}(z)+O(z^5)
\end{equation}
with
\begin{equation}\label{eq:wasdef}
w_{\text{as}}(z)\equiv -\gamma_{\text{E}}-\ln(4z)+4z\ln 2-\frac{2}{3}\,\zeta(3)\,z^3.
\end{equation}

To further evaluate  $J_{<,3}(x;\Omega)$, we  divide up $w(z)$ into its asymptotic contribution $w_{\text{as}}(z)$ introduced in Eq.~\eqref{eq:wasdef} and a remainder, writing
\begin{equation}
J_{<,3}(x;\Omega)=-\frac{1}{8\pi}\big[J_{<,3}^{(1)}(x;\Omega)+J_{<,3}^{(2)}(x;\Omega)\big]
\end{equation}
with
\begin{equation}
J_{<,3}^{(1)}(x;\Omega)=-\frac{4}{|x|}\int_0^{1/2}\rmd{z}\,w_{\text{as}}(z)\,R(z;\Omega)\end{equation}
and 
\begin{equation}
J_{<,3}^{(2)}(x;\Omega)=-\frac{4}{|x|}\int_0^{1/2}\rmd{z}\,[w(z)-w_{\text{as}}(z)]\,R(z;\Omega).
\end{equation}

In the last integral, we can replace $R(z;\Omega)$ by its asymptotic large-$z$ form   $3/(8z^4)$ to obtain
\begin{equation}
J_{<,3}^{(2)}(x;\Omega)=|x|^{-1}\big[d_{1,2}+O\big(\Omega^{-1/2}\big)\big],
\end{equation}
with
\begin{equation}\label{eq:d12def}
d_{1,2}=-\frac{3}{2}\int_0^{1/2}\rmd{z}\,z^{-4}[w(z)-w_{\text{as}}(z)].
\end{equation}
The large-$\Omega$ asymptotics of the remaining integral $J_{<,3}^{(1)}(x;\Omega)$ can be determined in a straightforward fashion. One finds
\begin{eqnarray}\label{eq:Jsma4}
|x|\,J_{<,3}^{(1)}(x;\Omega)&=&\Omega^{3/2}\bigg[\frac{2}{9}+\frac{2}{3}\ln\big(2/\sqrt{\Omega}\bigg]-2\Omega\ln 2\nonumber\\ &&\strut -\frac{4}{3}-4\gammaE+8\ln 2+\zeta(3)\bigg(\gammaE-\frac{11}{6}\nonumber\\&& \strut +\ln\sqrt{\Omega}\bigg)+O\Big(\Omega^{3/2}\rme^{-\sqrt{\Omega}}\Big).
\end{eqnarray}
The combination of Eqs.~\eqref{eq:Jgrt1first}--\eqref{eq:Jsma4} then yields
\begin{eqnarray}\label{eq:Jsma1res}
J_{<,1}(x;\Omega)&=&-\frac{1}{8\pi}\Bigg\{\frac{\zeta(3)}{2}+\frac{\Omega^{3/2}}{|x|}\bigg[\frac{8}{9}-\frac{2}{3}\ln\frac{\pi\Omega^{1/2}}{2|x|}\bigg]\nonumber\\&&\strut +\Omega\bigg[\frac{1}{2}-\ln\frac{\pi\Omega^{1/2}}{2|x|}\bigg]+\frac{1}{|x|}\bigg[d_{1,1}+d_{1,2}\nonumber\\ &&\strut
-\frac{4}{3}-4\gammaE+8\ln 2\bigg]+\frac{\zeta(3)}{|x|}\bigg[\gammaE-\frac{5}{6}\nonumber\\ &&\strut
+\ln\frac{\Omega^{1/2}|x|}{\pi}\bigg]\Bigg\}
+O(x^{-2}\ln|x|).
\end{eqnarray}

It remains to calculate the $x\to-\infty$ asymptotics of integral $J_>(x;\Omega)$ in Eq.~\eqref{eq:F12intermsofJs} to the order $\sim |x|^{-1}$.
 Upon making a change of variable $\omega\to y=x^{-2}\omega$, it becomes
\begin{align}\label{eq:Jgr}
J_>(x;\Omega)=&-x^2\int_{\Omega/x^2}^\infty\frac{\rmd y}{8\pi}\,\bigg[\chi(y;|x|,-1)+|x|\sqrt{y}\nonumber\\ &\strut +\frac{\mathcal{R}_*(1,x)}{2|x|\sqrt{y}}-\frac{2\ln (y\,x^2)}{\pi^2\sqrt{y}}-\frac{2}{\pi^2(y+x^{-2})}\bigg],
\end{align}
where $\mathcal{R}_*(1,x)$ can be read from Eq.~\eqref{eq:Rstartoxm2}.
The integrand in the square brackets is the (rescaled) analog of the one we encountered in Appendix~\ref{app:fs} in our calculation of $\mathcal{F}_1^{(2)}([v_*];L,m)$ to $o(L)$. There we could express the asymptotic from of the function $\chi(y m^2;L,m)$ for large $L$ in terms of the Jost function $F(\rmi\sqrt{y})$ for the semi-infinite case $L=\infty$. Here we must take into account the effects of the finite film thickness.  The differential equation from which the regular solution $\regsol(\zm,\sk;|x|,-1)$ is to be determined is  Eq.~\eqref{eq:SEwithzeta3}. Since we shall need the value of the potential coefficient $\beta_2$ introduced in Eq.~\eqref{eq:u3largezm}, we begin by proving our result stated in Eq.~\eqref{eq:beta2res}.

To this end, we shall compute the contributions of $\mathcal{R}_*(1,x)$ proportional to $x^{-2}\mod \ln|x|$ directly from the definition~\eqref{eq:Rdef} of $\mathcal{R}([v_*];1,x)$ and match them with those in Eq.~\eqref{eq:Rstartoxm2}. These terms  originate from the contribution $\overline{v_*^{\text{ns}}}$. Setting $L=1$ and $m=x$, we therefore split the corresponding integral into a sum of contributions from the boundary and inner region. Making the change of integration variables $\zL\to\zm=|x|\zL$ in the near-boundary integral, we arrive at
\begin{eqnarray}\label{eq:vnsavdec}
\overline{v_*^{\text{ns}}}&=&|x|\int_0^{\tilde{b}}\rmd\zm\bigg[v_*(\zm;|x|,-1)+\frac{1}{4\zm^2}\strut+\frac{4}{\pi^2\zm}\bigg]\nonumber\\
&&+\int_{\tilde{b}/|x|}^{1/2}\rmd\zL\bigg[v_*(\zL;1,x)+\frac{1}{4\zL^2}+\frac{4|x|}{\pi^2\zL}\bigg].
\end{eqnarray}

In the first integral, we insert the large-$x$ expansion~\eqref{eq:vstarlargex} of $v_*(\zm;|x|,-1)$. For the potential in the second integral we can substitute the corresponding expansion
\begin{equation}\label{eq:vstarlargexm}
v_*(\zL;1,x)=\frac{t_1(\zL)}{|x|}+\frac{t_2(\zL)}{|x|^2}+\ldots,
\end{equation}
where
\begin{equation}\label{eq:t1res}
t_1(\zL)=\frac{w'''(\zL)}{4}=\frac{\psi''(\zL)+\psi''(1-\zL)}{4}
\end{equation}
for $\tilde{b}/|x|\le\zL<1/2$ according to Eqs.~\eqref{eq:vinner} and \eqref{eq:wdef}. Furthermore, consistency with Eqs.~\eqref{eq:Aminlargezm}, \eqref{eq:Bminlargezm}, and \eqref{eq:u3largezm} requires that
\begin{equation}\label{eq:t2smallzL}
t_2(\zL\to 0)=-\alpha_2\,\zL^{-4}-\beta_2\,\zL^{-1}+O(1).
\end{equation}

The contributions linear in $|x|$ or proportional to $|x|\ln|x|$ of the first term in Eq.~\eqref{eq:vnsavdec} originate from the leading ${x=-\infty}$~term of the integrand. Their $\tilde{b}$-dependent part is canceled by analogous ones produced by the antiderivatives of the last two terms of the second integral's integrand at the lower integration limit. All contributions linear in $|x|$ and proportional to $|x|\ln|x|$ must add up to produce the contribution to $\mathcal{R}_*(1,x)$ given in the first line of Eq.~\eqref{eq:Rstartoxm2}. Since the latter equation is already established, there is no need to check this contribution again.

The contributions to $\mathcal{R}_*(1,x)$ of order $x^{-2}\mod \ln|x|$ arise from the sum of the integrals
\begin{eqnarray}
T_1&=&2|x|^{-1}\int_{\tilde{b}/|x|}^{1/2}\rmd\zL\,t_1(\zL)=(2|x|)^{-1}w''(\zL)|_{\tilde{b}/|x|}^{1/2},\nonumber\\
&=& \strut \frac{|x|}{2\tilde{b}}+\frac{2\tilde{b}\,\zeta(3)}{|x|^2}+O\big(|x|^{-4}\big),
\nonumber\\
T_2&=&2|x|^{-2}\int_{\tilde{b}/|x|}^{1/2}\rmd\zL\,t_2(\zL),\nonumber\\
T_3&=&2|x|^{-2}\int_0^{\tilde{b}}\rmd\zm\,u_3(\zm).
\end{eqnarray}

Assuming that $\tilde{b}>1$ and defining
\begin{eqnarray}
L_2(\zL)&=&\int_{\zL}^{1/2}\rmd\zL'\bigg[t_2(\zL')+\frac{\alpha_2}{\zL^{\prime 4}}+\frac{\beta_2}{\zL^{\prime }}\bigg],\nonumber\\
L_3(\zm)&=&\int_{0}^{\zm}\rmd\zm'\,\big[u_3(\zm')+\zeta(3)+\beta_2\,\Htheta(\zm'-1)/\zm'\big],\qquad
\end{eqnarray}
we can rewrite $T_2+T_3$ as
\begin{eqnarray}
T_2+T_3&=&\frac{2}{|x|^2}\bigg\{L_3(\tilde{b})+L_2(\tilde{b}/|x|) 
-\beta_2\ln\frac{|x|}{2}\nonumber\\ &&\strut+\frac{\alpha_2}{3}\bigg(8-\frac{|x|^3}{\tilde{b}}\bigg)-\tilde{b}\,\zeta(3)\bigg\}
\end{eqnarray}
by adding and subtracting appropriate terms to their integrands. We can now let $\tilde{b}\to\infty$ and $x\to-\infty$ such that $\tilde{b}/|x|\to 0$, i.e., we set $\tilde{b}=\text{const}\,|x|^\kappa$ with $0<\kappa<1$ and let $x\to-\infty$. Since the contributions $\propto \tilde{b}|x|^{-2}$ of $T_1$ and $T_2+T_3$ cancel each other, no contribution $\asprop |x|^{\kappa-2}$  to $\mathcal{R}_*$ results and we obtain
\begin{eqnarray}\nonumber
\mathcal{R}_*(1,x)&=&
-\frac{8|x|}{\pi^2}\bigg[1-\ln \frac{2|x|}{\pi}\bigg]+\frac{2}{x^2} \bigg[L_2(0) +L_3(\infty)\\
&&\strut -\beta_2 \ln(|x|/2) +\frac{8}{3}\alpha_2\bigg]+o(1/|x|^2).
\end{eqnarray}
Comparison  of this result with \eqref{eq:Rstartoxm2}  yields, besides the value of $\beta_2$ given in Eq.~\eqref{eq:beta2res}, the following relation
\begin{equation}\label{eq:magicrel}
{\zeta(3)\,\frac{2-d_1-\ln4}{4}}=L_2(0)+L_3(\infty)+\frac{8}{3}\,\alpha_2.
\end{equation}

Until now we have not taken into account the far-boundary correction $\propto u_3(\zm)$ of Eq.~\eqref{eq:vstarlargex} other than through its asymptotic value $u_3(\infty)$. In our calculation of $J_>(x;\Omega)$ we must be more careful and include also its correction $\propto \zm^{-1}$ specified in Eq.~\eqref{eq:u3largezm}. 

To this end, we express $\sk$ in the  Schr\"odinger equation~\eqref{eq:SEwithzeta3} again in terms of the shifted wavenumber $\sk_x$ defined in Eq.~\eqref{eq:kxdef}.  As a result, the potential term $u_3(\zm)$ gets replaced by the shifted one
\begin{equation}\label{eq:Delu3largezm}
\Delta u_3(\zm)=u_3(\zm)+\zeta(3)\mathop{\aseq}\limits_{\zm\to\infty}-\beta_2\,\zm^{-1},
\end{equation}
which vanishes at $\zm=\infty$. Let us introduce the notations
\begin{eqnarray}
f_+(\zm,\sk_x;x)&=&f(z,-\rmi\sk_x),\nonumber\\  F_+(\sk_x)&=&F(\rmi\sk_x),
\end{eqnarray}
for the Jost solution and Jost function on the right-hand sides. Exploiting Eq.~\eqref{eq:regsolJost}, we can express the regular solution introduced in Eq.~\eqref{eq:rescalingregsol} as
\begin{align}\label{eq:regsolomJost}
&\regsol(\zm,\sk ;|x|,-1)\nonumber\\ &=\frac{F_+(\sk_x)\,f_+(\zm,\sk_x;x)-F_+(-\sk_x)\,f_+(\zm,-\sk_x;x)}{2\sk_x}.\quad
\end{align}
Here $f_+(\zm;|x|,-1)$ is a solution to 
\begin{equation}\label{eq:SEfplus}
\left[-\partial_\zm^2+v(\zm;\infty,-1)+\frac{\Delta u_3(\zm)}{|x|^3}+\sk_x^2\right]f_+(\zm,\sk_x;x)=0
\end{equation}
that behaves asymptotically as
\begin{eqnarray}
f_+(\zm,\sk_x;x)\mathop{\asprop}\limits_{\zm\to\infty} \exp\left[\sk_x\zm-\frac{\beta_2}{2\sk_x|x|^3}\ln\zm\right],
\end{eqnarray}
where the $\ln\zm$ in the exponent is due to the slow Coulomb-like decay of $\Delta u_3(\zm)$ \cite{Tay12}.

To determine $f_+(\zm,\sk;x)$ in the boundary region for $1\ll \zm<\tilde{b}$, we replace $ v_*(\zm;|x|,-1)$  and  $\Delta u_3(\zm)$ in Eq.~\eqref{eq:SEfplus}  by their large-$\zm$  expansions $-(2\zm)^{-3}-\alpha_2\zm^{-4}+O(\zm^{-5})$ and \eqref{eq:Delu3largezm}, respectively, and use the ansatz
\begin{eqnarray}
f_+(\zm,\sk_x;x)&=&\mathcal{A}(\sk_x,x)\exp\Bigg\{\sk_x\zm\bigg[1+\sum_{j=2}^4\frac{f_j(\sk_x,x)}{\zm^j}\nonumber\\ && \strut+O\big(\zm^{-5}\big)\bigg]-\frac{\beta_2}{2\sk_x|x|^3}\ln\zm\Bigg\}.
\end{eqnarray}
Solving for $f_2,\dotsc,f_4$ yields
\begin{equation}
f_2(\sk_x,x)=\frac{\beta _2}{4\sk_x^3 |x|^3}
+\frac{\beta _2^2}{8 \sk_x^4 |x|^6},
\end{equation}
\begin{equation}
f_3(\sk_x,x)=\frac{1}{8\sk_x^2}+\frac{\beta_2}{8\sk_x^4|x|^3}+\frac{\beta_2^2}{8\sk_x^5
|x|^6}+\frac{\beta_2^3}{32\sk_x^6|x|^9},
\end{equation}
and
\begin{equation}
f_4(\sk_x,x)=\frac{1}{8\sk_x^3}+\frac{\alpha_2}{6\sk_x^2}+\beta_2\,\frac{1+O(1/\sk_x)}{24\sk_x^4|x|^3}+O\big(|x|^{-6}\big).
\end{equation}

We now express the exponent of the exponential of $f_+(\zm,\sk_x,x)$ in terms of $\sk$, expand in powers of $1/|x|$ and $1/\sk$, dropping the terms of higher than the orders $\sk^{-2}$, $|x|^{-3}$, and $\zm^{-3}$, and insert the result into Eq.~\eqref{eq:regsolomJost}. The contribution from $f_+(\zm,-\sk_x,x)$ can be omitted because it is exponentially small. We thus obtain
\begin{eqnarray}\label{eq:regsolbr}
\regsol(\zm,\sk;|x|,-1)&=&\frac{F_+(\sk_x,x)}{2\sk_x}\exp\bigg[\sk\,\zm+\frac{1}{8\sk\,\zm^2 }+\frac{\alpha_2}{6\sk\,\zm^3}\nonumber\\ 
&&\strut -\frac{\beta_2\ln\zm}{2\sk|x|^3}+\frac{1}{8\sk^2\zm^3}+\frac{\beta_2}{4\sk^2|x|^3\zm}\nonumber\\ 
&&\strut -\frac{\zm\,\zeta(3)}{2\sk|x|^3}+\ldots\bigg].
\end{eqnarray}

Having determined the regular solution in the boundary region, we now turn to its calculation in the inner region. Since we need the asymptotic behavior of $\chi(\omega;1,x)$ for large $\omega=\kL^2$ ,  we can again use the semiclassical expansion to determine the behavior of $\regsol(\zL,\kL;1,x)$ in the inner regime. We start from the analogs of Eqs.~\eqref{eq:Zdecoddeven}--\eqref{eq:b2ofz},
\begin{eqnarray}
\label{eq:regsollsemcl2}
\regsol(\zL,\kL ;1,x)&=&\mathcal{A}(\kL,x)\,\exp\bigg\{\kL\zL-\frac{1}{2\kL}\int_{\zL}^{1/2}\rmd\zL'\Bigg[\frac{t_1(\zL')}{|x|}\nonumber\\ && \strut +\frac{t_2(\zL')}{|x|^2}\bigg]
-\frac{1}{4\kL^2}\bigg[\frac{t_1(\zL)}{|x|}+\frac{t_2(\zL)}{|x|^2}\bigg]\nonumber\\ &&\strut
+O\big(\kL^{-3}\big)\Bigg\},
\end{eqnarray}
perform the integral involving $t_1(\zL')$ using Eq.~\eqref{eq:t1res}, make a Laurent expansion in $\zL$,  and express the integral of $t_2(\zL')$ in terms of $L_2(\zL)$. This gives
\begin{eqnarray}\label{eq:phi5}
\frac{\regsol(\zL,\kL ;1,x)}{\mathcal{A}(\kL,x)}&=&\exp\bigg\{\kL\zL+\frac{1}{\kL}\bigg[\frac{1}{8|x|\zL^2}-\frac{\zeta(3)\zL}{2|x|}+O(\zL^2)\nonumber\\&&\strut -\frac{L_2(\zL)+\beta_2\ln(2\zL)}{2|x|^2}-\frac{\alpha_2}{6|x|^2}\bigg(8-\frac{1}{\zL^3}\bigg)\bigg]\nonumber\\ &&\strut +\frac{1}{\kL^2
}\bigg[\frac{1}{8|x|\zL^3}+\frac{\zeta(3)}{4|x|}+\frac{\beta_2}{4|x|^2\zL}\nonumber\\ &&\strut+\frac{\alpha_2}{4|x|^2\zL^4}+O(\zL^2)\bigg]+O(\kL^{-3})\bigg\}.
\end{eqnarray}

Matching this with Eq.~\eqref{eq:regsolbr} at  $\zL_{\text{mp}}\equiv\zm_{\text{mp}}|x|=\tilde{b}/|x|$ then yields for the amplitude the result
\begin{eqnarray}\label{eq:mathcalAres}
\mathcal{A}(\kL,x)&=&\frac{F_+(\sk_x,x)\,\sqrt{|x|}}{2\kL}\exp\bigg\{\frac{\zeta(3)}{4\kL^2|x|}+\frac{4\alpha_2}{3\kL|x|^2}\nonumber\\ &&\strut -\frac{\zeta(3)}{4\kL|x|^2}\ln\frac{|x|}{2}+\frac{L_2(0)}{2\kL|x|^2}+o(x^{-2})
\bigg\}.
\end{eqnarray}
where $1/\sk_x=\sk^{-1}\exp[\zeta(3)/(2\kL^2|x|^3)+O(|x|^{-6})]$ was used.

The result must be inserted into the analog of Eq.~\eqref{eq:chiomegaAq}, namely
\begin{equation}
\chi(\kL^2;1,x)=-\kL-\ln\big[2\pi\kL\,\mathcal{A}(\kL,x)^2\big].
\end{equation}
To this end, we expand $\ln F_+(\sk_x,x)$ in its second argument about $\ln F_+(\sk,\infty)$ and subsequently its first argument to linear order in $\sk_x-\sk=O(|x|^{-3})$, obtaining
\begin{eqnarray}\label{eq:X3def}
\ln F_+(\sk_x,x)&=&\ln F_+(\sk_x,\infty)+\frac{X_3(\sk_x)}{|x|^3}+O\big(|x|^{-6}\big) \nonumber\\
&=&\ln F_+(\sk,\infty)-\frac{\zeta(3)}{2\sk |x|^3}\frac{\partial\ln F_+(\sk,\infty)}{\partial\sk}\nonumber\\ &&\strut +\frac{X_3(\sk)}{|x|^3}+o\big(|x|^{-3}\big),
\end{eqnarray}
where the term $\propto X_3(\sk_x)$ accounts for the change $\propto |x|^{-3}$ induced by the  far-boundary correction $\Delta u_3(\zm)$ to $v_*(\zm;\infty,-1)$ [cf.~Eq.~\eqref{eq:vstarlargex}].

We can now substitute these results into $J_>(x,\Omega)$,  change to the integration variable $y=\omega/x^2$, and use the fact that the contributions to the integrand resulting from the first term inside the curly brackets in Eq.~\eqref{eq:mathcalAres} and the term involving $\partial_\sk F_+(\sk,\infty)$ add up to a derivative $\partial_y$ of a function that vanishes at $y=\infty$. It follows that $J_>(x,\Omega)$ can be decomposed as
\begin{equation}
 \label{eq:Jgrdec}
J_>(x,\Omega)=J_{>,1}(x,\Omega/x^2,\infty)+J_{>,2}(x,\Omega)+\frac{Q(\Omega/x^2)}{|x|},
\end{equation}
where $J_{>,1}(x,y_1,y_2)$ denotes the integral
\begin{eqnarray}
\lefteqn{J_{>,1}(x,y_1,y_2)}&&\nonumber\\ &=&\frac{x^2}{8\pi}\int_{y_1}^{y_2}\rmd{y}\Bigg\{2\ln F_+(\sqrt{y},\infty)-\ln\frac{2\sqrt{y}}{\pi} \nonumber\\ && \strut+\frac{4}{\pi^2\sqrt{y}}\bigg[1+\ln\frac{\pi \sqrt{y}}{2}\bigg]+\frac{2}{\pi^2(y+|x|^{-2})}\Bigg\},\quad
\end{eqnarray}
while
\begin{eqnarray}\label{eq:Jgtr2interm}
J_{>,2}(x,\Omega)&=&-\frac{\zeta(3)}{4\pi|x|}\int_{\Omega/x^2}^\infty\rmd{y}\frac{\partial}{\partial y}\ln\frac{F_+(y^{1/2},\infty)}{y^{1/4}}\nonumber\\
&=&\frac{\zeta(3)}{4\pi|x|}\,\ln\frac{F_+(y^{1/2},\infty)}{(2/\pi)^{1/2}y^{1/4}}\bigg|_{y=\Omega/|x|^2}
\end{eqnarray}
and
\begin{equation}\label{eq:Qdef}
Q(y_0)\equiv\frac{1}{8\pi}\int_{y_0}^\infty\rmd{y}\big[2X_3(\sqrt{y})-L_3(\infty)\,y^{-1/2}\big].
\end{equation}

The integral $J_{>,1}(x,\Omega/x^2,\infty)$ can be decomposed into the $\Omega$-independent term $J_{>,1}(x,0,\infty)$ and the remainder $-J_{>,1}(x,0,\Omega/x^2)$.  The value of the former can be gleaned from Eq.~\eqref{eq:Qpartintegral},
\begin{equation}
J_{>,1}(x,0,\infty)=\frac{x^2}{4\pi^3}\left[
-1+2\ln\frac{2|x|}{\pi}
\right].
\end{equation}
 Since the upper integration limit of the latter integral $J_{>,1}(x,0,\Omega/x^2)$
 is small, $\Omega/x^2\ll1$, we can replace  the logarithm of the Jost function in its integrand
by the asymptotic small-$y$ form
\begin{equation}\label{eq:logFsmally}
\ln F_+(\sqrt{y},\infty)=\ln \sqrt{y}-\frac{\sqrt{y}}{2}\bigg[1-\ln\frac{\pi \sqrt{y}}{2}\bigg]+O(y\ln y),
\end{equation}
which follows by integrating the analog of Eq.~\eqref{eq:dlnFres} for $F_+(\sqrt{y},\infty)$ with respect to $y$ and expanding the result for small $y$. One obtains
\begin{eqnarray}
J_{>,1}(x,0,\Omega/x^2)&=&\frac{x^2}{4\pi^3}\ln(1+\Omega)+\frac{|x|\Omega^{1/2}}{\pi^3}
\ln\frac{\pi \Omega^{1/2}}{2|x|}\nonumber\\ && \strut -\frac{\Omega}{8\pi}\bigg[\frac{1}{2}-\ln\frac{\pi \Omega^{1/2}}{2|x|}\bigg]
\nonumber\\ &&\strut -\frac{\Omega^{3/2}}{|x|\pi}\bigg[\frac{1}{9}-\frac{1}{12}\ln\frac{\pi \Omega^{1/2}}{2|x|}\bigg]\nonumber\\ &&\strut +O\big(x^{-2}\ln|x|\big),
\end{eqnarray}
and substitution of Eq.~\eqref{eq:logFsmally} into Eq.~\eqref{eq:Jgtr2interm} yields
\begin{equation}
J_{>,2}(x,\Omega)=\frac{\zeta(3)}{16\pi|x|}\ln\frac{\pi^2\Omega}{4|x|^2}+O\big(x^{-2}\ln|x|\big).
\end{equation}

For large negative $x$, the last term on the right-hand side of \eqref{eq:Jgrdec} behaves  as
\begin{equation}
\frac{Q(\Omega/x^2)}{|x|}=\frac{Q(0)}{|x|}+O(|x|^{-3}).
\end{equation}
Hence, to the required order in $1/|x|$, all $\Omega$-dependent terms of $J_>(x,\Omega)$ are contained in $J_{>,1}(x,\Omega)+J_{>,2}(x,\Omega)$. Substitution of our results for this sum along with $Q(0)$ and those for $J_{<,1}(x,\Omega)$ and $J_{<,2}(x,\Omega)$ into Eq.~\eqref{eq:F12intermsofJs} yields
\begin{eqnarray}\nonumber
\mathcal{F}_1^{(2)}([v_*];1,x)=\frac{x^2}{4\pi^3}\left(-1+2\ln\frac{2 |x|}{\pi} \right)\\
+\frac{\zeta(3)}{8\pi}
\left[
-\frac{1}{2}-\frac{d_1+2\ln|x|}{|x|}
\right]+o(|x|^{-1})\label{eq:F},
\end{eqnarray}
with
\begin{eqnarray}\label{eq:d1intermediate}
d_1&=&\frac{1}{\zeta(3)}\bigg(d_{1,1}+d_{1,2}-\frac{4}{3}-4\gamma_{\text{E}}+8\ln2\bigg)\nonumber\\ &&\strut +\frac{1}{6}+\gamma_{\text{E}}+\ln\frac{2}{\pi^2}
-\frac{8\pi}{\zeta(3)}\,Q(0).
\end{eqnarray}
As one should expect, all $\Omega$-dependent terms cancel in \eqref{eq:F}.

Insertion of Eq.~\eqref{eq:F} into Eq.~\eqref{eq:LTr} then gives a differential equation for  $Y(x)$, which can be solved to determine the asymptotic large-$x$ form of this function. The result is [cf. \eqref{eq:fsres}]. 
\begin{equation}
Y(x)=\frac{x^2}{\pi^3}\left[
\frac{3}{4}+\frac{7\zeta(3)}{\pi^2}-\frac{1}{2}\ln \frac{2|x|}{\pi}
\right]+\Theta(x),
\end{equation}
where $\Theta(x)$ has the asymptotic series expansion~\eqref{eq:Thetalargemx}.

To establish the result for $d_1$ given in Eqs.~\eqref{eq:d1res}, we first prove that
\begin{equation}
Q(0)=0,
\end{equation}
if $\beta_2$ and $\alpha_2$ take their self-consistent values, given in Eq.~\eqref{eq:beta2res} and unknown, respectively. Subsequently, we shall complete the derivation of $d_1$ by computing the sum of integrals appearing in $d_{1,1}+d_{1,2}$.

The first step of the proof is to show that
\begin{equation}\label{eq:X3intermsofGa}
\frac{L_3(\infty)}{2\sk}-X_3(\sk)=\int_0^\infty\rmd\zm\bigg[\mathcal{G}_\infty(\zm,\zm;-\sk^2)+\frac{1}{2\sk}\bigg]\,\Delta u_3(\zm),
\end{equation}
where $\mathcal{G}_\infty(\zm,\zm';-\sk^2)=-[\sk^2+\mathcal{H}_{v_*}]^{-1}$ is the  Green's function for $L=\infty$ and $m=-1$. To do this, we can proceed along lines similar to those followed in Appendix~\ref{app:scatphase}. Let $f_\pm(\zm,\sk)$ be the Jost solutions associated with the Schr\"odinger equation  $\mathcal{H}_{v_*}f_\pm=0$ on the half-line that vary $\asprop \rme^{\pm\sk\zm}$ as $\zm\to\infty$. By analogy with Eqs.~\eqref{eq:Jostsolsmallzm} and \eqref{eq:Wzmzero}, one finds that $f_-(\zm\to 0,\sk)\aseq-F_+(\sk,\infty)\sqrt{\zm}\ln\zm$. Using the analog of Eq.~\eqref{eq:Wronskpartialz},  namely $(\sk^{\prime 2}-\sk^2)\regsol(\zm,\sk)\,f_-(\zm,\sk')=\partial_\zm W[\regsol(\zm,\sk),f_-(\zm,\sk^\prime)]$, one can derive
\begin{align}\label{eq:X3intermsofG}
&\int_{0}^{\infty}\rmd\zm\bigg[\lim_{\sk'\to\sk}\frac{\regsol(\zm,\sk)\,f_-(\zm,\sk')}{F_+(\sk,\infty)}-\frac{1}{2\sk}\bigg]\nonumber\\ &=\partial_{\sk^2}\ln \left[\frac{F_+(\sk,\infty)}{\sqrt{\sk}}\right]=-\int_0^\infty\rmd\zm\bigg[\mathcal{G}_\infty(\zm,\zm;-\sk^2)+\frac{1}{2\sk}\bigg]
\end{align}
in a straightforward manner, where the last result follows both from Eq.~\eqref{eq:GreenF} or else from the fact that the limit in square brackets is nothing other than $-\mathcal{G}_\infty(\zm,\zm;-\sk^2)$ [cf.\ Eq.~\eqref{eq:GEimE0}]. We now replace $v_*$ by $v_*+\delta v$ and denote the implied change of $\ln F_+(\sk,\infty)$ as $\delta \ln F_+(\sk)$. A calculation analogous to the one in Eq.~\eqref{eq:delG} yields $-\partial_{\sk^2}\Tr( \mathcal{G}_\infty\,\delta v)$ for the variation of the Green's function term on the right-hand side of Eq.~\eqref{eq:X3intermsofG}. If $\delta v(z)$ decayed faster than $\zm^{-1}$ as $\zm\to\infty$, we could integrate with respect to $\sk^2$, taking into account that the integration constant must be zero, to conclude that
\begin{equation}\label{eq:dellnFvR}
\delta \ln F_+(\sk)=-\int_0^\infty\rmd\zm\,\mathcal{G}_\infty(\zm,\zm;-\sk^2)\,\delta v(\zm).
\end{equation}
However, we are interested in  the variation $\delta v(\zm)=\Delta u_3(\zm)/|x|^3$ which has  the slower Coulombic decay~\eqref{eq:Delu3largezm}. We therefore replace $u_3(\zm)$ by $u_R(\zm)=\Htheta(R-\zm)\,u_3(\zm)$ in $\delta v$. The associated regular solution varies as $\regsol_R(\zm,\sk)\asprop \exp[\sk \zm-\beta_2(2\sk|x|^3)^{-1}\ln\zm]$ and $\asprop \rme^{\sk\zm}$ for  $1\ll \zm<R$ and $1\ll R<\zm$, respectively. Taking into account that $\regsol_R$ and its $\zm$-derivative must both be continuous at $\zm=R$, one sees that the corresponding Jost function must behave as  $F_{+,R}(\sk,x)=F_+(\sk,x)\exp[-\beta_2(2\sk|x|^3)^{-1}\ln R]$. Using Eq.~\eqref{eq:dellnFvR} for $\delta \ln F_{+,R}$, we get
\begin{align}
&\delta\ln F_+(\sk,x)=\lim_{R\to\infty}\bigg[\delta\ln F_{+,R}(\sk,x)+\frac{\beta_2}{2\sk|x|^3}\ln R\bigg]\nonumber\\
&=\lim_{R\to\infty}\bigg\{-\int_0^R\rmd\zm\bigg[\mathcal{G}_\infty(\zm,\zm;-\sk^2)+\frac{1}{2\sk}\bigg]\frac{\Delta u_3(\zm)}{|x|^3}\nonumber\\
&+\frac{1}{2\sk|x|^3}\int_0^R\rmd\zm\big[\Delta u_3(\zm)+\beta_2\,\Htheta(\zm-1)/\zm\big]\bigg\}.
\end{align}
The limit $R\to\infty$ of the second integral is $L_3(\infty)$, and the left-hand side is $X_3(\sk)/|x|^3$ according to Eq.~\eqref{eq:X3def}. Thus Eq.~\eqref{eq:X3intermsofGa} is established.

To proceed, we multiply the self-consistency Eq.~\eqref{eq:Gscepmges} by $\Delta u_3(\zm)$ and integrate it over the half-line. The result enables us to express the right-hand side of Eq.~\eqref{eq:X3intermsofG} in terms of the square of the semi-bound state $\regsol_0(\zm)$. It conjunction with Eq.~\eqref{eq:Qdef}, it yields
\begin{equation}\label{eq:Jgr3itofsbs}
Q(0)=\frac{-1}{4}\left\{\frac{\zeta(3)}{2}+\int_0^\infty\rmd\zm\,\Delta u_3(\zm)\,[\regsol^2_0(\zm)-1]\right\}
\end{equation}
with
\begin{equation}\label{eq:J3def}
J_3\equiv \int_0^\infty\rmd\zm\,\Delta u_3(\zm)\,[\regsol^2_0(\zm)-1],
\end{equation}
where $\Delta u_3(0)=\zeta(3)$ was used. 
Hence we must show that
\begin{equation}\label{eq:J3res}
J_3=-\zeta(3)/2,
\end{equation}

By adding and subtracting appropriate terms to the integrand and choosing a finite value $R$ of the upper integration limit whose limit $R\to\infty$ must be considered, one can rewrite $J_3$ as
\begin{eqnarray}\label{eq:J3}
J_3&=&\zeta(3)\,J_\varphi-L_3(\infty)\nonumber\\ &&\strut +\lim_{R\to\infty}\bigg\{\zeta(3)\,R+\int_0^R\rmd\zm\,u_3(\zm)\,\regsol_0^2(\zm)\bigg\}.
\end{eqnarray}

To evaluate the integral on the right-hand side, we can exploit the exponential decrease~\eqref{eq:E1asres}
 of the lowest eigenvalue $E_1(x)/L^2$ in the limit $x\to-\infty$ [cf.\ Eq.~\eqref{eq:eigvalfctscal}]. This means that $[\mathcal{H}_{v_*}\oneigfct_1(\zL,x)]/\oneigfct_1(\zL,x)=O(|x|\rme^{-|x|})$ as $x\to-\infty$. From Eqs.~\eqref{eq:varphipsi1} and \eqref{eq:varphiinf}, we know that the associated eigenfunction $\oneigfct_1(\zm/|x|,x)$ approaches the semi-bound state $\regsol_0(\zm)$ when $x\to-\infty$ at fixed $\zm$ so that $\oneigfct_1(\zL,-\infty)=\regsol_0(\infty)=1$. The leading large-$|x|$ corrections to the latter limiting value result from the potential terms given on the right-hand side of Eq.~\eqref{eq:vstarlargexm} and hence are of the same orders $|x|^{-1}$ and $|x|^{-2}$. Allowing for a change of the  normalization of the eigenfunction, we therefore make the ansatz 
\begin{eqnarray}\label{eq:ansf1hat}
 \hat{\oneigfct}_1(\zL,x)&=&\left[1+b_1/|x|+O\big(|x|^{-2}\big)\right]\oneigfct_1(\zL,x)\nonumber\\ &=&1+\frac{g_1(\zL)}{|x|}+\frac{g_2(\zL)}{|x|^2}+\ldots
\end{eqnarray}
for the modified eigenfunction $\hat{\oneigfct}_1(\zL,x)$. For convenience, we choose $b_1$ such that
\begin{equation}
g_1(\zL)=[\psi(\zL)+\psi(1-\zL)+2\gamma_{\text{e}}].
\end{equation}
This choice implies that
\begin{equation}
g_1(\zL)\mathop{\aseq}\limits_{\zL\to0}-\frac{1}{4\zL}-\frac{\zeta(3)}{2}\,\zL^2+O(\zL^4),
\end{equation}
ensuring the absence of a $\zL^0$  term in this expansion.

Substituting of the ansatz~\eqref{eq:ansf1hat}  and the large-$|x|$ form~\eqref{eq:vstarlargexm} of $v_*$ into the differential equation
$\mathcal{H}_{v_*} \hat{\oneigfct}_1(\zL,x)=O(|x|\rme^{-|x|})$,  we can solve for $g_1''$ and $g_2''$ to obtain
\begin{eqnarray}
g_1''(\zL)&=&t_1(\zL),\\
g_2''(\zL)&=&t_1(\zL)\,g_1(\zL)+t_2(\zL).\label{eq:g22der}
\end{eqnarray}
We  now integrate Eq.~\eqref{eq:g22der} over the inner region $[\tilde{b}/|x|,1/2]$ and add and subtract terms to the integrand to obtain finite integrals over $[0,1/2]$. This gives 
\begin{equation}
g_1'(\tilde{b}/|x|)=\bigg(\frac{1}{8}-\alpha_2\bigg)\bigg(\frac{8}{3}-\frac{|x|^3}{3\tilde{b}^3}\bigg)-L_2(0)-d_{1,3}
\end{equation}
with
\begin{equation}
d_{1,3}=\int_0^{1/2}\rmd\zm\left[g_1''(\zL)\,g_1(\zL)-\frac{1}{8\zL^4}-\frac{\zeta(3)}{2\zL}\right].
\end{equation}
The latter integral can be written in terms of the parameters $d_{1,1}$ and $d_{1,2}$ introduced in Eqs.~\eqref{eq:d11def} and \eqref{eq:d12def}. Upon expressing the integrand of the combination $d_{1,3}+(d_{1,1}+d_{1,2})/4$ in terms of the function $w(\zm)$ and its derivatives, one can determine its antiderivative and compute the integral. The result yields the relation
\begin{equation}\label{eq:d13res}
d_{1,3}=\frac{d_{1,1}+d_{1,2}}{4}-\gamma_{\text{E}}+\ln4-\frac{\zeta(3)}{4}\bigg(\frac{1}{3}+\gamma_{\text{E}}+\ln2\bigg).
\end{equation}

Turning to the rescaled analog of $\hat{\oneigfct}_1$,
\begin{equation}
\tilde{\oneigfct}_1(\zm,x)= \hat{\oneigfct}_1(\zm/|x|,x),
\end{equation}
we note that the leading large-$|x|$ correction  to its ${x\to-\infty}$ limit is due to the $|x|^{-3}$ term of $v_*(\zm;|x|,-1)$ in Eq.~\eqref{eq:u3smallzm}. We therefore write
\begin{equation}\label{eq:tildef1largex}
\tilde{\oneigfct}_1(\zm,x)=\regsol_0(\zm)+\frac{h_1(\zm)}{|x|^3}+\ldots
\end{equation}
and substitute this ansatz into the differential equation for $\tilde{\oneigfct}_1$. Taking into account the large-$\zm$ behavior  $v_*(\zm,\infty,x)\aseq -(2z^3)^{-1}-\alpha_2\zm^{-4}$ implied by Eq.~\eqref{eq:Aminlargezm}] and that of  $u_3$ given in Eq.~\eqref{eq:u3largezm}, one concludes that $h_1(\zm)$ behaves as
\begin{equation}\label{eq:h1largezm}
h_1(\zm)\mathop{\aseq}\limits_{\zm\to\infty} -\frac{\zeta(3)}{2}\,\zm^2+a_1\zm
+O(\ln^2\zm),
\end{equation}
where the absence of a contribution $\propto \zm\ln\zm$ requires that $\beta_2$ takes its self-consistent value~\eqref{eq:beta2res}.

To determine $a_1$, we match the limiting behavior of the derivative $\partial_\zm\tilde{\oneigfct}_1(\tilde{b},x)$ for large $\tilde{b}$ that follows from Eqs.~\eqref{eq:tildef1largex}, \eqref{eq:regsol0as}, and  \eqref{eq:h1largezm} with the behavior of $\hat{\oneigfct}_1(\zL,x)$ at the matching point $\zL_{\text{mp}}=\tilde{b}/|x|$. One finds that the $x$-independent term $\propto \tilde{b}^{-2}$ and $\tilde{b}^{-3}$ as well as the ones $\propto \tilde{b}|x|^3$ of both expressions are consistent and that the coefficient of the $\tilde{b}$-independent $|x|^{-3}$ term is given by
\begin{equation}\label{eq:a1}
a_1=\frac{1}{3}(1-8\alpha_2)-L_2(0)-d_{1,3}.
\end{equation}

We are now ready to express the integral in Eq.~\eqref{eq:J3} in terms of this coefficient. To this end we equate the $|x|^{-3}$ terms of  $\mathcal{H}_{v_*}\tilde{\oneigfct}_1(\zm,x)$ to zero. This yields $u_3(\zm)\,\regsol_0(\zm)=-\mathcal{H}_{v_\infty}\,h_1(\zm)$. Multiplying the result by $\regsol_0$, integrating by parts, one is led to
\begin{eqnarray}\label{eq:usregsol02}
\int_0^R\rmd\zm\, u_3(\zm)\,\regsol_0^2(\zm)&=&\left[\regsol_0(\zm)\,h_1'(\zm)-\regsol'_0(\zm)\,h_1(\zm)\right]_0^R\nonumber\\ &=&\zeta(3)\left[\frac{3}{8}-R\right]+a_1+O(1/R),\nonumber\\
\end{eqnarray}
where we have again used Eqs.~\eqref{eq:regsol0as} and \eqref{eq:h1largezm}. Note that the contribution from the lower integration limit $\zm=0$ vanishes because $\regsol_0(\zm)\asprop \sqrt{\zm}$ and $h_1(\zm)\asprop\zm^{5/2}$ as $\zm\to0$, respectively. To see the latter, note that $u_3(\zm\to0)=o(1)$ and hence does not contribute to the pole part of $v_*$. However, inserting the ansatz $\oneigfct_1(\zm,x)=\sqrt{\zm}[1+r_1\zm+r_2\zm^2+\ldots]$ into the differential equation for $\oneigfct_1$, one finds that the regular parts of $v_*$ do not effect $r_1$, which implies the stated small-$\zm$ behavior of $h_1$.

Combining Eqs.~\eqref{eq:J3}, \eqref{eq:a1} and \eqref{eq:usregsol02} gives
\begin{eqnarray}\label{eq:J3d13}
J_3&=&\zeta(3)\left[J_\varphi+\frac{3}{8}\right]-L_2(0)-L_3(\infty)-d_{1,3}+\frac{1-8\alpha_2}{3}\nonumber\\
&=&\zeta(3)\left[J_\varphi-\frac{1}{8}+\frac{d_1}{4}+\frac{1}{2}\ln2\right]-d_{1,3}+\frac{1}{3}.
\end{eqnarray}
To obtain the second line, we eliminated $L_2(0)+L_3(0)+8\alpha_2/3$ by exploiting the relation~\eqref{eq:magicrel}.

We can now use Eqs.~\eqref{eq:Jgr3itofsbs} and \eqref{eq:J3def}  to eliminate $Q(0)$ on the right-hand side of Eq.~\eqref{eq:d1intermediate} in favor of $J_3$.  This gives us $d_1$ expressed in terms of $d_{1,1}$, $d_{1,2}$, and $J_3$. Upon inserting the result into Eq.~\eqref{eq:J3d13} and  using the exact result for $J_\varphi$ given in Eq.~\eqref{eq:Jvarphires} and Eq.~\eqref{eq:d13res}, we can solve for $J_3$. All contributions involving $d_{1,1}$ and $d_{1,2}$ cancel out. One obtains the result for $J_3$ given in Eq.~\eqref{eq:J3res}, which implies that $Q(0)=0$. Inserting this result into Eq.~\eqref{eq:d1intermediate}, we combine the integrals that the coefficients $d_{1,2}$ and $d_{1,2}$ involve into a single integral and integrate by parts using $[f(z)f'''(z)-f'(z)f''(z)]'=f(z)f^{(4)}(z)-[f''(z)]^2$,
 with $f(z)=\ln \Gamma(z)-\ln \Gamma(1-z)$.  This gives \footnote{We are indebed to  A.\ Hucht for pointing out to us that our original result for $d_1$ given by the right-hand side of Eq.~\eqref{eq:d1intermediate} can be rewritten in this fashion.}
\begin{equation}\label{eq:AHd1}
d_1=\frac{2}{3\zeta(3)}+2[\gammaE+\ln(2/\pi)]-\frac{J_4}{4 \zeta(3)}
\end{equation}
with 
\begin{equation}\label{eq:J4def}
J_4\equiv \int_0^{1/2}\rmd{z}\{[\psi'(z)-\psi'(1-z)]^2-s(z)\},
\end{equation}
where
\begin{equation}\label{eq:sdef}
s(z)=\frac{1}{z^4}-\frac{8\,\zeta(3)}{z}.
\end{equation}
The square of the difference of $\psi'$ functions in  the integrand of $J_4$ can be decomposed into  the square of the sum of these functions,  which reduces to $\pi^4/\sin^4(\pi z)$, and a remainder. As a consequence, we arrive at
\begin{eqnarray}\nonumber
J_4&=&\int_0^{1/2}\rmd{z}\,\left[
\frac{\pi^4}{\sin^4(\pi z)}-\frac{1}{z^4}-\frac{2\pi^2}{3 z^2}\right]-2J_5\\
&=&\frac{4}{3}\,(2+\pi^2)-2\,J_5,\label{eq:J4new}
\end{eqnarray}
where
\begin{equation}\label{J5def}
J_5=2  \int_0^{1/2}\rmd{z}\,[\psi'(z)\psi'(1-z)-s_1(z)]
\end{equation}
with
\begin{equation}\label{eq:s1def}
s_1(z)=\frac{\pi^2}{6z^2}+\frac{2}{z}\,\zeta(3).
\end{equation}

It is straightforward to rewrite the last integral as
\begin{equation}\label{eq:J5}
J_5=J_6+\frac{\pi^2}{3}+4\,\zeta(3)\ln 2,
\end{equation}
where 
\begin{equation}
J_6=\int_0^{1}\rmd{z}\,[\psi'(z)\psi'(1-z)-s_1(z)-s_1(1-z)].
\end{equation}
The integral $J_6$, in turn, can be represented as the  limit $\epsilon\to0+$ of the contour integral 
\begin{equation}\label{eq:J6}
J_6=-\lim_{\epsilon\to0+}\frac{J_7(\epsilon)}{2\pi \rmi\epsilon}
\end{equation}
with
\begin{equation}
J_7(\epsilon)=\oint_C\rmd{z}\,\left(\frac{z}{1-z}\right)^\epsilon  \psi'(z)\,\psi'(1-z),
\end{equation}
where the integration contour $C$ runs around the branch cut $[0,1]$ of the integrand in the 
clock-wise direction. Recasting the integral $J_7(\epsilon)$ into the sum of residua of the integrand at the poles located in the exterior of the integration path $C$, one obtains 
\begin{eqnarray}\label{eq:J7}
\lefteqn{\frac{J_7(\epsilon)}{2\pi \rmi}}&&\nonumber\\
&=&\rme^{\rmi \pi \epsilon}\sum_{n=2}^\infty
\bigg\{
\left(\frac{n}{n-1}\right)^\epsilon\bigg[\psi''(n)+\epsilon\,\psi'(n)\left(
\frac{1}{n}-\frac{1}{n-1}
\right)\bigg]\nonumber\\
&&\strut +
\left(\frac{n-1}{n}\right)^\epsilon\bigg[-\psi''(n)+\epsilon\,\psi'(n)\left(
\frac{1}{n}-\frac{1}{n-1}
\right)\bigg]
\bigg\}.\nonumber\\
\end{eqnarray}
After substitution of this result into \eqref{eq:J6} and proceeding to the limit $\epsilon\to0+$, one finds
\begin{eqnarray}\nonumber
J_6&=&2\psi'(2)+2\sum_{n=3}^\infty
\frac{-1+2 \ln(n-1)}{(n-1)^3}\\
&=&\frac{\pi^2}{3}-2\,\zeta(3)-4\,\zeta'(3).\label{eq:J7a}
\end{eqnarray}

In deriving the first line of Eq.~\eqref{eq:J7a} we have rearranged the summation on the 
right-hand side of Eq.~\eqref{eq:J7}, and used the first and the second derivatives of the well-known 
relation $\psi(z)-\psi(z-1)=1/(z-1)$. Substitution of this result into Eqs.~\eqref{eq:J5}, \eqref{eq:J4new}, and \eqref{eq:AHd1} finally leads to the result given in Eq.~\eqref{eq:d1res}.


\end{document}